\documentclass[a4paper,11pt]{article}
\pdfoutput=1 

\usepackage{jcappub} 
\usepackage{soul}

\usepackage{amsfonts}
\usepackage{amsmath}
\usepackage{amsthm}
\usepackage{amssymb}
\usepackage{array}
\usepackage{dcolumn}
\usepackage{placeins}
\usepackage[svgnames]{xcolor}
\usepackage{graphics}
\usepackage{tikz}
\usepackage{graphicx}
\usetikzlibrary{positioning}
\usepackage{caption}
\usepackage{subcaption}
\usepackage{adjustbox}
\usepackage{gensymb}
\usepackage{breqn}
\usepackage{braket}
\usepackage{enumitem}
\usepackage{makecell}
\usepackage{ragged2e}
\usepackage{comment}
\usepackage{url}
\usepackage{placeins}
\usepackage{blindtext}
\usepackage{float}
\usepackage[colorlinks=true,citecolor=blue,urlcolor=blue]{hyperref}
\usepackage{fontawesome}
\usepackage{cleveref}
\usepackage{upgreek}
\crefname{section}{section}{Sections}
\crefname{figure}{figure}{Figures}
\crefname{equation}{eq.}{eqs.}
\crefname{table}{table}{tables}
\crefname{appendix}{appendix}{appendices}
\usepackage{tikz-feynman}
\usepackage{amsmath}

\arxivnumber{2606.25050}

\newcommand\be{\begin{equation}}
\newcommand\ee{\end{equation}}
\newcommand\bea{\begin{eqnarray}}
\newcommand\eea{\end{eqnarray}}

\definecolor{nicegreen}{rgb}{0., 0.75, 0.46}

\title{Probing scalar non-standard neutrino interactions using high-energy astrophysical neutrinos}

\author[a,b]{Ankur Verma,}
\author[c]{Carlos A. Arg\"{u}elles,}
\author[d,e]{P. S. Bhupal Dev,}
\author[b]{Bhaskar Dutta,}
\author[f]{Ivan Martinez-Soler}

\affiliation[a]{Department of Physics, University of South Dakota, Vermillion, SD 57069, USA}

\affiliation[b]{Mitchell Institute for Fundamental Physics and Astronomy, Department of Physics  and Astronomy, 
Texas A$\&$M University, College Station, TX  77843,  USA}

\affiliation[c]{Department of Physics \& Laboratory for Particle Physics and Cosmology,
Harvard University, Cambridge, MA 02138, USA}

\affiliation[d]{Department of Physics and McDonnell Center for the Space Sciences,
Washington University, St. Louis, MO 63130, USA}

\affiliation[e]{PRISMA$^{++}$ Cluster of Excellence \& Mainz Institute for Theoretical Physics, 
Johannes Gutenberg-Universit\"{a}t Mainz, 55099 Mainz, Germany}

\affiliation[f]{Institute for Particle Physics Phenomenology, Durham University, South Road, DH1 3LE, Durham, UK}

\emailAdd{averma1@tamu.edu}
\emailAdd{carguelles@g.harvard.edu}
\emailAdd{bdev@wustl.edu}
\emailAdd{dutta@physics.tamu.edu}
\emailAdd{ivan.j.martinez-soler@durham.ac.uk}

\abstract{Scalar non-standard interaction (SNSI) of neutrinos contributes as modifications to the neutrino mass matrix in the oscillation Hamiltonian and can induce a small active-sterile mass splitting due to the matter effect induced by the relic neutrino background via a Majorana-type interaction. 
This framework leads to pseudo-Dirac behavior of neutrinos, introducing rich phenomenology in neutrino oscillations, particularly for high-energy astrophysical neutrinos.
We show that these hyperfine active-sterile splittings imprint themselves in two complementary ways on high-energy astrophysical neutrino flux, namely, in modifying the flavor composition and energy distribution.
In this work, we perform both flavor and spectral analyses of the high-energy astrophysical neutrino flux to probe SNSI.
We confront the predicted flavor ratios with current IceCube measurements and with the projected reach of next-generation detectors such as IceCube-Gen2.
For the spectral analysis, we use the diffuse-flux ESTES (tracks) and cascade data sets, together with point-source spectral shape analysis based on a recent catalog of neutrino-bright sources.
The regions excluded by the combined flavor and spectral analyses are translated into limits on the underlying SNSI parameters, namely, Yukawa couplings and scalar mass, providing new sensitivities on the SNSI parameter space for ultra-light mediators.
}  

\begin{document} 
\maketitle
\flushbottom

\section{Introduction}

Various solar, atmospheric, reactor, and accelerator neutrino experiments have now firmly established the 3-neutrino oscillation paradigm. To explain the observed solar and atmospheric mass-squared differences,   at least two of the Standard Model (SM) neutrinos must have nonzero mass, thereby necessitating the existence of some beyond-the-SM (BSM) physics. 
%Solar and atmospheric experiments have precisely measured the differences in their squared masses~\cite{Super-Kamiokande:1998kpq, SNO:2002tuh, KamLAND:2002uet, T2K:2011ypd, DoubleChooz:2011ymz, DayaBay:2012fng, RENO:2012mkc,IceCubeCollaboration:2024ssx}.
However, the nature of neutrino mass  remains unknown. %Given that neutrino masses are at least six orders of magnitude smaller than the mass of the lightest charged fermion, presenting a puzzle that requires physics beyond the Standard Model to explain these minuscule masses and their mixing patterns.

Fundamentally, neutrinos may admit either Dirac or  Majorana mass terms, or a combination of both, depending on the mechanism behind their mass generation.
A Dirac mass requires the existence of right-handed neutrino fields. 
In contrast, a Majorana mass term for the left-handed neutrino fields can be realized by introducing an $SU(2)_L$-triplet scalar or by including higher-dimensional, non-renormalizable terms in the Lagrangian, potentially originating from new physics at higher energy scales. 
Alternatively, it is also possible to have a Lagrangian that generates both Dirac and Majorana mass terms, leading to a hybrid mass structure for neutrinos. 
Typically, these mass-generation mechanisms, especially those involving extensions of the Higgs sector, naturally introduce new types of interactions for neutrinos beyond the standard weak interactions, which are commonly referred to as non-standard interactions (NSIs)~\cite{Wolfenstein:1977ue, Farzan:2017xzy, 
Proceedings:2019qno}.
While there is currently no concrete experimental evidence for such NSIs~\cite{MINOS:2016sbv,Super-Kamiokande:2017yvm,ANTARES:2021crm,IceCube:2022ubv,Super-Kamiokande:2022lyl, DayaBay:2024hya,NOvA:2024lti,KM3NeT:2024pte, BOREXINO:2026owb,RES-NOVA:2026fii}, they remain well motivated because they can arise naturally in many neutrino mass-generation frameworks~\cite{Babu:2019mfe} and their effects on neutrino oscillation and scattering observables~\cite{Coloma:2019mbs,Dutta:2020che,Coloma:2023ixt,Coloma:2024ict, Gehrlein:2024vwz} make them a useful phenomenological probe of the underlying new physics in a wide range of energies and baselines. Such interactions have also been discussed in the context of possible long-baseline oscillation anomalies and the NOvA-T2K tension~\cite{Chatterjee:2020kkm,Denton:2020uda, Cherchiglia:2023ojf,Chatterjee:2024kbn}.

Most of the NSI studies so far have focused on the vector current, which is inspired by the SM weak interaction and is parametrized by the dimension-6 Wolfenstein operator of the form   $\varepsilon_{\alpha\beta}G_F(\bar{\nu}_\alpha\gamma^\mu P_L\nu_\beta)(\bar{f}\gamma_\mu P_{L,R}f)$~\cite{Wolfenstein:1977ue}, where $P_{L,R}=(1\mp \gamma_5)/2$ are the chirality projection operators, $f=\{e,u,d\}$ denotes the matter field, and   $\varepsilon_{\alpha\beta}$ governs the size of the neutral-current vector NSI with respect to the Fermi coupling $G_F$. It induces additional matter effects for neutrino propagation in a medium. A similar charged-current vector NSI term $\varepsilon_{\alpha\beta}G_F(\bar{\nu}_\alpha\gamma^\mu P_L\ell_\beta)(\bar{f}\gamma_\mu P_{L,R}f')$~\cite{Grossman:1995wx} affects the neutrino production and detection at experiments. These vector NSIs usually arise due to either vector mediators (similar to the SM case) or charged-scalar mediators (after Fierz transformation)~\cite{Babu:2019mfe}.  

On the other hand, NSI induced by a neutral scalar mediator leads to an effective dimension-6 operator of the form~\cite{Bergmann:1999rz,Ge:2018uhz,Babu:2019iml}
\begin{align}
    -{\cal L}^{\rm scalar}_{\rm eff} = \frac{y_fy_{\alpha\beta}}{m_\phi^2}(\bar{\nu}_\alpha\nu_\beta)(\bar{f}f) \, , 
\end{align}
where $y_f$ and $y_{\alpha\beta}$ are the Yukawa couplings of the scalar mediator $\phi$ to (matter) fermions and neutrinos, respectively. This operator cannot be Fierzed into a vector current, and therefore, does not contribute to the matter potential. Instead, it appears as a medium-dependent correction to the neutrino mass which, in the non-relativistic limit, is given by~\cite{Babu:2019iml}
\begin{align}
    (\delta M_\nu)_{\alpha\beta} \simeq \frac{y_{\alpha\beta} y_f}{m_\phi^2} (n_f+n_{\bar{f}}) \, ,
    \label{eq:nu_mass}
\end{align}
where $n_{f(\bar{f})}$ is the total number density of the fermions (anti-fermions) in the medium. Such scalar NSIs (SNSIs) lead to interesting phenomenological consequences~\cite{Ge:2018uhz, 
Khan:2019jvr,
Babu:2019iml,
Smirnov:2019cae,
Venzor:2020ova,
Escrihuela:2021mud,
Chaves:2021kxe,
Medhi:2021wxj,
Schwemberger:2022fjl, 
Medhi:2022qmu,
Dutta:2022fdt,
Denton:2022pxt,
Gupta:2023wct, 
Medhi:2024rsi,
Schwemberger:2023hee,
Singha:2023set,
Sarker:2023qzp,
ESSnuSB:2023lbg,
Dutta:2024hqq,
Sarker:2024ytu,
Denton:2024upc,
Bezboruah:2024yhk, 
Pusty:2024hcn, DeRomeri:2024iaw,
Yadav:2024qav, Das:2025zts, Dutta:2025rxh, Choubey:2026jiq,
Alves:2026ydc,
Flores:2026vbx,
Akita:2023iwq}, significantly
different from those induced by vector NSIs.

The SNSI studies so far have mostly considered the correction to neutrino mass [cf.~Eq.~\eqref{eq:nu_mass}] induced by matter fermions (electrons and/or nucleons) in the propagating medium, which is relevant for neutrino propagation in the Earth and the Sun, and to some extent, also in supernovae and in the early Universe. In this paper, we consider a minimal scenario where the scalar $\phi$ only interacts with neutrinos at leading order. Such neutrinophilic scalars can naturally occur in many BSM scenarios, such as in models of neutrino self-interactions~\cite{Bialynicka-Birula:1964ddi,Bardin:1970wq,Berryman:2022hds} and Majoron models~\cite{Chikashige:1980qk, Chikashige:1980ui,Gelmini:1980re,Schechter:1981cv}, and lead to many interesting laboratory, astrophysical and cosmological signatures~\cite{Barger:1981vd, 
Kolb:1987qy,
Bilenky:1992xn,
Bilenky:1994ma,
Beacom:2002cb,
Farzan:2002wx,
Beacom:2004yd,
Lessa:2007up,
Laha:2013xua,
Cyr-Racine:2013jua,
Ioka:2014kca,
Ng:2014pca,
Forastieri:2015paa,
Araki:2015mya,
Pasquini:2015fjv,
Shoemaker:2015qul,
Heurtier:2016otg,
Das:2017iuj,
Huang:2017egl,
Berryman:2018ogk,
Blum:2018ljv,
Brune:2018sab,
Kreisch:2019yzn,
Barenboim:2019tux,
Murase:2019xqi,
Park:2019ibn,
Forastieri:2019cuf,
Blinov:2019gcj,
Lei:2019nma,
deGouvea:2019qaz,
DeGouvea:2019wpf,
Shalgar:2019rqe, EscuderoAbenza:2020cmq,
Bustamante:2020mep,
Grohs:2020xxd,
Brdar:2020nbj,
Lyu:2020lps,
Deppisch:2020sqh,
Creque-Sarbinowski:2020qhz,
Das:2020xke,
Brinckmann:2020bcn,
Esteban:2021tub,
Dev:2021axj,
Ge:2021lur,
Hufnagel:2021pso,
Chichiri:2021wvw,
Smirnov:2022sfo,
deGouvea:2022cmo,
Das:2022xsz,
Akita:2022etk,
Chang:2022aas,
Kreisch:2022zxp,
Taule:2022jrz,
RoyChoudhury:2022rva,
Chen:2022kal,
Abbar:2022jdm,
Fiorillo:2022cdq,
Coyle:2022bwa,
Das:2023npl,
Venzor:2023aka,
Doring:2023vmk,
Sandner:2023ptm,
Li:2023puz,
Fiorillo:2023ytr,
Fiorillo:2023cas,
Wu:2023twu,
Camarena:2023cku,
He:2023oke,
Camarena:2024daj,
Franklin:2024amy, 
Agashe:2024owh,
Craig:2024tky,
Telalovic:2024cot,
Huang:2024tbo,
Dev:2024ygx,
Bai:2024kmt,
Suliga:2024nng,
Benso:2024qrg,
Zhang:2024meg,
deLima:2024ohf,
Ehring:2024mjx,
Liu:2024ywd,
Kaplan:2024ydw,
Foroughi-Abari:2025upe,
Wang:2025qap,
Poudou:2025qcx,
He:2025jwp,
Leal:2025eou,
Libanore:2025ack,
He:2025bex,
Noriega:2025ulc,
Das:2025asx,
Natwariya:2025ftu,
Dev:2025tdv,
Montefalcone:2025ibh,
Foroughi-Abari:2025mhj,
Whitford:2025dmq,
PandaX:2025tls,
Boudjema:2025okq,
deVries:2025hqa,
Machado:2025ltu,
Perez-Castro:2026muj,
Parashari:2026dxo,
Pal:2026cgj}. As we show in this paper, in the presence of such scalar-mediated neutrino self-interactions, a correction to neutrino mass can be induced for propagating neutrinos in the cosmic relic neutrino background (C$\nu$B), which could have interesting observable consequences.     
 
In particular, our study focuses on high-energy astrophysical neutrinos and on assessing how their flavor composition and energy spectra get affected by SNSI. 
We assume that neutrinos have a baseline Dirac mass, while a Majorana-type neutrino–scalar coupling generates a tiny Majorana mass through the C$\nu$B matter effect during propagation. 
This generates a small active-sterile mass splitting and makes the neutrinos pseudo-Dirac~\cite{Wolfenstein:1981kw,Petcov:1982ya,Valle:1983dk,Doi:1983wu}, thereby inducing active-sterile oscillations during their propagation over astrophysical  baselines~\cite{Giunti:1992hk,Crocker:1999yw,
Chang:1999pb,
Nir:2000xn,
Kobayashi:2000md,
Crocker:2001zs,
Balaji:2001fi,
Keranen:2003xd,
Beacom:2003eu,
Cirelli:2004cz,
deGouvea:2009fp, 
Esmaili:2009fk,
Esmaili:2012ac,
Joshipura:2013yba,
%Shoemaker:2015qul,
Anamiati:2017rxw,
Brdar:2018tce,
Anamiati:2019maf,
DeGouvea:2020ang,
Martinez-Soler:2021unz,
deGouvea:2021ymm,
Ansarifard:2022kvy,
Rink:2022nvw,
Carloni:2022cqz,
Franklin:2023diy,
Perez-Gonzalez:2023llw,
Dixit:2024ldv,
Fong:2024msb,
Dev:2024yrg,
Carloni:2025dhv,
Jana:2025vyb,
MacDonald:2025jbm}.
The resulting pseudo-Dirac behavior manifests in two complementary observables for high-energy astrophysical neutrinos: (i) flavor composition and (ii) spectral shape.
Previous studies of SNSIs with astrophysical neutrinos have primarily focused on spectral distortions arising from scattering between high‑energy neutrinos and the C$\nu$B~\cite{Esteban:2021tub,Franklin:2024amy}.

The detection of these high-energy astrophysical neutrinos requires extremely large detectors due to the tiny fluxes involved. IceCube~\cite{IceCube:2016zyt}, the largest neutrino telescope currently operating in the world, serves as a prime example of such a detector. 
Located in the deep Antarctic ice, IceCube spans $\sim 1~\text{km}^3$ instrumented volume and is comprised of photomultiplier tubes embedded in pressure-resisting vessels~\cite{IceCube:2010dpc} that detect Cherenkov radiation produced when high-energy neutrinos interact with nucleons in the ice. 
These interactions primarily occur through deep-inelastic scattering~\cite{Gandhi:1995tf}, which can take two forms: neutral-current (NC), where the neutrino scatters without changing flavor, and charged-current (CC), where the neutrino interacts and changes into a charged lepton. %The final state of these interactions produces a shower of particles, which emits Cherenkov radiation detectable by IceCube's photomultipliers. 

IceCube is sensitive to the neutrino flavor by sorting the events into three distinct morphological categories: tracks, cascades, and double cascades~\cite{IceCube:2015rro}.
Track events, which are primarily produced by CC interactions of muon neutrinos ($\nu_{\mu}$), offer excellent angular resolution but modest energy resolution. 
They are particularly valuable for identifying the location of astrophysical neutrino sources. 
Cascades, on the other hand, result from CC interactions of electron or tau neutrinos ($\nu_e, \nu_{\tau}$) and NC interactions of neutrinos of all flavors. 
While they provide good energy resolution, their angular resolution is poorer than that of track events. 
Double cascades are a specific subset of cascade events, typically produced by CC interactions of tau neutrinos ($\nu_{\tau}$)~\cite{Stachurska:2019wfb,IceCube:2020fpi,IceCube:2024nhk}.
These interactions generate two distinct showers: one from the hadrons produced at the interaction vertex and another from the decay of the energetic tau lepton some distance away. 

The flavor composition of the neutrino flux can be inferred by analyzing these different types of events~\cite{IceCube:2015rro}. 
High-Energy Starting Events (HESE), which include showers, tracks, and double cascades, provide crucial information for understanding the distribution of neutrino flavors from astrophysical sources~\cite{IceCube:2020wum,IceCube:2023fgt}.
By combining data from HESE events with through-going muons, IceCube can estimate the flavor composition of the detected neutrinos, providing deeper insights into the sources and mechanisms behind the observed neutrino flux. 
In addition, pseudo-Dirac splittings (from SNSIs) may introduce additional oscillation phases that can deplete or enhance individual flavors, shifting these ratios~\cite{Beacom:2003eu,Keranen:2003xd, Dev:2024yrg, Fong:2024msb}. 
Ten years of IceCube-Gen2 data are expected to shrink the corresponding likelihood contours in the flavor triangle dramatically, delimiting the region of SNSI parameter space compatible with future flavor measurements~\cite{Song:2020nfh}. 
The same ultra-fine splittings also modulate the energy dependence of the flux via $\sin ^2\left(\delta m^2 L / 4 E\right)$ factors, imprinting energy-dependent “wiggles” on
\begin{itemize}
    \item the diffuse spectra observed in IceCube’s ESTES (tracks) and cascade data sets~\cite{IceCube:2020acn,IceCube:2024fxo,IceCube:2025dlr}; see Ref.~\cite{Carloni:2025dhv} for recent analysis; 
    {\sloppy\item reconstructed spectra of bright point sources, including NGC~1068, PKS~1424+240, TXS~0506+056 and several additional bright candidates~\cite{IceCube:2025lev}; see Ref.~\cite{Carloni:2022cqz} for a sensitivity analysis.\par}
\end{itemize}
In this paper, we show that flavor and spectral information are complementary, providing two semi-independent, high-precision probes of SNSIs in the pseudo-Dirac regime. This is the first analysis that combines flavor triangle information with spectral information (from both diffuse and point source analyses) from high-energy astrophysical neutrino data to constrain SNSI.   

%Combining spectral and flavor information provides two complementary, and partially independent, probes of light SNSIs
The rest of this article is organized as follows:
In \cref{sec:Scalar NSI}, we discuss the SNSI framework and show how it can give rise to pseudo-Dirac mass splittings in the presence of matter effects sourced by the C$\nu$B. 
In \cref{sec:AstroNu}, we review the production mechanisms, standard flavor evolution, and detection channels of high-energy astrophysical neutrinos, setting the stage for the new oscillation effects from SNSI considered here. 
\Cref{sec:analysis} describes the statistical methodology along with the three complementary channels we analyze: the diffuse flux flavor ratios, the diffuse spectral shapes (ESTES and cascade), and a point-source spectral-shape analysis combining track-like and cascade-like events from a small catalog of bright sources.
Our main results and projected sensitivities for IceCube and IceCube-Gen2 are presented in \cref{sec:results}.
We conclude in \cref{sec:conclusion} with a summary and outlook. \Cref{app:inmedium_bounds} discusses the in-medium neutrino mass bounds on SNSI from BBN and supernovae. \Cref{app:exchange-self-energy} derives the finite-density exchange self-energy in the C$\nu$B and
compares it with the tadpole contribution responsible for the pseudo-Dirac splitting.

\section{Pseudo-Dirac framework from SNSI} 
\label{sec:Scalar NSI}
The SNSI contribution to the neutrino oscillation Hamiltonian manifests as a medium-dependent correction to the neutrino mass term, uniquely affecting neutrino oscillation probabilities.
%, which also depends on neutrino absolute mass values.
In this work, we consider a scenario where neutrinos are fundamentally Dirac particles; however, SNSIs can induce effective Majorana mass terms through interactions with background fermions. 
As shown below, the Majorana-type interaction term generates a small Majorana mass for the astrophysical neutrinos via matter effects from C$\nu$B during propagation. 
This mechanism leads to a small active-sterile mass splitting, inducing pseudo-Dirac behavior during the propagation of astrophysical neutrinos, and leading to  observable signatures in high-energy astrophysical neutrino fluxes. 
%For this mechanism to work without assuming that the medium already carries a Majorana condensate, the scalar must have both a Dirac-type scalar interaction and a Majorana-type scalar interaction.%

To model this scenario, we consider the following Lagrangian with Majorana-type Yukawa coupling terms for the left-handed neutrino fields in addition to Dirac-type scalar interaction  and Dirac neutrino mass:
\begin{equation}
\mathcal{L} \supset
-M_{\alpha \beta}^D \bar{\nu}_{R \alpha} \nu_{L \beta}+y_{\alpha \beta}^D \phi \bar{\nu}_{R \alpha} \nu_{L \beta}+\frac{1}{2} y_{\alpha \beta}^M \phi \nu_{L \alpha}^T C \nu_{L \beta}
-\frac{m_\phi^2}{2}\phi^2
+{\rm H.c.}\, ,
\label{eq:lag}
\end{equation}
where $\alpha,\beta=e,\mu,\tau$ are the neutrino flavor indices, $M^D$ is the Dirac mass matrix, $y^D$ and $y^M$ are the Dirac- and Majorana-type scalar Yukawa couplings, $C$ is the charge-conjugation matrix, and $m_\phi$ is the mass of scalar $\phi$.
The Dirac scalar interaction sources a scalar tadpole from the C$\nu$B through the Dirac bilinear, and the Majorana-type scalar interaction converts the resulting scalar expectation value into a small effective Majorana mass.
%In this case, the C$\nu$B acts as a source for $\phi$ through the Dirac bilinear. 
The Euler-Lagrange equation for the scalar field  becomes
\begin{align}
(\Box+m_\phi^2)\phi
=-\left[y^D_{\alpha\beta} \bar{\nu}_{R\alpha}\nu_{L\beta}+\frac{1}{2}y^M_{\alpha\beta}\nu^T_{L\alpha}C\nu_{L\beta}
+{\rm H.c.}\right] .
\label{eq:Euler}
\end{align}
In a homogeneous cosmological background, $\Box\phi\approx 0$, and taking the expectation values on both sides of Eq.~\eqref{eq:Euler}, we get \footnote{Although $\phi$ is ultralight, throughout the parameter range considered
in this work, finite-size effects associated with the long-range interaction are negligible when compared to the cosmological extent of the C$\nu$B}
\begin{equation}
\langle\phi\rangle_{\mathrm{bg}} \simeq-\frac{1}{m_\phi^2}  y_{\alpha\beta}^D\left\langle\bar{\nu}_{R \alpha} \nu_{L \beta}\right\rangle_{\mathrm{C} \nu \mathrm{B}}+\text { H.c. }
\label{eq:phi_med_general}
\end{equation}
Diagrammatically, the background value of $\phi$ comes from the scalar one-point function: the Dirac-type vertex is closed on a C$\nu$B neutrino line, producing a finite-density tadpole proportional to the scalar density $\langle\bar\nu\nu\rangle_{\mathrm{C}\nu\mathrm{B}}$ \cite{Babu:2019iml}.
The medium expectation value is given by 
\begin{align}
    \langle \bar\nu\nu\rangle = \int \frac{d^3{\bf p}}{(2\pi)^3}\frac{m_\nu}{E_p}[f_\nu(p)+f_{\bar\nu}(p)] \, ,
\end{align}
where $f_{\nu,\bar\nu}$ are the phase-space occupation numbers, and $E_p=\sqrt{{\bf p}^2+m_\nu^2}$. If the background neutrinos are relativistic (e.g. supernova or early Universe), $\langle \bar\nu\nu\rangle \propto m_\nu/E_p\ll 1$, which is suppressed by the neutrino mass. However, for a non-relativistic C$\nu$B background, $E_p\approx m_\nu$, and hence, $\langle \bar\nu\nu\rangle\simeq n_\nu+n_{\bar\nu}$.\footnote{Note that the neutrino self-energy correction term due to scalar interaction in Eq.~\eqref{eq:lag} is different from the vector case. For instance, SM weak interactions yield a contribution proportional to $G_F(n_\nu-n_{\bar{\nu}})$~\cite{Notzold:1987ik}. This is because of the vector current in the weak interaction, which makes the medium density go like $\langle \bar\nu\gamma^\mu\nu\rangle$, which in the medium rest frame (assuming nonrelativistic C$\nu$B background) is simply $\langle \bar\nu\gamma^0\nu\rangle=\langle \nu^{\dagger}\nu\rangle=n_\nu-n_{\bar\nu}$. In contrast, for the Dirac-type scalar interaction in Eq.~\eqref{eq:lag}, the medium density goes like $\langle \bar\nu \nu\rangle\simeq n_\nu+n_{\bar\nu}$ in the non-relativistic limit. The Majorana-type term does not contribute to the expectation value in Eq.~\eqref{eq:phi_med_general}, since $\langle \nu^T_{L}C\nu_L\rangle_{\mathrm{C} \nu \mathrm{B}}=0$. % Moreover, for Majorana neutrinos, the vector current identically vanishes, $\bar{\nu}_M\gamma^\mu\nu_M=0$, whereas the scalar part survives, $\bar\nu_M\nu_M\neq 0$.
}
Here $n_\nu$ and $\bar n_\nu$  are the standard cosmological mean C$\nu $B neutrino and antineutrino number densities. More generally, this expression can be recast in terms of the local C$\nu$B overdensity, if such an overdensity exists~\cite{Dev:2024yrg}.

Schematically, from Eq.~\eqref{eq:phi_med_general}, we thus get 
\begin{equation}
\langle\phi\rangle_{\mathrm{bg}} \simeq -\frac{y^{D}_{\rm eff}}{m_\phi^2}\left(n_\nu+n_{\bar{\nu}}\right),
\end{equation}
where we have defined the effective scalar source coupling 
\begin{equation}
y^D_{\text {eff }}\left(n_\nu+n_{\bar{\nu}}\right) \simeq \sum_i\left(U_R^{\dagger} y^D U_L\right)_{i i} \frac{m_i}{E_i}\left(n_{\nu_i}+n_{\bar{\nu}_i}\right)+\text {H.c. }
\end{equation}
Then a propagating neutrino feels an effective mass term due to the scalar interaction from both Majorana and Dirac bilinears. The Majorana and Dirac scalar interaction terms in Eq.~\eqref{eq:lag} give medium-dependent Majorana and Dirac mass corrections given by\footnote{This should be distinguished from the finite-density exchange self-energy contribution discussed in \cref{app:exchange-self-energy}. In the ultralight-mediator regime, its masslike contribution is proportional to $n_\nu-n_{\bar{\nu}}$, while its refractive potential contribution is proportional to $n_\nu+n_{\bar{\nu}}$ but is controlled by the scale $2Em_\nu$ and therefore lacks the $m_\phi^{-2}$ enhancement of the tadpole.
Ref.~\cite{Nieves:2018vxl} instead considered the refractive potential in the heavy mediator limit with the leading contribution proportional to $n_{\bar{\nu}}-n_\nu$.
}
% \begin{align}
% \frac{1}{2} y_{\alpha \beta}^M\langle\phi\rangle_{\mathrm{bg}} \nu_{L \alpha}^T C \nu_{L \beta} \equiv &\frac{1}{2} \delta M_{\alpha \beta}^L \nu_{L \alpha}^T C \nu_{L \beta} \\
% %\quad \text{and} \quad 
% y_{\alpha \beta}^D\langle\phi\rangle_{\mathrm{bg}} \bar{\nu}_{R \alpha} \nu_{L \beta} \equiv & \delta M_{\alpha \beta}^D \bar{\nu}_{R \alpha} \nu_{L \beta},
% \end{align}
% so that 
\begin{align}
\delta M_{\alpha \beta}^L=& y_{\alpha \beta}^M\langle\phi\rangle_{\mathrm{bg}} \simeq-\frac{y_{\alpha \beta}^M y^D_{\text {eff }}}{m_\phi^2}\left(n_\nu+n_{\bar{\nu}}\right) , \\
% \end{align}
% and
% \begin{equation}
%  \quad
\delta M_{\alpha \beta}^D=&y_{\alpha \beta}^D\langle\phi\rangle_{\mathrm{bg}}\simeq-\frac{y_{\alpha \beta}^D y^D_{\text {eff }}}{m_\phi^2}\left(n_\nu+n_{\bar{\nu}}\right).
\end{align}

For Dirac neutrinos with bare mass matrix $m_D$, the full $6\times6$ neutrino mass matrix in the presence of SNSI is then given by
\begin{equation}
\mathcal{M}_{\mathrm{eff}}=\left(\begin{array}{cc}
\delta M^L & M^D+\delta M^D \\
\left(M^D+\delta M^D\right)^T & 0
\end{array}\right).
\label{eq:mass_matrix}
\end{equation}
%Here $\delta M^L$ is the medium-induced Majorana mass matrix generated by the Majorana-type scalar interaction, while $\delta M^D$ is the corresponding medium-induced correction to the Dirac mass matrix. 
Both corrections are assumed to be small compared to the vacuum Dirac mass matrix:
\begin{equation}
\left|\delta M^L\right|,\left|\delta M^D\right| \ll\left|M^D\right| .
\end{equation}
The Majorana block $\delta M^L$ is responsible for splitting each Dirac neutrino into a quasi-Dirac pair, whereas $\delta M^D$ mainly shifts the common Dirac mass of the pair.\footnote{In principle, a small correction to the right-handed neutrino Majorana mass $\delta M^R$, i.e., the (2,2) block entry in Eq.~\eqref{eq:mass_matrix} also leads to a quasi-Dirac neutrino scenario, but not SNSI in the active sector. Therefore, we do not consider a Majorana-type coupling of $\nu_R$ to $\phi$ in the Lagrangian~\eqref{eq:lag}.} 

To obtain the pseudo-Dirac mass splittings, we first diagonalize the Dirac mass matrix as 
\begin{equation}
U_R^{\dagger} M^D U_L=D_m=\operatorname{diag}\left(m_1, m_2, m_3\right)
\end{equation}
ignoring $\delta  M^D$ compared to $M^D$.
The Majorana perturbation in the vacuum Dirac mass basis is then \begin{equation}
\delta M_{\mathrm{mass}}^L=U_L^T \delta M^L U_L \, .
\end{equation}
At first order in perturbation theory,  the pseudo-Dirac splitting of the $i$-th pair is controlled by the diagonal element 
\begin{equation}\delta m_i \equiv\left(U_L^T \delta M^L U_L\right)_{i i}
\label{eqn:mi},
\end{equation}
where in the charged-lepton mass basis, $U_L$ is identified with the PMNS matrix. 
For $\left|\delta m_i\right| \ll m_i$, each Dirac mass eigenstate is split into two quasi-degenerate Majorana states,
\begin{equation}
m_{i+} \simeq m_i+\frac{\delta m_i}{2}, \quad m_{i-} \simeq m_i-\frac{\delta m_i}{2}.
\end{equation}
%Diagonalizing Eq.~\eqref{eq:mass_matrix}
 %yields mass eigenstates revealing the resulting mass splittings of the active and sterile states:
Equivalently, at the level of mass-squared eigenvalues, 
\begin{equation}
\begin{aligned}
& m_{i +}^2=m_i^2+\delta (m_i^2)/2 =m_i^2+m_i\left|\delta m_i\right|,\\
& m_{i -}^2=m_i^2-\delta (m_i^2)/2 =m_i^2-m_i\left|\delta m_i\right|.
\end{aligned}
\end{equation}

A non-zero SNSI term thus turns every Dirac mass eigenstate into a quasi-Dirac pair.
The small splitting $\delta (m_i^2)\simeq 2m_i\delta m_i$ drives active-sterile oscillations with astrophysical baselines. Beyond their oscillation phenomenology~\cite{
Kobayashi:2000md,
Keranen:2003xd,
Beacom:2003eu,
deGouvea:2009fp, 
Esmaili:2009fk,
Esmaili:2012ac,
Martinez-Soler:2021unz,
deGouvea:2021ymm,
Ansarifard:2022kvy,
Rink:2022nvw,
Carloni:2022cqz,
Franklin:2023diy,
Perez-Gonzalez:2023llw,
Dixit:2024ldv,
Fong:2024msb,
Dev:2024yrg,
Carloni:2025dhv,
MacDonald:2025jbm}, quasi-Dirac neutrinos have also been connected to mechanisms that could account for the observed baryon asymmetry of the Universe~\cite{Ahn:2016hhq, Fong:2020smz}. They have also been proposed as a possible explanation for the excess diffuse radio background~\cite{Chianese:2018luo,Dev:2023wel,Dev:2025ufo}.

\subsection{UV-complete models of SNSI}
Due to various laboratory, astrophysical and cosmological constraints, the scalar field $\phi$ mediating SNSI interactions must be ultralight and the corresponding Yukawa coupling $y^M$ must be extremely small~\cite{Babu:2019iml}.
Some ultraviolet (UV)-complete realizations of such ultralight scalars in the context of SNSI and quasi-Dirac neutrinos have been discussed in the literature~\cite{Lindner:2001hr,McDonald:2004qx,Ahn:2016hhq,Babu:2019iml, Babu:2022ikf, Dutta:2022fdt, Carloni:2022cqz}.

One possible origin of an ultralight $\phi$ is an extended scalar sector containing additional triplet, doublet, or singlet fields. In this framework, the Higgs potential, through appropriate parameter choices and minimization, can yield an ultralight state as the lightest scalar degree of freedom~\cite{Dutta:2022fdt}. In this setup, the required SNSI interaction is realized through an $SU(2)_L$-triplet scalar with hypercharge $Y=1$. The triplet couples to left-handed lepton doublets $L=(\nu_L,e_L)^T$ via the interaction: 
\begin{equation}
\mathcal{L}\supset\bar{L^{\prime}}_{\alpha}^c\left(y^{\prime}\right)_{\alpha \beta} i \sigma_2 \Delta L_{\beta}^{\prime},
\end{equation}
where $\alpha$ and $\beta$ are the flavor indices and the primed fermions denote interaction states.
After electroweak symmetry breaking, the neutral component of the triplet can mix with other CP-even scalars in the model, so that diagonalization of the scalar sector yields an ultralight physical scalar $\phi$ mediating the SNSIs.

Another possibility is that the $\phi$ mass arises as a thermal correction in a medium where neutrino interactions are relevant.
These corrections depend on the couplings of $\phi$ to other fermions and can naturally yield masses in the sub-eV range~\cite{Babu:2019iml}.

Another well-motivated scenario is that $\phi$ serves as a Majoron, associated with the spontaneous breaking of the  global lepton number symmetry~\cite{Chikashige:1980qk,Chikashige:1980ui,Gelmini:1980re}. 
In this case, the ultralight $\phi$ mass can be protected by an underlying gauge symmetry, which restricts the allowed operators to specific orders, thereby stabilizing the mass scale~\cite{Rothstein:1992rh}.

It is noteworthy that some UV constructions within the string-theory landscape, including Swampland-inspired scenarios, suggest that neutrinos may be fundamentally Dirac particles~\cite{Ooguri:2016pdq,Ibanez:2017kvh,Gonzalo:2021zsp,Casas:2024clw}. Since quantum-gravity effects are generally expected to violate global symmetries, including lepton number, neutrinos that are Dirac at leading order can acquire tiny lepton-number-violating mass terms, effectively becoming quasi-Dirac states. More broadly, any model framework with initially Dirac neutrinos (see e.g., Refs.~\cite{Mohapatra:1987hh,Babu:1988yq,Saad:2019bqf,Babu:2022ikf}) and an exact lepton-number symmetry may be subject to higher-dimensional, non-renormalizable operators induced by quantum gravity. These operators, suppressed by the Planck scale, can generate small mass-squared splittings, providing a natural origin for quasi-Dirac neutrinos.

\section{Flavor evolution of astrophysical neutrinos}
\label{sec:AstroNu}
Astrophysical neutrinos are believed to originate from highly energetic processes mostly occurring in extragalactic environments, such as active galactic nuclei (AGNs)~\cite{Stecker:1991vm,Jain:2026jdh}.
These AGNs act as natural particle accelerators, where neutrinos are primarily produced through two mechanisms: proton-proton ($pp$) collisions and photohadronic ($p\gamma$) interactions. 
In $pp$ interactions, accelerated protons collide with other protons in the medium, while in $p\gamma$ interactions, they interact with high-energy photons present in the surrounding environment. 
Both processes result in the production of high-energy pions, which subsequently decay to yield neutrinos.
The canonical expectation from the complete pion decay chain predicts a flavor composition of (1/3, 2/3, 0) for neutrinos at the source, corresponding to electron, muon, and tau neutrinos, respectively.
In certain astrophysical environments, the flavor composition of neutrinos at the source can deviate from the canonical expectation due to environmental effects.
One notable scenario is the muon-damped case, which arises when the muon produced in pion decay rapidly loses energy through interactions with the surrounding medium, such as synchrotron radiation, before it has a chance to decay. 
In this case, the decay chain halts at the pion stage, resulting in a flavor composition at the source of (0, 1, 0), with only muon neutrinos being produced.

\subsection {Standard oscillation probability}
As the neutrinos propagate to Earth, the source flavor ratios get modified due to standard vacuum oscillations (VO).
Since the standard 3-neutrino oscillations are rapid enough over astrophysical distances, we can only detect the averaged transition probabilities between neutrino flavors, \( \nu_{\alpha} \to \nu_{\beta} \):
\begin{equation}
P_{\alpha \beta}^{\mathrm{VO}}=\sum^3_{i}\left|U_{\alpha i}\right|^2\left|U_{\beta i}\right|^2,
\label{eq:VO}
\end{equation}
where $U$ is the PMNS matrix.
For various source flavor ratios $f_\alpha^s$, the flavor composition at the earth $f_\alpha^\oplus$ is then computed as,
\begin{equation}
\left(f_e^s, f_\mu^s, f_\tau^s\right) \longrightarrow f_\beta^{\oplus}=\sum_{\alpha=e, \mu, \tau} P_{\alpha \beta}^{\mathrm{VO}} f_\alpha^s.
\end{equation}

The flavor ratios measured on Earth depend on the values of the PMNS mixing angles and the CP phase.
For the standard case and assuming pion production, the expected flavor composition at Earth corresponds to $\left(f_e^\oplus, f_\mu^\oplus, f_\tau^\oplus\right)\sim (1/3,1/3,1/3)$.
IceCube's published figures of the flavor ratios of the diffuse astrophysical neutrino flux align with the standard expectation of a $(1/3,1/3,1/3)$ composition as well as with alternative compositions predicted by various new physics scenarios~\cite{IceCube:2025uyt,Abbasi:2025fjc}.
Improved measurements by the next generation of experiments, such as IceCube-Gen2~\cite{IceCube-Gen2:2023rds}, are expected to refine these flavor ratio constraints.

\subsection{Pseudo-Dirac oscillations in the presence of SNSI matter effects}

In the SNSI framework, the matter potential generated by the C$\nu$B induces a hyperfine active–sterile mass splitting that modifies the standard oscillation pattern of astrophysical neutrinos. 
The flavor–transition probability for a neutrino produced in flavor state $\alpha$ and detected as flavor $\beta$ becomes~\cite{Carloni:2022cqz}
%The matter effect due to SNSI would affect the flavor ratio predictions from the high-energy astrophysical neutrinos when compared to the standard oscillation flavor ratio measurements.
\begin{equation}
P_{\alpha \beta}^{\text{SNSI}}
%=\frac{1}{2} \sum_j\left|U_{\alpha j}\right|^2\left|U_{\beta j}\right|^2\left[1+ \cos \left(\int d x \frac{\delta m_j^2}{4 E_\nu}\right)\right]
=\frac{1}{2}\sum_j\left|U_{\alpha j}\right|^2\left|U_{\beta j}\right|^2\left[1+ \cos \left( \frac{2m_j\left|\delta m_j\right|L_{\text{eff}}}{4 E_\nu}\right)\right],
\label{eq:PD_SNSI}
\end{equation}
where $m_j$ is the mass of the $j^{\text{th}}$ mass eigenstate, $\delta m_j$ is as defined in~\cref{eqn:mi}, $E_\nu$ is the neutrino energy as measured at Earth, and
\begin{equation}
L_{\mathrm{eff}}=\int_0^{z_{\rm src}} \frac{d z}{H(z)(1+z)^2},
\end{equation}
is the redshift-corrected baseline from the source at redshift $z_{\rm src}$ to Earth that accounts for the adiabatic expansion of the Universe, with $H(z)$ being the Hubble expansion rate.

In the limit $\delta m_{j}\to0$,~\cref{eq:PD_SNSI} reduces to the familiar three-flavor VO probability of Eq.~\eqref{eq:VO}. 
Conversely, when the phase $2m_j|\delta m_j|L_{\rm eff}/4E_\nu\gg\pi$, the cosine term averages to zero and the flavor ratios become indistinguishable from the standard prediction; maximal deviations occur when that phase is $\mathcal{O}(\pi)$, which determines the preferred range of sensitivity to $\delta m_j$ for a given baseline and energy.

%If $\delta m_j=0$  then there we get back the standard oscillation scenario with three active neutrinos. \\
%-If $\int dx2m_j\left|\delta m_j\right|/(4 E_\nu)>>\pi$, then the cosine term just averages to $0$ leaving no effective difference in the flavor ratio analysis when compared to the standard case. 
%- If $\int dx2m_j\left|\delta m_j\right|/(4 E_\nu)\sim\pi$, it may cause a significant influence on the flavor composition.
To connect this phase directly to the relic background, recall that the pseudo-Dirac splitting for each mass eigenstate is set by the diagonal entries of the SNSI-induced Majorana mass correction in the vacuum-mass basis,
$\delta m_j(z)=\left(U_L^t\,\delta M^L(z)\,U_L\right)_{jj}$ (Eq.~\eqref{eqn:mi}), with $\delta M_{\alpha\beta}(z)\simeq y^M_{\alpha\beta}y^D\,[n_\nu(z)+n_{\bar{\nu}}(z)]/m_\phi^2$.
Therefore $\delta m_j$ inherits the redshift dependence of the C$\nu$B number density $n_\nu(z)$, and because the C$\nu$B scales as $n_\nu(z)=n_{\nu,0}(1+z)^{3}$ with $n_{\nu,0}\simeq 56~\mathrm{cm^{-3}}$, the effective splitting and hence the oscillation phase in~\cref{eq:PD_SNSI} grows with distance. 
Consequently, compared to the standard pseudo-Dirac case with a redshift-independent splitting, the phase accumulates more rapidly for higher-$z$ contributions, and after averaging over the source redshift distribution and the detector’s finite energy resolution, the oscillatory interference is more efficiently washed out than in the usual pseudo-Dirac scenario.
Throughout this work, we do not differentiate between neutrinos and antineutrinos, as current high-energy detectors cannot separate them event-by-event except in the narrow Glashow-resonance window~\cite{IceCube:2021rpz}, which occurs specifically with high-energy $\bar{\nu}_e$.

%-The relic neutrino density evolves with redshift as $n_\nu = n_{\nu,0}(1+z)^3$ making the effective mass splitting distance-dependent. Here
%$$n_{\nu, 0}=\frac{3}{4} \frac{\zeta(3)}{\pi^2} g_\nu T_{\nu, 0}^3 \simeq 112 \mathrm{~cm}^{-3}$$
%-In our analysis, we treat neutrinos and anti-neutrinos as indistinguishable, since high-energy neutrino telescopes typically cannot differentiate them on an event-by-event basis, with the exception of the Glashow resonance, which occurs specifically with high-energy $\bar{\nu}_e$.

%The oscillation and coherence lengths that control the onset and damping of the pseudo-Dirac interference can be written as
%\begin{equation}
%L_{\mathrm{osc}}=\frac{4 \pi E_\nu}{\delta \tilde{m}_j^2} \approx 15 \mathrm{Mpc}\left(\frac{E_\nu}{1 \mathrm{TeV}}\right)\left(\frac{5 \times 10^{-18} \mathrm{eV}^2}{\delta \tilde{m}^2}\right),
%\end{equation}

%and
%\begin{equation}
%L_{\mathrm{coh}}=\frac{4 \sqrt{2 }  E_\nu^2}{\left|\delta \tilde{m}^2\right|} \sigma_x \approx 10^6 \mathrm{Gpc}\left(\frac{E_\nu}{1 \mathrm{TeV}}\right)^2\left(\frac{5 \times 10^{-18} \mathrm{eV}^2}{\delta \tilde{m}^2}\right)\left(\frac{\sigma_x}{10^{-10} \mathrm{~m}}\right).
%\end{equation}
%
%For wavepacke size larger than the size of an atom, the decoherence length is larger than the oscillation length $L_{\mathrm{coh}}>>L_{\mathrm{osc}}$, making decoherence by spatial separation is irrelevant.
%

For intuition, it is useful to translate the SNSI-induced splitting into the corresponding oscillation and coherence lengths  in terms of the SNSI parameters $\delta M_{\alpha\beta}$. The oscillation length of the $j$th pseudo-Dirac pair can be written as
\begin{equation}
L_{\mathrm{osc}}^{(j)}
=\frac{2 \pi E_\nu}{m_j\left|\left(U_L^t\delta M^L U_L\right)_{jj}\right|}.
\end{equation}
In the single-parameter benchmarks used throughout this paper, where one diagonal entry $\delta M_{\alpha\alpha}$ dominates, this simplifies to
\begin{equation}
L_{\mathrm{osc}}^{(j)} \simeq \frac{2 \pi E_\nu}{m_j |U_{\alpha j}|^2 |\delta M_{\alpha\alpha}|}
\approx 80 \,\mathrm{Mpc}\left(\frac{E_\nu}{1 \mathrm{TeV}}\right)\left(\frac{0.01 \,\mathrm{eV}}{m_j}\right)\left(\frac{10^{-16}\,\mathrm{eV}}{|\delta M_{\alpha\alpha}|}\right)\left(\frac{0.5}{|U_{\alpha j}|^2}\right).
\end{equation}
The coherence length, on the other hand, is given by~\cite{Rink:2022nvw}
\begin{equation}
L_{\mathrm{coh}}^{(j)}=\frac{2 \sqrt{2 }  E_\nu^2}{m_j\left|\left(U_L^T\delta M^L U_L\right)_{jj}\right|} \sigma_x \approx  10^7 \mathrm{Gpc}\left(\frac{E_\nu}{1 \mathrm{TeV}}\right)^2\left(\frac{0.01 \mathrm{eV}}{m_j}\right)\left(\frac{10^{-16} \mathrm{eV}}{\left|\delta M_{\alpha \alpha}\right|}\right)\left(\frac{0.5}{\left|U_{\alpha j}\right|^2}\right)\left(\frac{\sigma_x}{10^{-10} \mathrm{~m}}\right).
\end{equation}
Here, $\sigma_x$ denotes the spatial width of the neutrino wave packet at production~\cite{Kersten:2015kio}. At TeV energies and for the splittings of interest in this paper, $L_{\mathrm{osc}}^{(j)}$ is typically of order Mpc-Gpc, precisely the range probed by extragalactic neutrino baselines. By contrast, for wave-packet sizes exceeding atomic scales, the coherence length is enormously larger, $L_{\text {coh }}^{(j)} \gg L_{\text {osc }}^{(j)}$. Consequently, loss of coherence from wave-packet separation is entirely negligible in the regime considered here.

\section{Analysis}
\label{sec:analysis}
Our central goal is to determine whether SNSI can imprint the TeV–PeV neutrino sky with the characteristic hallmarks of pseudo-Dirac oscillations, namely, energy-dependent spectral modulations and coherent shifts in the flavor composition at Earth.
Three complementary channels are exploited:
\begin{itemize}
\item Diffuse flux flavor ratios~\cite{IceCube:2025uyt,Abbasi:2025fjc}, 
\item Cascade and ESTES diffuse spectra~\cite{IceCube:2020acn,IceCube:2024fxo,IceCube:2025dlr}, 
\item Point‑source spectral‑shape analysis (tracks + cascades) for the most significant sources identified by IceCube~\cite{IceCube:2025lev}.
%\item Point-source spectral-shape analysis (tracks+cascades) for a small catalog of bright sources
\end{itemize}

Throughout this work, we assume that the astrophysical neutrino flux is generated solely through the canonical charged-pion decay chain with the source flavor ratio  $(1/3: 2/3: 0)_S$ . The predicted flavor ratios and spectral shapes obtained under the SNSI–induced pseudo-Dirac oscillation framework are then directly compared with the standard three-flavor expectations for pure pion decay in vacuum, thereby isolating the imprint of SNSI-driven pseudo-Dirac effects on the observable neutrino signal.
All standard three-flavor oscillation parameters are fixed to their global-fit best-fit values from NuFIT 6.0~\cite{Esteban:2024eli} and the lightest neutrino mass is assumed to be 0.01~eV for concreteness; see Refs.~\cite{Elbers:2025vlz,RoyChoudhury:2025dhe}
for recent cosmological constraints on the absolute neutrino-mass scale.

\subsection{Diffuse flux flavor analysis}
If the contributing point sources to the neutrino flux cannot be identified, the oscillation probabilities must be averaged over the redshift distribution of neutrino sources. 
IceCube currently lacks the statistical precision to reliably determine neutrino flavor ratios from specific point sources due to limited event counts, but the detector has proven highly effective in analyzing the diffuse astrophysical neutrino flux.
Since the current IceCube flavor ratio constraints and future IceCube-Gen-2 predictions are derived from a data set of neutrino events across various energies, we must also average over a suitable energy range corresponding to astrophysical neutrino energies. 
We therefore calculate the distance-averaged and energy-averaged expected flavor composition of high-energy astrophysical neutrinos at Earth to account for the finite range of energies and various sources at different redshifts that may contribute to the signal. We assume that the energy spectrum of these neutrinos $\frac{d\phi_{\nu}}{dE}(E)~(\text{in units of } \text{GeV}^{-1} {\rm s}^{-1})$ follows the same power law spectrum at the source (i.e., the energy dependence of the flux is assumed to be identical for all neutrino flavors), $\frac{d\phi_{\nu}}{dE}(E) \propto E^{-\gamma}$, with a fixed spectral index  $\gamma = 2.5 $, which is consistent with IceCube analyses that combine high-energy starting events (HESE)~\cite{IceCube:2020wum} and through-going tracks~\cite{Naab:2023xcz}. 

The diffuse flux of neutrinos observed on Earth is the sum of contributions from sources located at different cosmological distances and time scales.
While the precise identity of these sources remains unknown, it is our working hypothesis that their spatial distribution correlates with some redshift-dependent function $\rho_s(z)$ (e.g., star formation rate density).
The redshift distribution of sources then follows from the comoving volume element
\begin{equation}
\frac{dN(z)}{dz}=\frac{dN(z)}{dV}\frac{dV}{dz} \propto {\rho_s}(z) \times 4 \pi r(z)^2 \frac{d r(z)}{d z},
\end{equation}
where $r(z)=\int_0^{z} \frac{d z'}{H(z')}$ is the comoving distance.
Therefore, the redshift source distribution is given by
\begin{equation}
 \frac{dN(z)}{dz} \propto {\rho_s}(z) 4 \pi r(z)^2 \frac{c}{H(z)}.
\end{equation}

We take into account the distribution of the sources for several physically motivated source-evolution templates, corresponding to BL Lacs, flat-spectrum radio quasars (FSRQ),  star-formation-rate density (SFRD)~\cite{Elias-Chavez:2018dru}, low-luminosity AGN (LLAGN), and radio-quiet AGN (RQAGN) as done in Ref.~\cite{Carloni:2025dhv}; see Refs.~\cite{Capel:2020txc,Groth:2025aan} for a discussion of the preferred source populations in IceCube.
We evaluate the redshift integral over the redshift from $z_{\text{min}}=0$ up to $z_{\text{max}}=5$ beyond which we do not expect a significant population of astrophysical sources of high-energy neutrinos. 
We also calculate the average flux at the detectors by averaging over the neutrino energy distribution from 1~TeV to 1~PeV. 

The expected neutrino flux of flavor $\beta$ from the diffused source is thus given by,
\begin{equation}
\langle \phi_{\beta}^\oplus \rangle \propto \sum_{\alpha=e, \mu, \tau}\int_{E_{\min}}^{E_{\max}} dE \int_{z_{\min }}^{z_{\max }} dz~  P_{\alpha \beta}^{\text{SNSI}}(E,z) f_\alpha^s \frac{{\rho_s}(z)}{H(z)}\frac{d\phi_{\nu}}{dE}(E(1+z)),
\end{equation}
where the Hubble parameter $H(z)$ is given by $H(z)=H_0 \sqrt{\Omega_{\Lambda}+\Omega_m(1+z)^3}$. 
Here, $H_0 = 67.4 \, \mathrm{km\,s^{-1}\,Mpc^{-1}}$ is the Hubble constant, $\Omega_m = 0.315$ represents the matter density parameter, and $\Omega_\Lambda = 1 - \Omega_m$ denotes the vacuum energy density parameter. 
Here, $E(1+z)$ is the neutrino energy at the time of emission.
We then calculate the flavor ratios at Earth from the average fluxes as
\begin{equation}
 \langle f_{\alpha}^\oplus \rangle \approx \frac{\langle \phi_{\alpha}^{\oplus} \rangle}{\sum_{\beta = e,\mu,\tau} \langle \phi_{\beta}^{\oplus} \rangle}.
\end{equation}
Note that we may also define energy-averaged flavor ratios corresponding to a fixed point source by omitting the redshift integration and instead retaining only the integration over energy.

\subsection{Diffuse flux spectral analysis}
\label{sec:diffuse-spectral}

Following Ref.~\cite{Carloni:2025dhv}, we adapt the flux–spectral technique that has been employed for pseudo-Dirac oscillations to the case of SNSI. We base our study on two statistically independent IceCube datasets that probe distinct event topologies and are most sensitive to the astrophysical diffuse TeV–PeV flux~\cite{Naab:2023xcz}:
\begin{itemize}
\item Cascade: electromagnetic and hadronic showers accumulated over 6.5 years of IceCube data taking.
\item ESTES: ``starting" muon tracks ($\nu_\mu$ CC interactions with vertices inside the instrumented volume) collected over 10.3 years of IceCube data taking.
\end{itemize}
These samples have been used by IceCube to extract the diffuse astrophysical flux, reported in piece-wise energy bins under the assumption of an $E^{-2}$ spectrum and flavor equipartition at Earth. 
For each data set $S$ we fold the reported flux through the published, flavor-dependent, zenith-averaged effective areas $A_\alpha^S(E)$ to obtain binned event counts $N^S_i$~\cite{IceCube:2016tpw,IceCube:2024fxo,IceCube:2020acn}.

In the presence of SNSI, the flavor-dependent neutrino flux that reaches Earth at an observed energy $E$ can be written as
\begin{equation}
\phi_\beta^{\oplus}(E)=\sum_{\alpha=e, \mu, \tau} \int_{z_{\min }}^{z_{\max }} d z P_{\alpha \beta}^{\mathrm{SNSI}}(E, z) f_\alpha^s \frac{{\rho_s}(z)}{H(z)} \frac{d \phi_\nu}{d E}(E(1+z)),
\end{equation}
where we integrate over the redshift distribution of the sources ${\rho_s}(z)$.
Here, $P_{\alpha \beta}^{\text {SNSI }}(E, z)$ is the SNSI-modified oscillation probability for a neutrino produced in flavor $\alpha$ to be detected in flavor $\beta$ after propagating from redshift $z$; $f_\alpha^s$ denotes the intrinsic source flavor composition; 
%$H(z)$ is the Hubble expansion rate; 
and $d \phi_\nu / d E$ is the differential emission spectrum evaluated at the emitted energy $E(1+z)$. 
We consider two spectral hypotheses for the emitted spectrum:
\begin{itemize}
\item \textbf{Single power law (SPL):}
\begin{equation*}
\frac{d\phi_\nu}{dE}(E) \;=\; \Phi_0\left(\frac{E}{100~{\rm TeV}}\right)^{-\gamma}
\end{equation*}
with nuisance parameters $\tilde{\boldsymbol{\theta}}_{\rm SPL}=(\Phi_0,\gamma)$.

\item \textbf{Broken power law (BPL):}
\begin{equation*}
\frac{d\phi_\nu}{dE}(E) \;=\; \Phi_0 \, \mathcal{N}(E_b)\times
\begin{cases}
\left(\dfrac{E}{E_b}\right)^{-\gamma_1}, & E < E_b,\\[6pt]
\left(\dfrac{E}{E_b}\right)^{-\gamma_2}, & E \ge E_b,
\end{cases}
\end{equation*}
with nuisance parameters $\tilde{\boldsymbol{\theta}}_{\rm BPL}=(\Phi_0,\gamma_1,\gamma_2,E_b)$.
\end{itemize}

In each hypothesis, the observed flux at Earth is obtained by folding (i) the SNSI-modified oscillation probabilities,
(ii) the chosen source evolution $\rho(z)$, and (iii) either the SPL or BPL emission spectrum evaluated at the emitted energy $E(1+z)$.

The theoretically expected number of events in the $i$-th energy bin for a particular data sample $S$ is given by folding the flux with the effective area
\begin{equation}
\mu_i^S(\tilde{\boldsymbol{\theta}})=\Delta T^S \int_{E_i}^{E_{i+1}} dE \sum_\alpha \phi_\alpha^{\oplus}(E \mid \tilde{\boldsymbol{\theta}}) A_\alpha^S(E).
\end{equation}
Here, $\Delta T^S$ is the exposure of sample $S$; $A_\alpha^S(E)$ is the flavor-dependent effective area; and  $\phi_\alpha^{\oplus}(E \mid \tilde{\boldsymbol{\theta}})$ is the flavor-dependent neutrino flux at Earth, which depends on the flux nuisance parameters $\tilde{\boldsymbol{\theta}}$ and on the oscillation/SNSI parameters that enter through the probability $P_{\alpha \beta}^{\mathrm{SNSI}}$.
To account for finite energy resolution, we incorporate an approximate Gaussian smearing in $\ln E$ with $\sigma_{\ln E}\simeq 0.1$.

From the IceCube published piecewise flux points (and their asymmetric uncertainties) we reconstruct an effective ``observed'' binned event content $N_i^S$ and corresponding asymmetric errors $(\sigma_{i,+}^S,\sigma_{i,-}^S)$ by folding with the same effective area and exposure used for $\mu_i^S$.
We then define a bin-wise log likelihood $-2\log\mathcal{L}_i$  such that for $E<100~{\rm TeV}$ we use an asymmetric Pearson-$\chi^2$,
while for $E\ge 100~{\rm TeV}$ we use a calibrated Poisson deviance rescaled such that
$-2\log\mathcal{L}_i(\mu_i=N_i)=0$ and $-2\log\mathcal{L}_i(\mu_i=N_i\pm\sigma_i)=1$ on the appropriate side.
The total test statistic is
\begin{equation}
\mathrm{TS}(\delta M)=
\min_{\tilde{\boldsymbol\theta}}
\sum_i \left[-2\log \mathcal{L}_i\big(N_i^S\,\big|\,\mu_i^S(\delta M;\tilde{\boldsymbol\theta})\big)\right].
\end{equation}
We finally report $\Delta\mathrm{TS}(\delta M)\equiv \mathrm{TS}(\delta M)-\mathrm{TS}(0)$ and interpret it using a $\chi^2$ distribution with one degree of freedom.

In addition to the separate cascade and ESTES analyses above, we also perform a combined diffuse-flux analysis based on the published IceCube~\emph{CombinedFit}~\cite{IceCube:2025ewu} that merges track-like and cascade-like events into a single sample. 
The IceCube public result reports only the total all-flavor flux fit from the combined track+cascade dataset and does not provide separate event counts or fluxes for the two morphologies.
We therefore construct a single total expected event rate for each $\delta M_{\alpha\alpha}$ by summing the SNSI-modified track and cascade contributions, and we compare it to the total number of neutrino events inferred from the IceCube combined all-flavor flux.
The test statistic is then a single $\chi^2$ built from this combined total, rather than a sum of separate $\chi^2$ terms for tracks and cascades.
Note that because we do not have access to topology-resolved likelihoods, the combined analysis cannot exploit flavor-dependent differences between track-dominated ($\nu_\mu$) and cascade-dominated ($\nu_e$, $\nu_\tau$) channels. In particular, when the three pseudo-Dirac splittings are non-degenerate, $\delta m_1^2\neq\delta m_2^2\neq\delta m_3^2$ (as is the case when we have a single non-zero diagonal SNSI), flavor-dependent spectral distortions are not resolvable with a single combined $\chi^2$, since only the total event count enters the test statistic.

\subsection{Point-source spectral analysis}
To assess the sensitivities and projected constraints on SNSI from IceCube point-source observations, we consider the most significant sources from Table 3 of Ref.~\cite{IceCube:2025lev}
%that includes neutrino events from the combined sample of tracks and cascades data collected by IceCube
%between $04 / 06 / 2008 \text { to } 05 / 23 / 2022$. 
which includes neutrino events from the combined sample of track and cascade data collected by IceCube over 14 years of track data and 10 years of cascade data.
Specifically, we include the extragalactic sources: NGC~1068, PKS~1424+240, TXS~0506+056, GB6~J1542+6129, and PMN~J1603-4904, as well as the galactic sources: G343.1-2.3 and MGRO J2019+37.
The redshifts/distances assumed for these sources in our analysis are taken from various references and are listed in \cref{tab:pt_source}.
We do not include the source PMN J1650-5044 for our analysis since its redshift is poorly constrained. Because two of these sources are galactic and the rest extragalactic, they probe a broad range of baselines from few kpc to Gpc and are therefore sensitive to complementary ranges of $\delta m_i^2$.
For each source, the neutrino spectrum is assumed to follow a power law, with best-fit spectral indices $\hat{\gamma}$ and total signal counts $\hat{n}_s$ taken from Table 3 of Ref.~\cite{IceCube:2025lev}, as reported in \cref{tab:pt_source}.

\begin{table}[t]
\centering
\small
\setlength{\tabcolsep}{4pt}
\begin{tabular}{|c|l|c|p{2.8cm}|c|c|c|c|}
\hline
 & Source Name & Class & Redshift $z$ & $\delta\,[^{\circ}]$ & $\hat{n}_s$ & $\hat{\gamma}$ & $-\log_{10}(p_{\text{local, combined}})$ \\
\hline
1 & NGC 1068        & SBG & 0.003793~\cite{Meyer:2004hr}                          & $-$0.01 & 92.32 & 3.14 & 5.902 \\
2 & PKS 1424+240    & BLL & 0.6010~\cite{Rovero:2016igo}                          & 23.8    & 75.79 & 3.50 & 3.534 \\
3 & PMN J1650-5044  & BLL & (unknown)                                             & $-$50.75& 88.46 & 2.93 & 2.856 \\
4 & GB6 J1542+6129  & BLL & 0.507~\cite{Marcha:2013tt}                            & 61.5    & 50.79 & 3.34 & 2.709 \\
5 & TXS 0506+056    & BLL & 0.3365~\cite{Paiano:2018qeq}                          & 5.7     & 9.67  & 0.87 & 2.663 \\
6 & G343.1-2.3      & PWN & Galactic~\cite{Dodson:2002pz, HESS:2011zks}& $-$44.3 & 77.60 & 2.92 & 2.085 \\
7 & PMN J1603-4904  & BLL & 0.2321~\cite{Goldoni:2015jua}                         & $-$49.06& 84.59 & 3.10 & 2.046 \\
8 & MGRO J2019+37   & GAL & Galactic~\cite{Abdo:2006fq}                           & 36.8    & 43.35 & 2.99 & 2.000 \\
\hline
\end{tabular}
%\captionsetup{justification=RaggedRight, labelfont=bf}
\caption{Point-source candidates and their corresponding redshifts. The $\hat{n}_s$, $\hat{\gamma}$, and $-\log_{10}(p_{\text{local, combined}})$ values are taken from Ref.~\cite{IceCube:2025lev}. 
Redshifts are obtained from references cited in the table. PMN~J1650-5044 is excluded from our analysis since its redshift is unknown.}
\label{tab:pt_source}
\end{table}

Following the analysis done in Refs.~\cite{IceCube:2022der,Carloni:2022cqz}, for each point-source we assume an unbroken power-law spectrum beginning at $E_{\min }=100 \mathrm{GeV}$: 
\begin{align}
\Phi_\alpha^{s}(E)={\Phi}_0\left(\frac{E}{E_0}\right)^{-\gamma} f_\alpha^{s} \, ,
\end{align}
with $E_0=1~\mathrm{TeV}$.
We compute reconstructed-energy spectra by folding the flux with topology- and hemisphere-dependent effective areas and with an energy-migration kernel. 
For morphologies $t\in\{\mathrm{track},\mathrm{cascade}\}$ and source $s$, the expected number of events in reconstructed-energy bin $i$ is given by
\begin{equation}
\mu^{t}_{i,s}(\delta M)=\Delta T_t  \sum_{\alpha,\beta=e, \mu, \tau}\int dE_{\rm true}\;
\Phi^{s}_{\alpha}(E_{\rm true})\,
P_{\alpha \beta}^{\text {SNSI }}\left(E_{\text {true }}, z_s\right)\,
A^{t}(E_{\rm true})\,
\mathcal{M}^{t}_{i}(E_{\rm true})\,,
\label{eq:pt_mu}
\end{equation}
where $\Delta T_t$ is the live-time of the corresponding sample and $A^{t}(E)$ denotes the effective area appropriate to the source hemisphere and topology.
The migration kernel $\mathcal{M}^{t}_{i}(E_{\rm true})$ maps true to reconstructed energy; we model it as a Gaussian in $\log_{10}E$ with widths $\sigma_{\log_{10}E}=0.30$ (tracks) and $\sigma_{\log_{10}E}=0.15$ (cascades), reflecting the different energy resolutions of the two event classes~\cite{IceCube:2013dkx}.

We quantify sensitivity using the Poisson deviance summed over reconstructed-energy bins and (when available) over both topologies. 
For a given source $s$, we define
\begin{equation}
-2\ln \mathcal{L}_s\left(\delta M\right)=\min_{\Phi_0,\gamma}\sum_{t\in\{\mathrm{track},\mathrm{cascade}\}}\sum_i
2\left[\mu^{t}_{i,s}(\delta M)-N^{t}_{i,s}+N^{t}_{i,s}\ln\left(\frac{N^{t}_{i,s}}{\mu^{t}_{i,s}(\delta M)}\right)\right],
\label{eq:pt_Xi}
\end{equation}
where $N^{t}_{i,s}$ are the observed counts and $\Phi_0$ is the per-source overall normalization applied coherently to the track and cascade spectra of the same source (while $\gamma$ controls the spectral index).
At every test value of the SNSI parameter $\delta M$, we treat the flux parameters $\Phi_0$ and $\gamma$ as nuisance quantities, finding the values that maximize the likelihood before we evaluate the test statistic.
The total test statistic is then the sum over sources, $-2\ln \mathcal{L}\left(\delta M\right)=-2\sum_s \ln \mathcal{L}_s\left(\delta M\right)$.
%No external priors are imposed on the flux normalization ${\Phi}_{0}$ and spectral index $\gamma$, they are profiled out at every trial $\delta M_{\alpha\beta}$ value

%%%%%%%%%%%%%%%%%%%%%%%%%%%%%%%%%%%%%%%%%%%%%%%%%%%%%%%%%%%%%%%%%%%%%%%
%%%%%%%%%%%%%%%%%%%%%%%%%%%%%%%%%%%%%%%%%%%%%%%%%%%%%%%%%%%%%%%%%%%%%%%
%%%%%%%%%%%%%%                 Results                   %%%%%%%%%%%%%%
%%%%%%%%%%%%%%%%%%%%%%%%%%%%%%%%%%%%%%%%%%%%%%%%%%%%%%%%%%%%%%%%%%%%%%%
%%%%%%%%%%%%%%%%%%%%%%%%%%%%%%%%%%%%%%%%%%%%%%%%%%%%%%%%%%%%%%%%%%%%%%%
\section{Results} \label{sec:results}

We present the impact of SNSIs, $\delta M_{\alpha\beta}$, that induce pseudo-Dirac mass splittings, on high-energy astrophysical neutrinos detected by IceCube. Three complementary analyses underpin our study: First,
we map the allowed diffuse flavor ratios on the flavor triangle, showing that IceCube-Gen2 can exclude large portions of parameter space in the range $\delta M\!\sim\!(10^{-18}\text{-}10^{-16})\,$eV.
Second, using the most recent cascade and ESTES diffuse-flux samples, we show the existing constraints and preferred windows on the parameter space by a diffuse flux spectral-shape fit analysis.
Finally, spectral analysis using the differing baselines and spectral indices of the most significant point sources enable a probe of $\delta M$ between $10^{-19}$ to $10^{-14}\,\text{eV}$ from extragalactic sources and $ 10^{-12} $ to $10^{-10}$~eV from galactic sources.
The detailed outcomes of each sub-analysis are discussed below:

\subsection{Diffuse flux flavor analysis}

\begin{figure}[tbp]
\centering
\begin{minipage}[c]{0.4\textwidth}
  \centering
  \includegraphics[width=\textwidth]{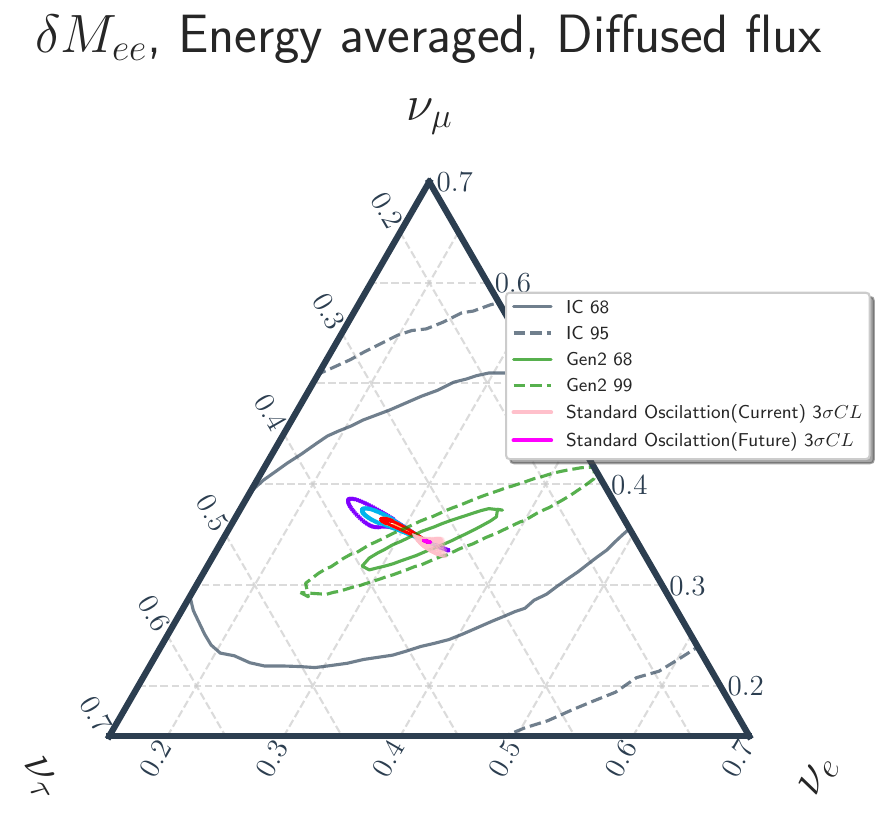}
\end{minipage}
\quad
\begin{minipage}[c]{0.4\textwidth}
  \centering
  \includegraphics[width=\textwidth]{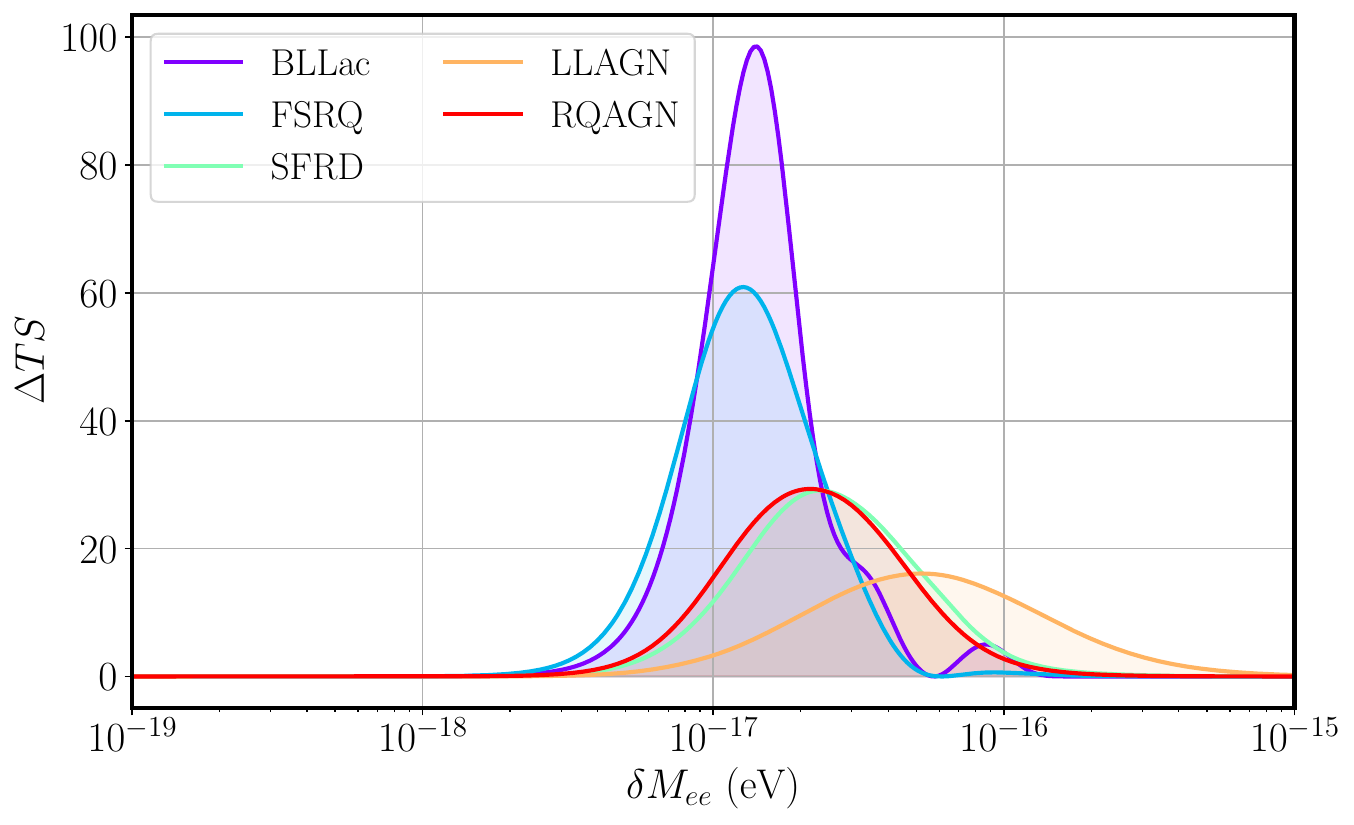}
\end{minipage}
\begin{minipage}[c]{0.4\textwidth}
  \centering
  \includegraphics[width=\textwidth]{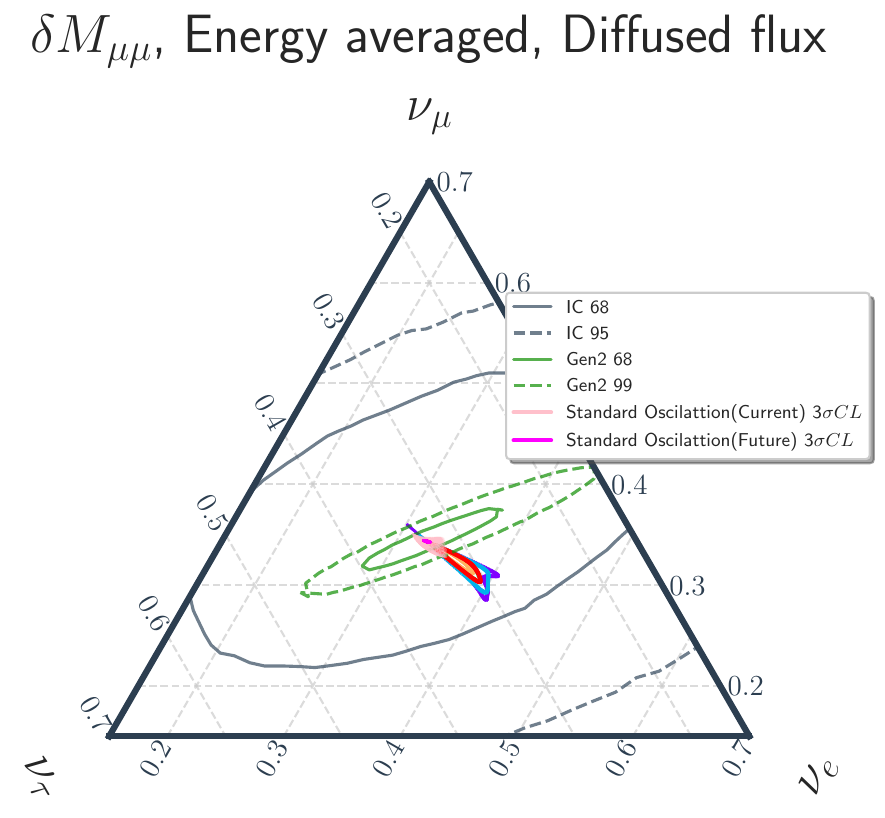}
\end{minipage}
\quad
\begin{minipage}[c]{0.4\textwidth}
  \centering
  \includegraphics[width=\textwidth]{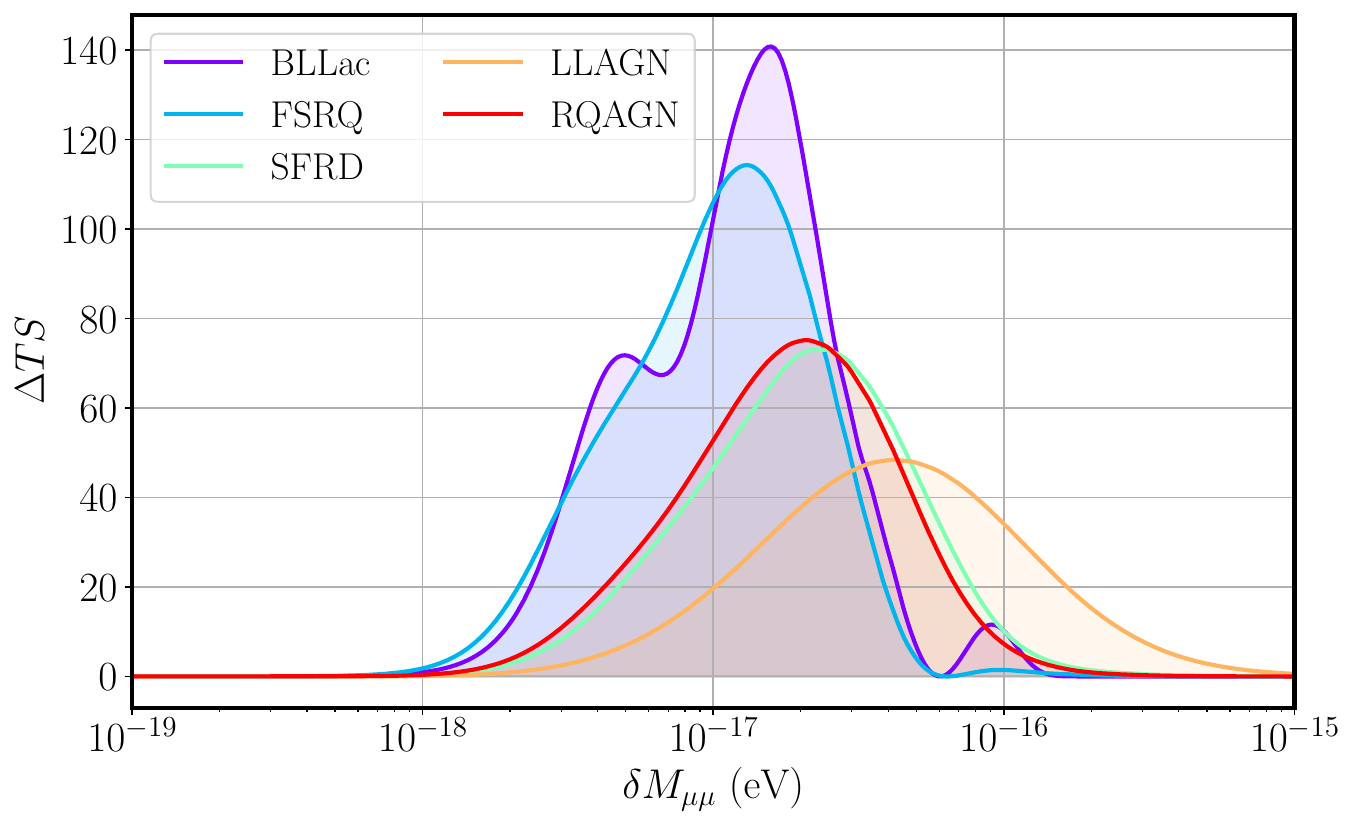}  
\end{minipage}
\begin{minipage}[c]{0.4\textwidth}
  \centering
  \includegraphics[width=\textwidth]{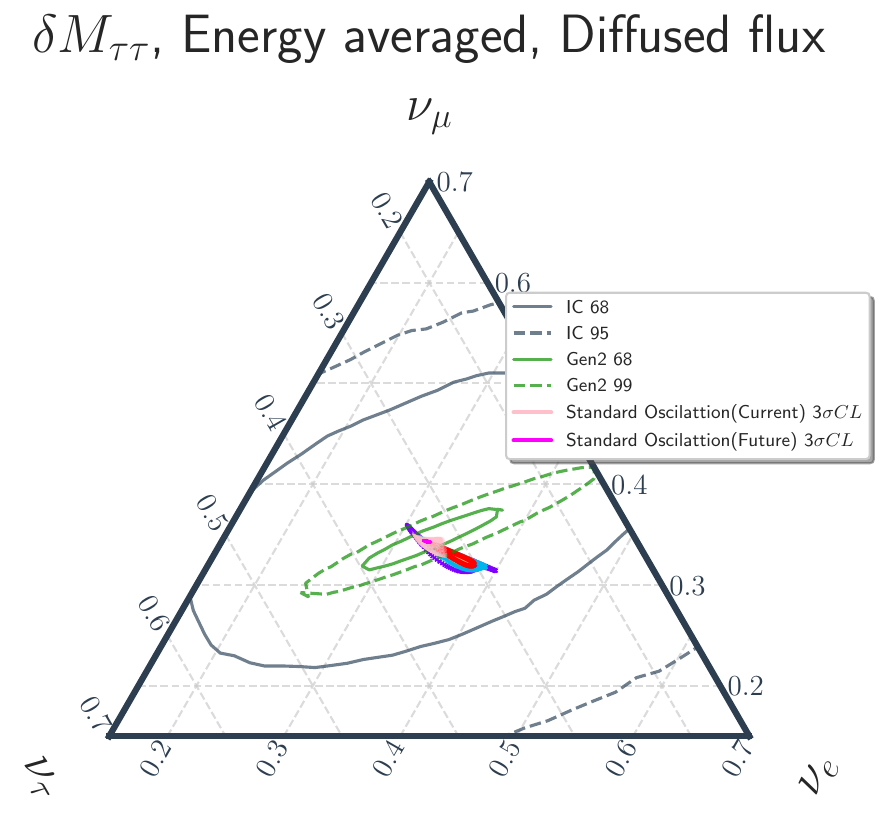}
\end{minipage}
\quad
\begin{minipage}[c]{0.4\textwidth}
  \centering
  \includegraphics[width=\textwidth]{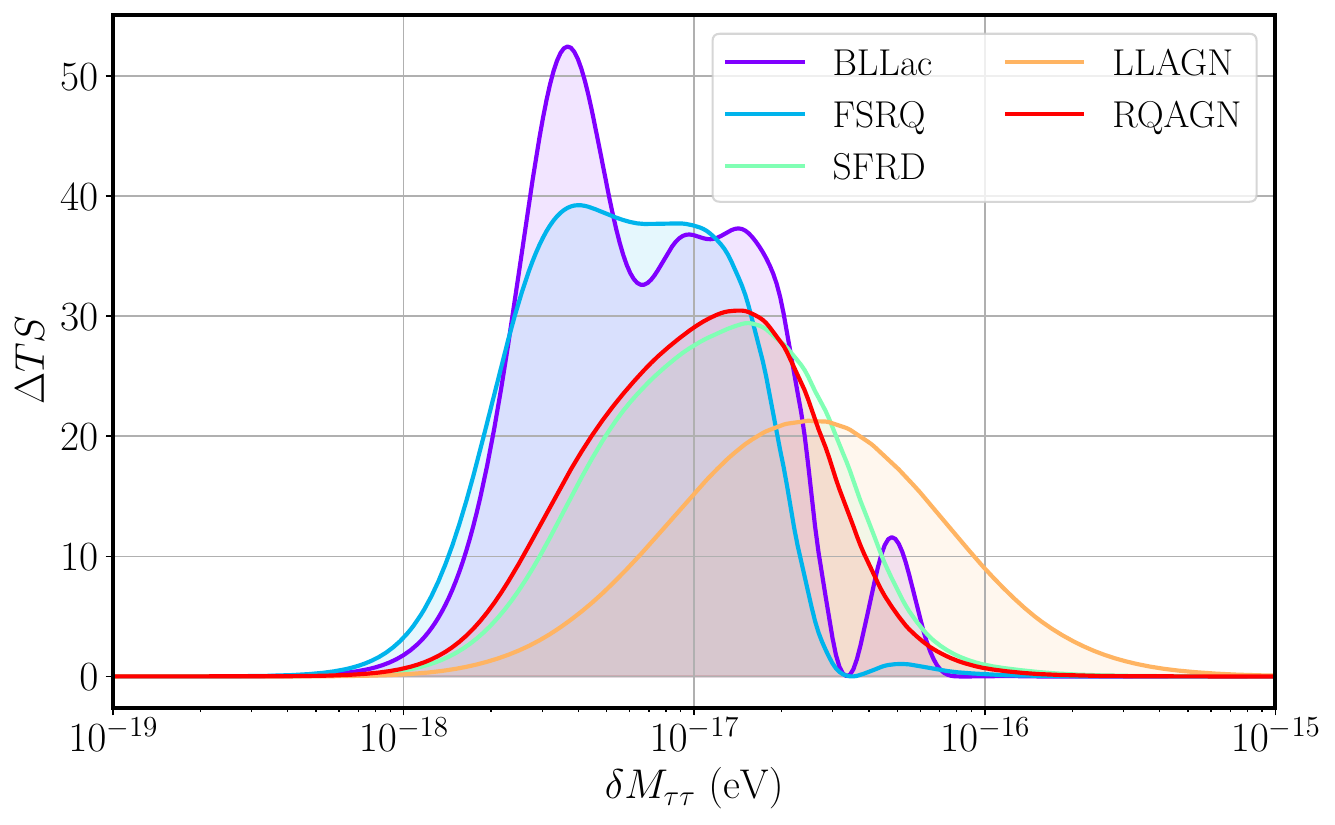}  
\end{minipage}
\caption{\label{fig:diff_flav} Energy-averaged ternary plot of the detected neutrino flavor ratios from diffused source of neutrino flux assuming standard pion-decay source and SPL with initial flavor ratio $(\nu_e : \nu_\mu : \nu_\tau) = (\frac{1}{3} : \frac{2}{3} : 0)$. The lightest neutrino mass $m_1$ is taken to be $0.01$ eV, assuming normal mass ordering and best fit values of oscillation parameters. Various colored curves show the region explored by $\delta M_{\alpha\beta}$ parameter points inside the flavor triangle corresponding to various source distributions. The gray curves illustrate the current IceCube (IC) and green curves show projected 10-year IceCube-Gen2 (Gen2) sensitivity contours. In addition, the pink and magenta regions indicate the 
$3\sigma$ spread of the standard-oscillation prediction arising from current and future oscillation-parameter uncertainties, respectively.
Right column shows the corresponding sensitivities to $\delta M_{\alpha\alpha}$ from IceCube-Gen2, assuming various source distributions. To get the approximate likelihood from the flavor triangle, IceCube contours are interpreted as constant $\Delta \mathrm{TS}$ levels for 2 dof $(2.30$ for $68\%$, $9.21$ for $99\%)$ CL.}
\end{figure}
%\clearpage
\begin{figure}[tbp]
\centering
\begin{minipage}[c]{0.48\textwidth}
  \centering
  \includegraphics[width=\textwidth]{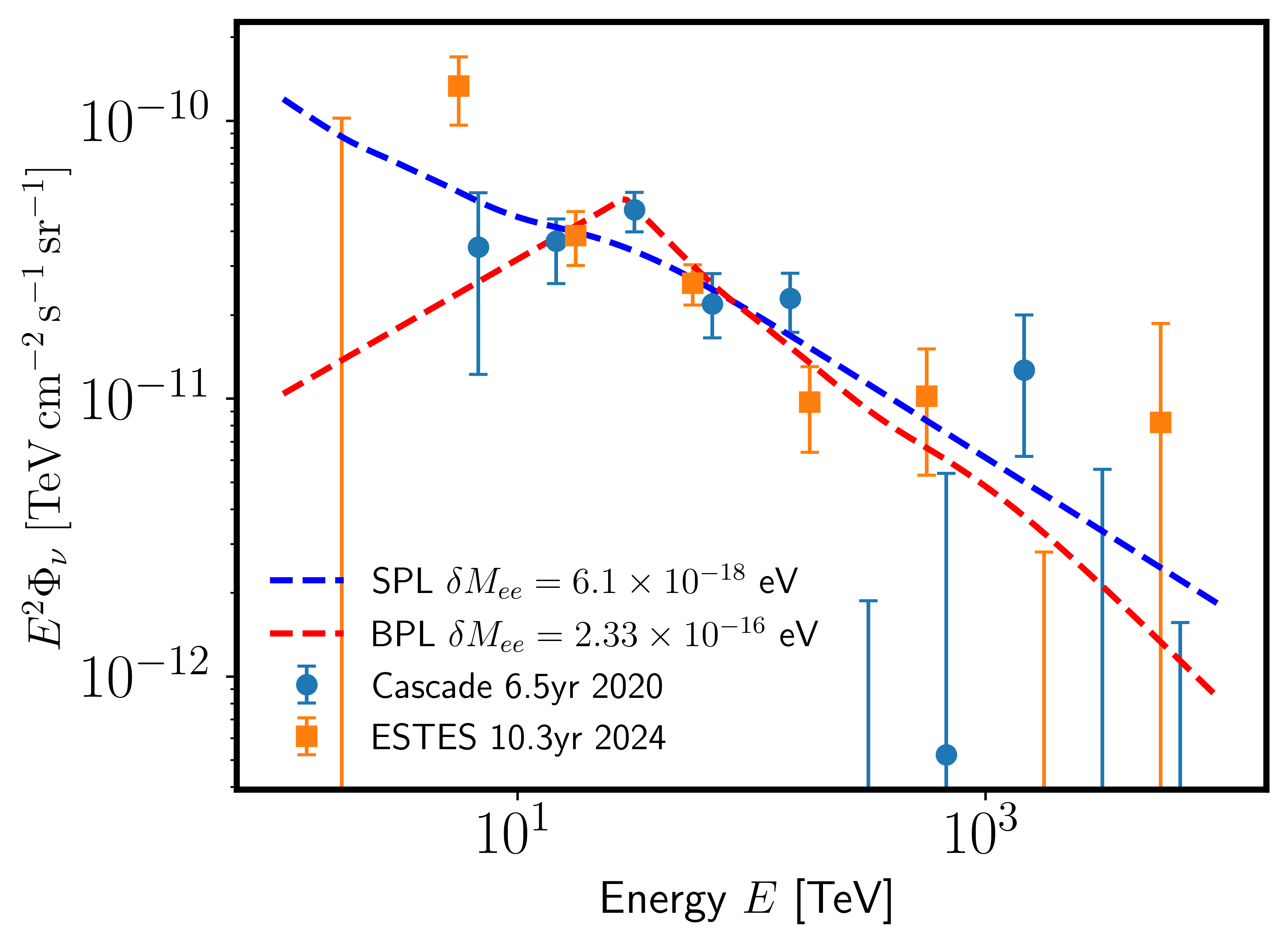}
\end{minipage}
\quad
\begin{minipage}[c]{0.48\textwidth}
  \centering
  \includegraphics[width=\textwidth]{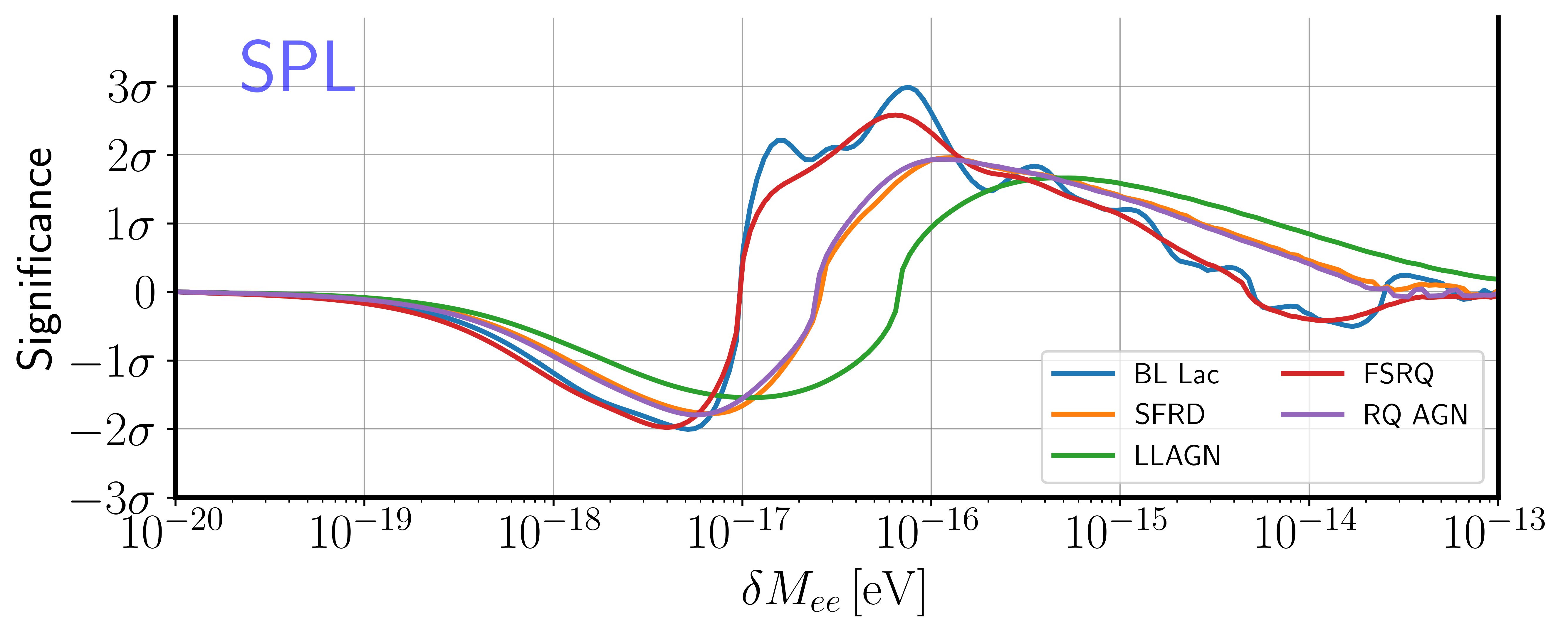}
  \includegraphics[width=\textwidth]{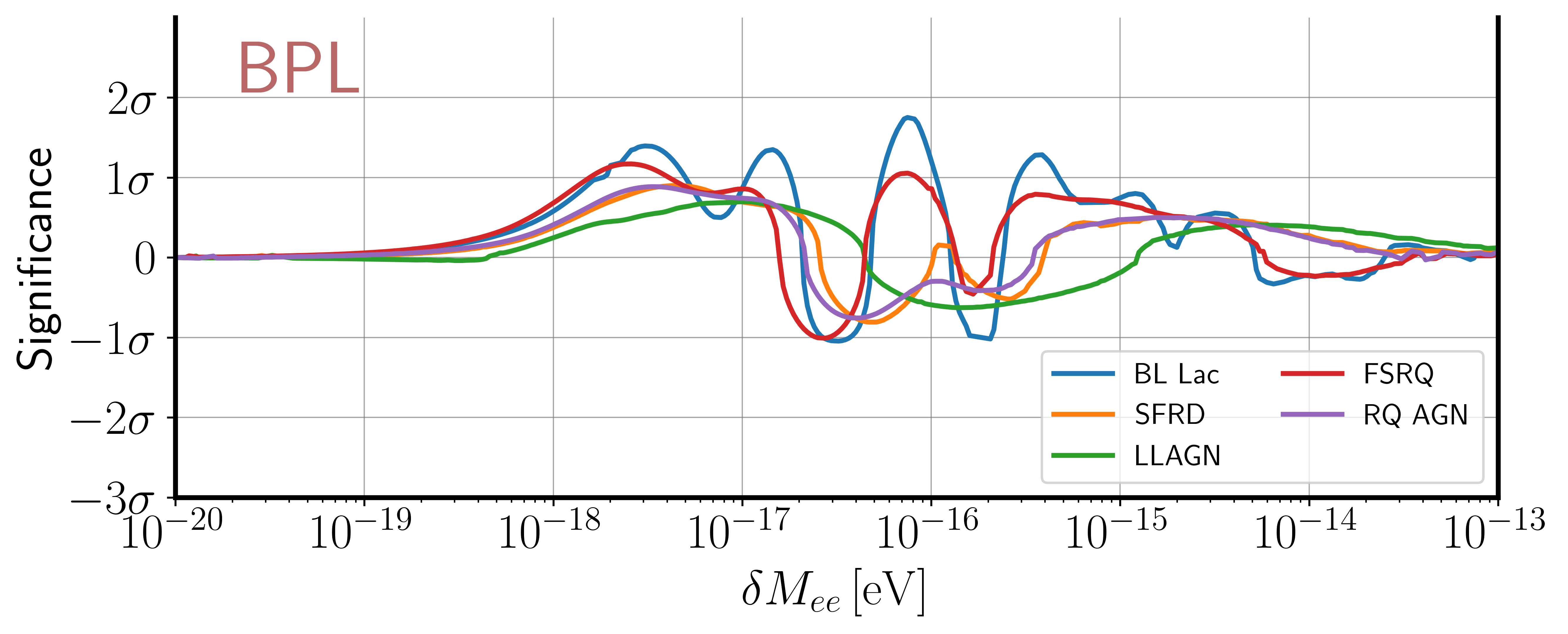}
\end{minipage}
\begin{minipage}[c]{0.48\textwidth}
  \centering
  \includegraphics[width=\textwidth]{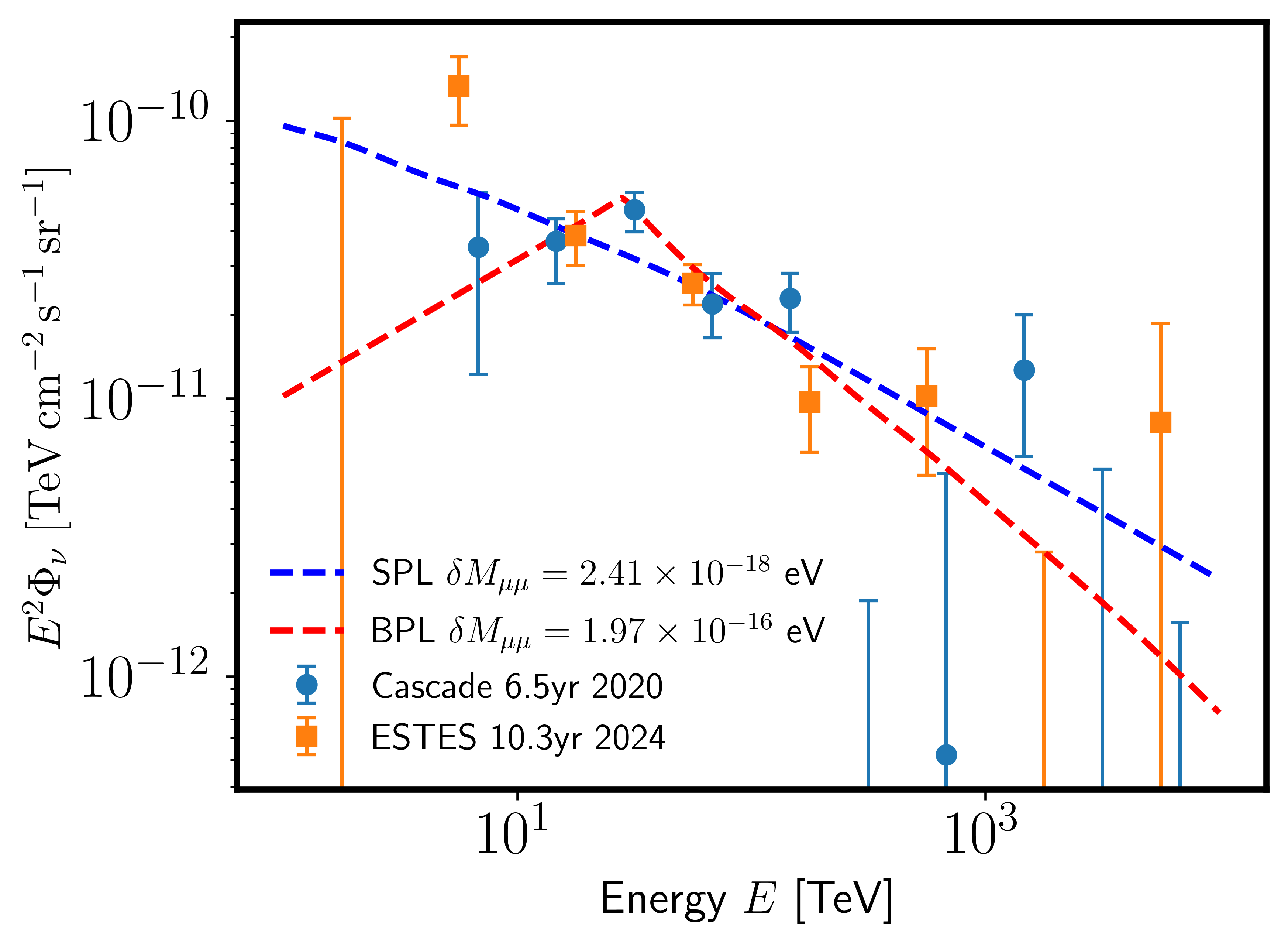}
\end{minipage}
\quad
\begin{minipage}[c]{0.48\textwidth}
  \centering
  \includegraphics[width=\textwidth]{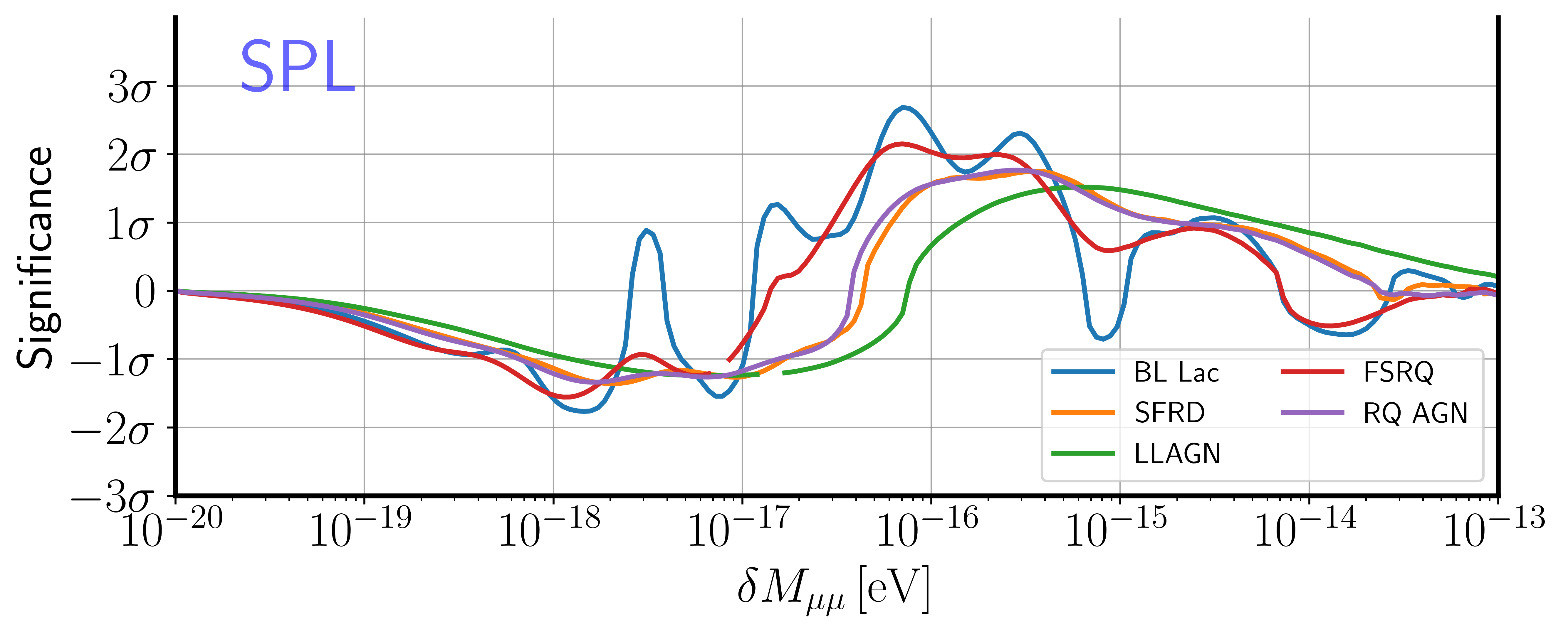} 
  \includegraphics[width=\textwidth]{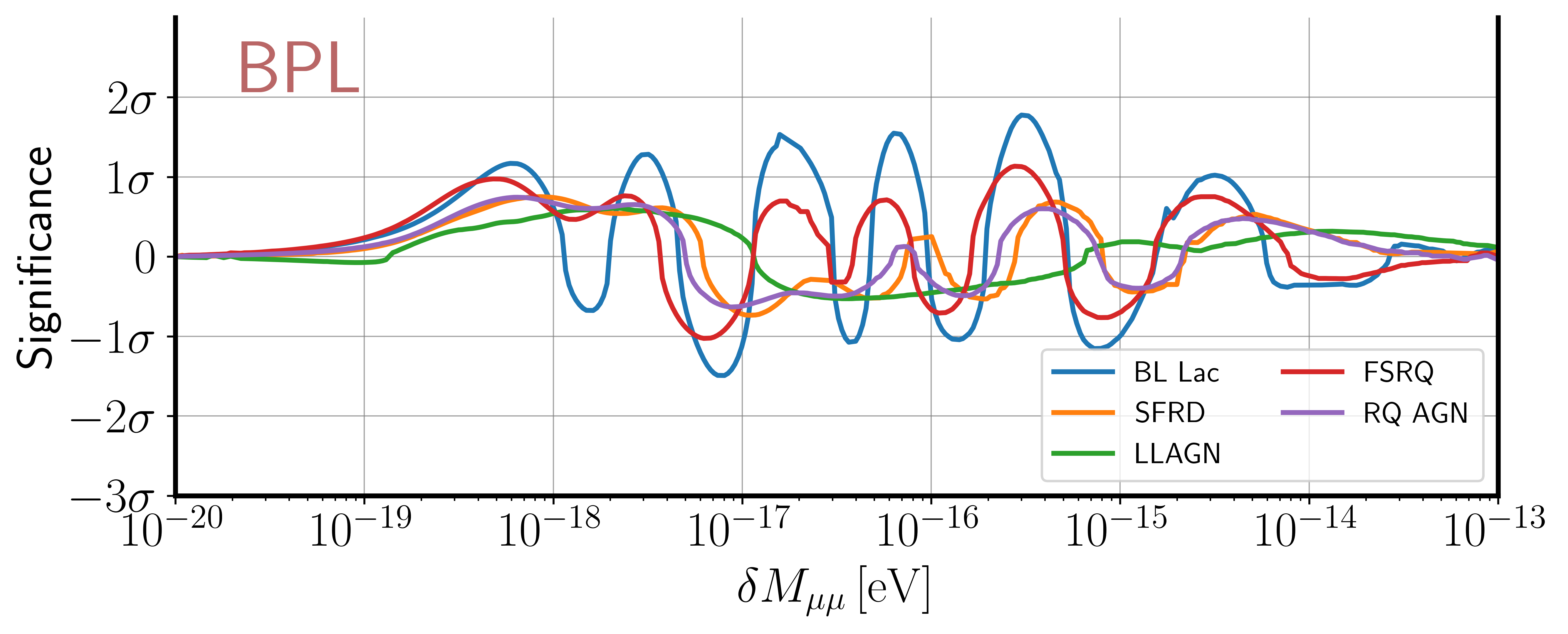}
  \end{minipage}
\begin{minipage}[c]{0.48\textwidth}
  \centering
  \includegraphics[width=\textwidth]{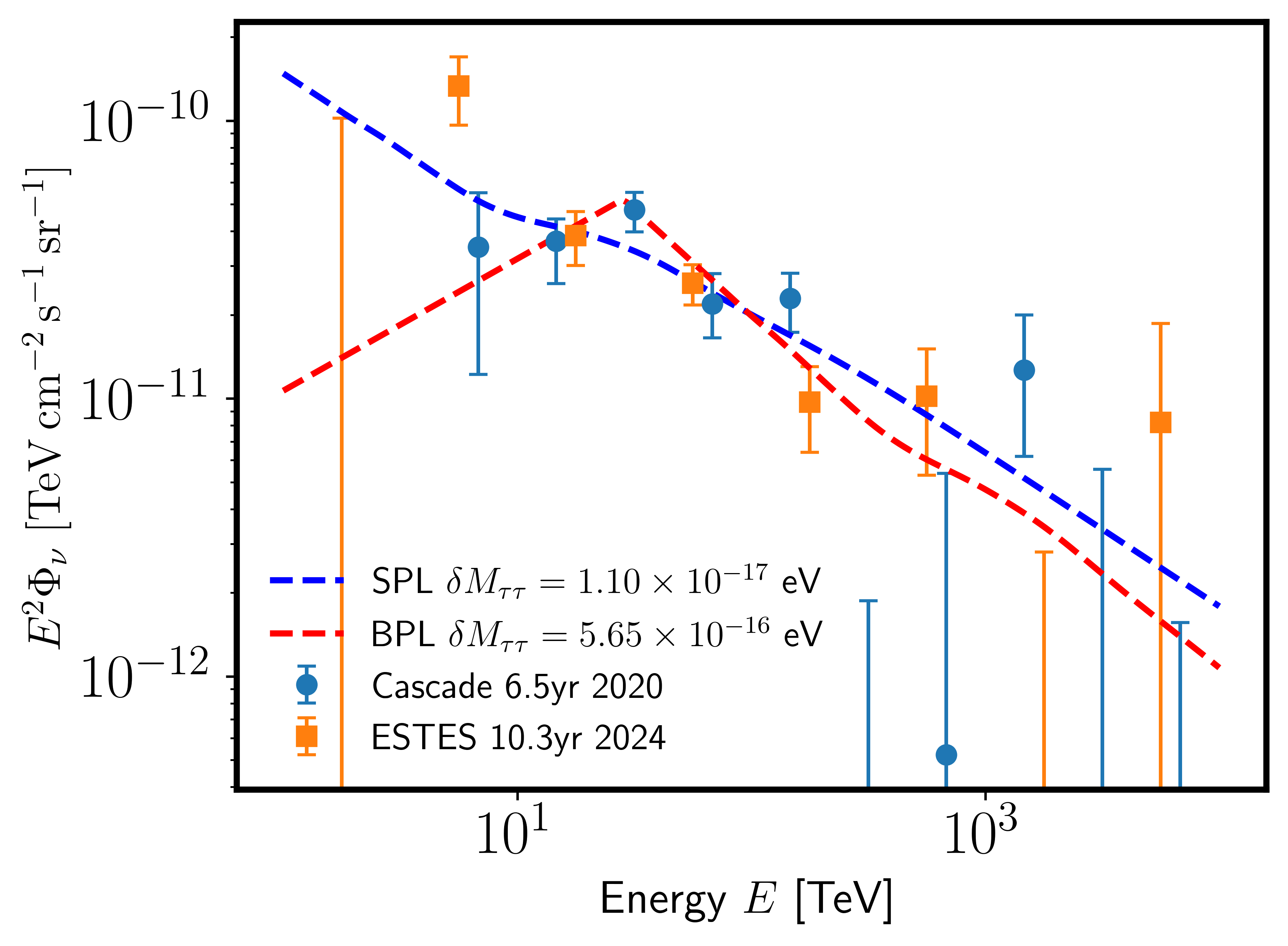}
\end{minipage}
\quad
\begin{minipage}[c]{0.48\textwidth}
  \centering
  \includegraphics[width=\textwidth]{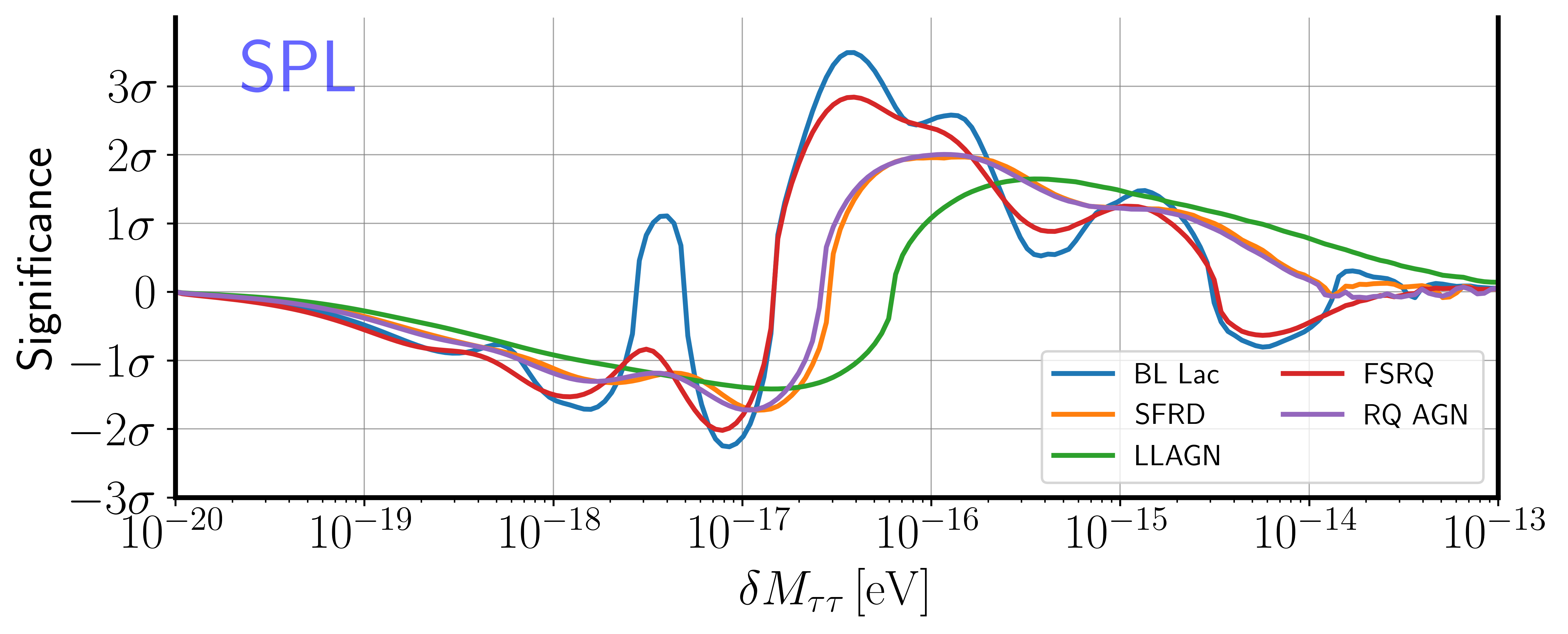}  
    \includegraphics[width=\textwidth]{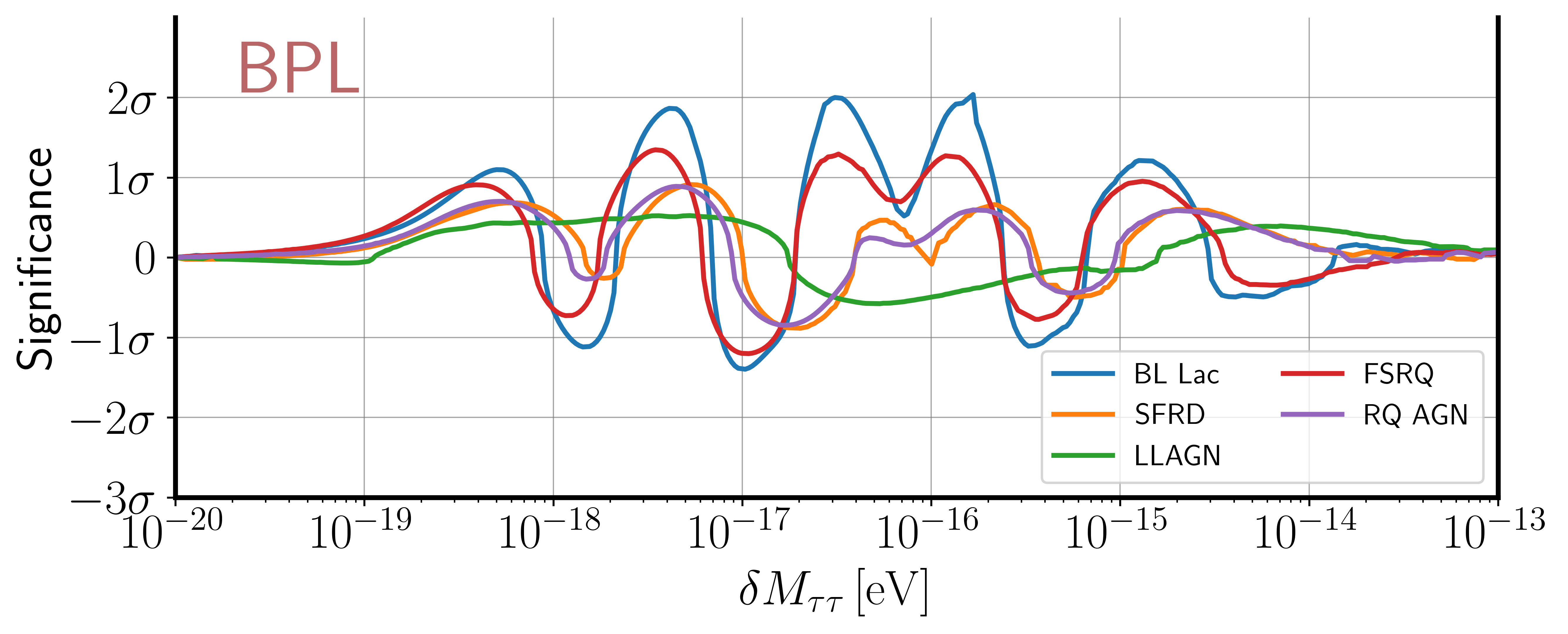}
    \end{minipage}

\caption{\label{fig:diff_shape} 
Diffuse spectral-shape analysis for a single diagonal SNSI parameter $\delta M_{\alpha \alpha}$ using the cascade and ESTES samples. In each row, the left panel shows the reconstructed diffuse spectrum representing SNSI pseudo-Dirac distortion (dashed) for the SPL and BPL hypotheses. The right panel shows the corresponding one-dimensional signed Gaussian significance versus $\delta M_{\alpha \alpha}$, evaluated separately for SPL and BPL and for each source-evolution model. Negative (positive) values indicate an improved (degraded) fit relative to the SM.}
\end{figure}

\begin{figure}[tbp]
\centering
\begin{minipage}[c]{0.48\textwidth}
  \centering
  \includegraphics[width=\textwidth]{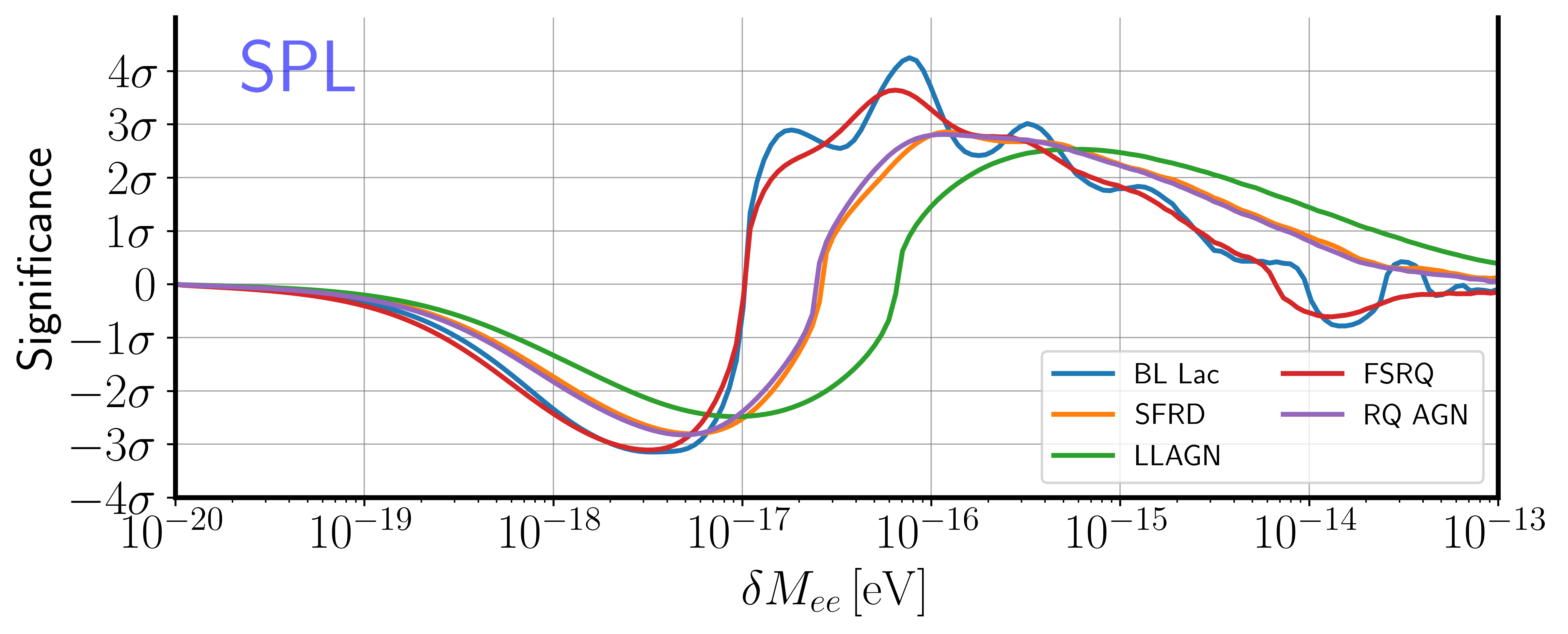}
\end{minipage}
\quad
\begin{minipage}[c]{0.48\textwidth}
  \centering
  \includegraphics[width=\textwidth]{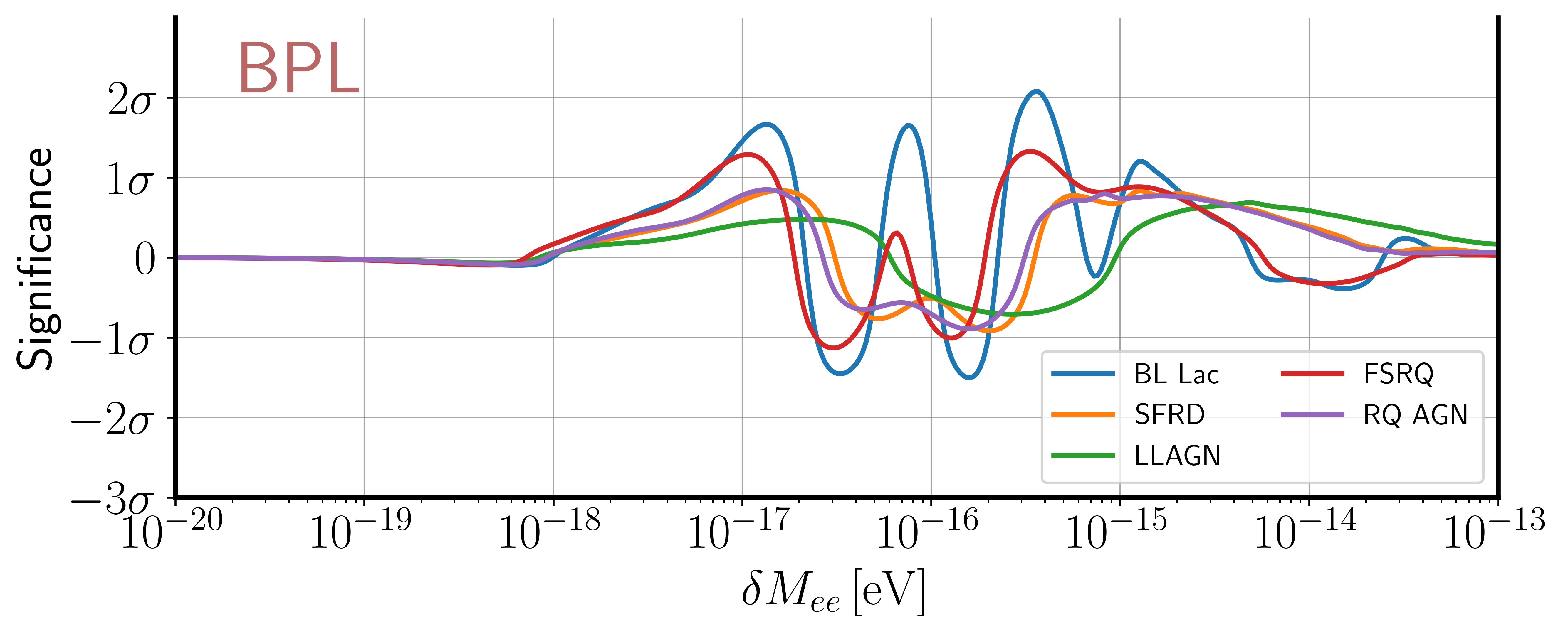}
\end{minipage}
\begin{minipage}[c]{0.48\textwidth}
  \centering
  \includegraphics[width=\textwidth]{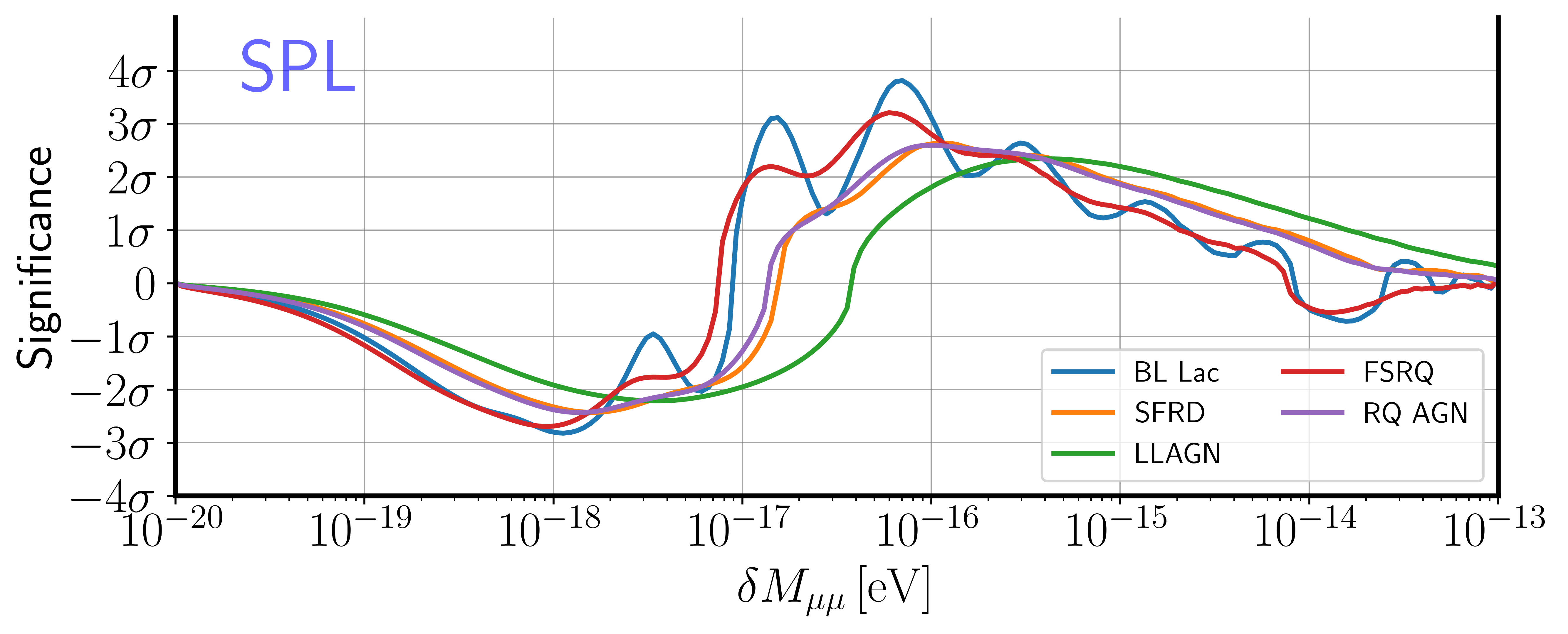}
\end{minipage}
\quad
\begin{minipage}[c]{0.48\textwidth}
  \centering
  \includegraphics[width=\textwidth]{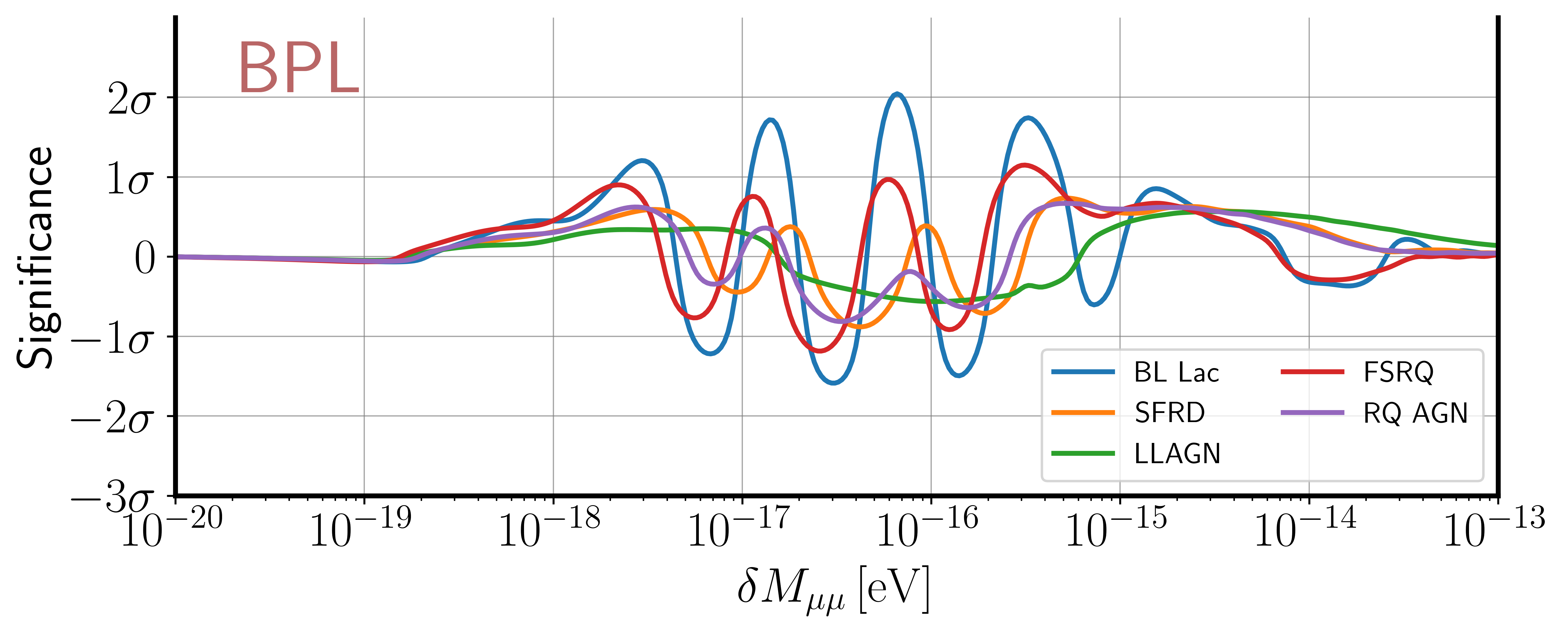}  
\end{minipage}
\begin{minipage}[c]{0.48\textwidth}
  \centering
  \includegraphics[width=\textwidth]{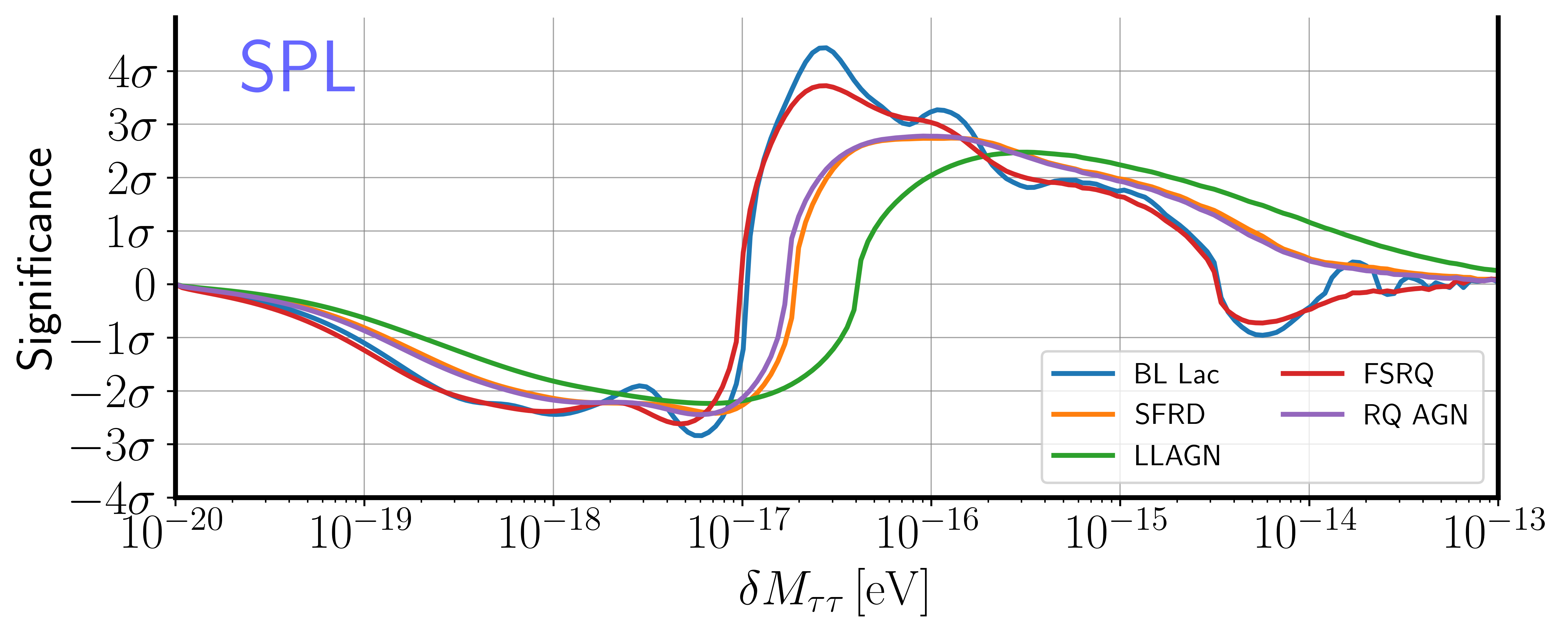}
\end{minipage}
\quad
\begin{minipage}[c]{0.48\textwidth}
  \centering
  \includegraphics[width=\textwidth]{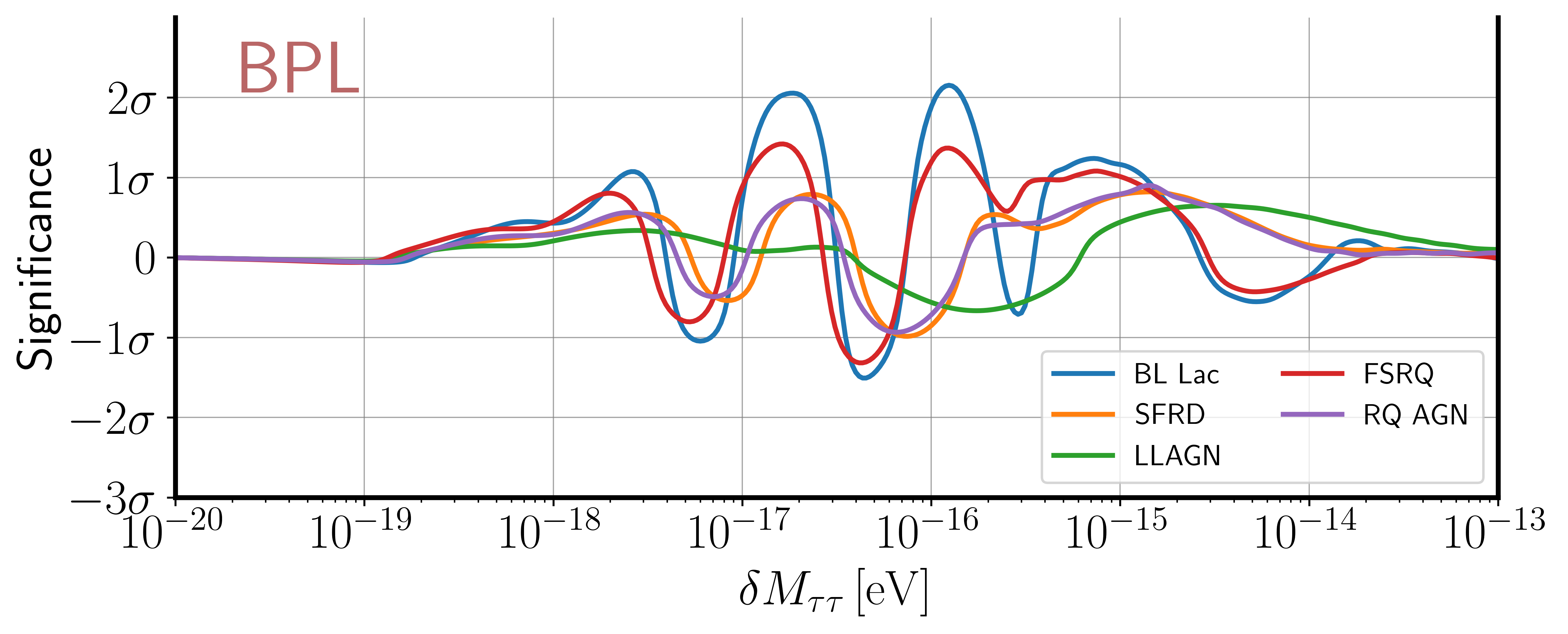}  
\end{minipage}

\caption{\label{fig:diff_shape_combined} 
Significance curves for a single diagonal SNSI mass parameter $\delta M_{\alpha \alpha}$ obtained from the diffuse-flux IceCube \emph{CombinedFit}: cascades $+$ tracks sample. In each row the left panel assumes an SPL flux, while the right panel assumes a BPL flux. For each choice of $\delta M_{\alpha\alpha}$ we construct a single total event sample by summing the expected track-like and cascade-like contributions in the presence of SNSI and comparing to the total number of neutrino events inferred from the published IceCube \emph{CombinedFit} all-flavor flux.
}
\end{figure}

\Cref{fig:diff_flav} shows the energy-averaged flavor composition at Earth for an isotropic \emph{diffuse} astrophysical neutrino flux in the presence of SNSIs and assuming source flavor ratios corresponding to pion-decay (1/3:2/3:0).
We visualize the predicted flavor ratios at Earth on the ternary (flavor-triangle) plane, where each vertex corresponds to a pure flavor state and each axis gives the relative fraction of $(\nu_e,\nu_\mu,\nu_\tau)$.
We show our results for diagonal SNSI parameters $\delta M_{ee},\delta M_{\mu\mu},\delta M_{\tau\tau}$ taken one at a time. 
Scanning over $\delta M_{\alpha\alpha}$ traces out a connected region in the flavor triangle that encodes the imprint of SNSI-induced pseudo-Dirac oscillations on the flavor ratios measured at Earth.
Also shown in the figure are the 68\% and 95\% confidence level (CL) contours obtained from IceCube’s analysis of the flavor composition of high-energy astrophysical neutrinos~\cite{Abbasi:2025fjc}.
The existing IceCube limits on the flavor ratio lack the precision needed to distinguish standard oscillations from possible new-physics effects, preventing any firm conclusions.
Next-generation observatories such as IceCube-Gen2~\cite{IceCube-Gen2:2020qha}, which will collect far more events and feature enhanced instrumentation, should greatly improve flavor-composition sensitivity. 
Accordingly, \cref{fig:diff_flav} also overlays the forecast 68\% and 99\% CL regions expected from IceCube-Gen2 after ten years of data, based on the projections in Ref.~\cite{IceCube-Gen2:2023rds}.
Parameter points that fall outside the dashed green 99\% contour would be testable (and potentially excluded) by IceCube-Gen2 at 99\% CL via flavor measurements. 
%We see that with IceCube-Gen2, it will become possible to differentiate the standard oscillation picture from SNSI induced pseudo-Dirac model predictions.

The right panels in each row display the corresponding projected sensitivities of IceCube-Gen2 in $\delta M_{\alpha \beta}$ parameter space corresponding to various assumed source distributions.
The sensitivity is primarily across $(10^{-18},10^{-16})$ eV range, where the flavor composition is displaced away more from the standard-oscillation expectation while remaining resolvable by the experiment.
Varying the source-evolution function $\rho(z)$ changes the strength with which some intervals in $\delta M_{\alpha\alpha}$ can be disfavored, but the locations of the main sensitivity regions in parameter space remain broadly stable.
We also observe the anticipated trends for extreme values of $\delta M_{\alpha \beta}$. For low $\delta M_{\alpha \beta}$, the system aligns with conventional three-flavor neutrino oscillations. 
On the other hand, high values of $\delta M_{\alpha \beta}$ lead to rapid oscillations which get averaged out due to the finite energy resolution of the detector. 
This averaging effectively halves the observed flux of active neutrinos while leaving the flavor ratios unchanged. This observation is valid for all three parameters $\delta M_{ee}, \delta M_{\mu\mu}$ and $\delta M_{\tau\tau} $. 

Overall, these results demonstrate that diffuse-flux flavor composition provides a sensitive and complementary probe of SNSI induced pseudo-Dirac phenomenology.
While current IceCube data are not yet precise enough to decisively test these effects, IceCube-Gen2 has the potential to discriminate standard oscillations from SNSI-induced pseudo-Dirac predictions over a broad and well-motivated range of $\delta M_{\alpha\alpha}$.

\subsection{Diffuse flux spectral analysis}

\Cref{fig:diff_shape} displays the signed significance obtained when the pseudo-Dirac hypothesis with a single SNSI parameter, $\delta M_{\alpha\alpha}$, is fitted to the binned all-flavor flux extracted from the cascade and ESTES samples.
In each row, the left panel shows the diffuse-flux spectrum for the cascade and ESTES data together with two representative best-fit pseudo-Dirac distortions obtained by turning on a single diagonal parameter $\delta M_{\alpha\alpha}$: one example each for the SPL and BPL hypotheses (both shown as dashed curves). 
The right panel shows the corresponding one-dimensional signed Gaussian significance as a function of $\delta M_{\alpha\alpha}$, evaluated separately for SPL and BPL and for each source-evolution model. We adopt the convention that a negative (positive) signed significance indicates an improved (degraded) fit relative to the standard-oscillation case.
For $\delta M_{ee}$, the SPL case exhibits a mild preference at the level of $\sim 2\sigma$, with the minimum occurring around $\delta M_{ee}\sim  \mathrm{few}\times 10^{-18}\,\mathrm{eV}$ (with mild source model variations). In contrast, a broad region at $\delta M_{ee}\sim \mathrm{few}\times 10^{-17}$--$10^{-16}\,\mathrm{eV}$ is disfavored, reaching up to $\sim (2.5$--$3)\sigma$ for BL Lac and FSRQ source evolution models.
For $\delta M_{\mu\mu}$, the SPL curves show a mild preference (typically $\lesssim 1$--$1.5\sigma$) around $\delta M_{\mu\mu}\sim \mathrm{few}\times 10^{-18}$--$10^{-17}\,\mathrm{eV}$, followed by a stronger degradation of the fit near $\delta M_{\mu\mu}\sim 10^{-16}\,\mathrm{eV}$, where the signed significance peaks at $\sim (2$--$3)\sigma$ depending on the evolution model.
For $\delta M_{\tau\tau}$, the SPL case shows the clearest preference among the three diagonal parameters: the curves reach a minimum of about $-2\sigma$ to $-2.5\sigma$ near $\delta M_{\tau\tau}\sim 10^{-17}\,\mathrm{eV}$. A pronounced disfavored region follows at $\delta M_{\tau\tau}\sim \mathrm{few}\times 10^{-17}$--$10^{-16}\,\mathrm{eV}$, with maxima approaching $\sim 3\sigma$ (again with the strongest excursions for the more structured evolution models). 
For all three in the BPL case, the significance curves become more oscillatory, and the overall strengths of significance are reduced, with multiple local extrema at the $\lesssim \mathcal{O}(1$--$2)\sigma$ level.
Across the different source evolution $\rho(z)$  model templates, the height of these significance excursions changes.
Also note that for $\delta M_{\alpha\alpha}\gtrsim \mathrm{few}\times 10^{-15}\,\mathrm{eV}$ in all channels, the oscillation features become too rapid to be resolved once the experimental energy resolution is taken into account, and the signed significance smoothly returns toward zero.

We also show in \cref{fig:diff_shape_combined}, the signed significance from IceCube's 2025 \emph{CombinedFit}~\cite{IceCube:2025ewu} as a function of a single diagonal SNSI parameter $\delta M_{\alpha \alpha}$, for the same set of source-evolution models. The left (right) column corresponds to the SPL (BPL) flux hypothesis. 
The curves reflect the total sensitivity of the combined track+cascade sample to each $\delta M_{\alpha\alpha}$; they do not resolve flavor-dependent effects as only the combined $\chi^2$ is available.
Overall, the combined-fit curves closely track the structures seen in the cascade$+$ESTES analysis and do not qualitatively alter the allowed/preferred pseudo-Dirac windows inferred from that study.
Instead, the combined fit modestly sharpens the sensitivity, yielding somewhat more significant excursions in signed significance and more clearly resolved preferred/disfavored windows in $\delta M_{\alpha \alpha}$.

\subsection{Point-source spectral analysis}
\begin{figure}[tbp]
\centering
\begin{minipage}[c]{0.48\textwidth}
  \centering
  \includegraphics[width=\textwidth]{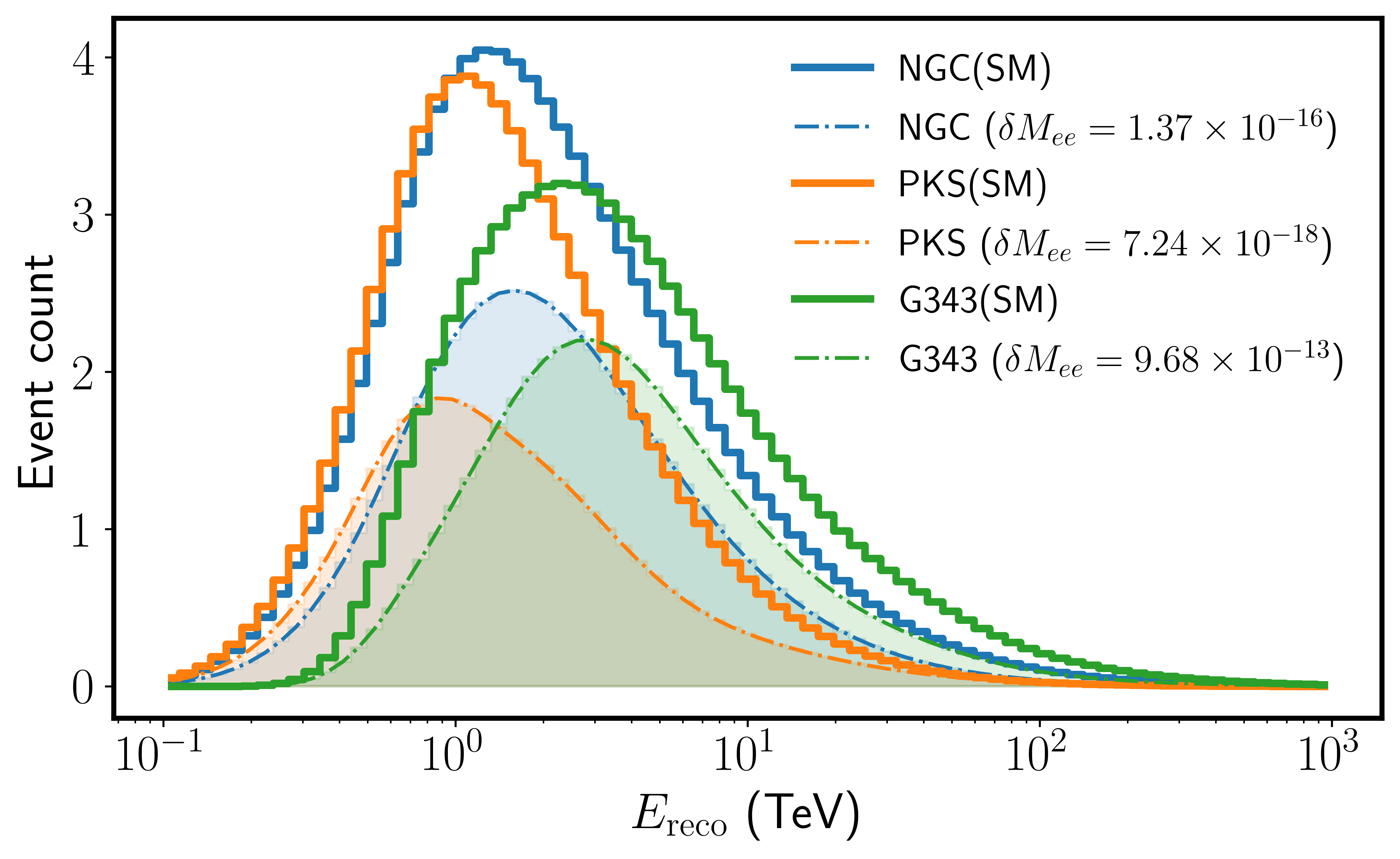}
\end{minipage}
\quad
\begin{minipage}[c]{0.48\textwidth}
  \centering
  \includegraphics[width=\textwidth]{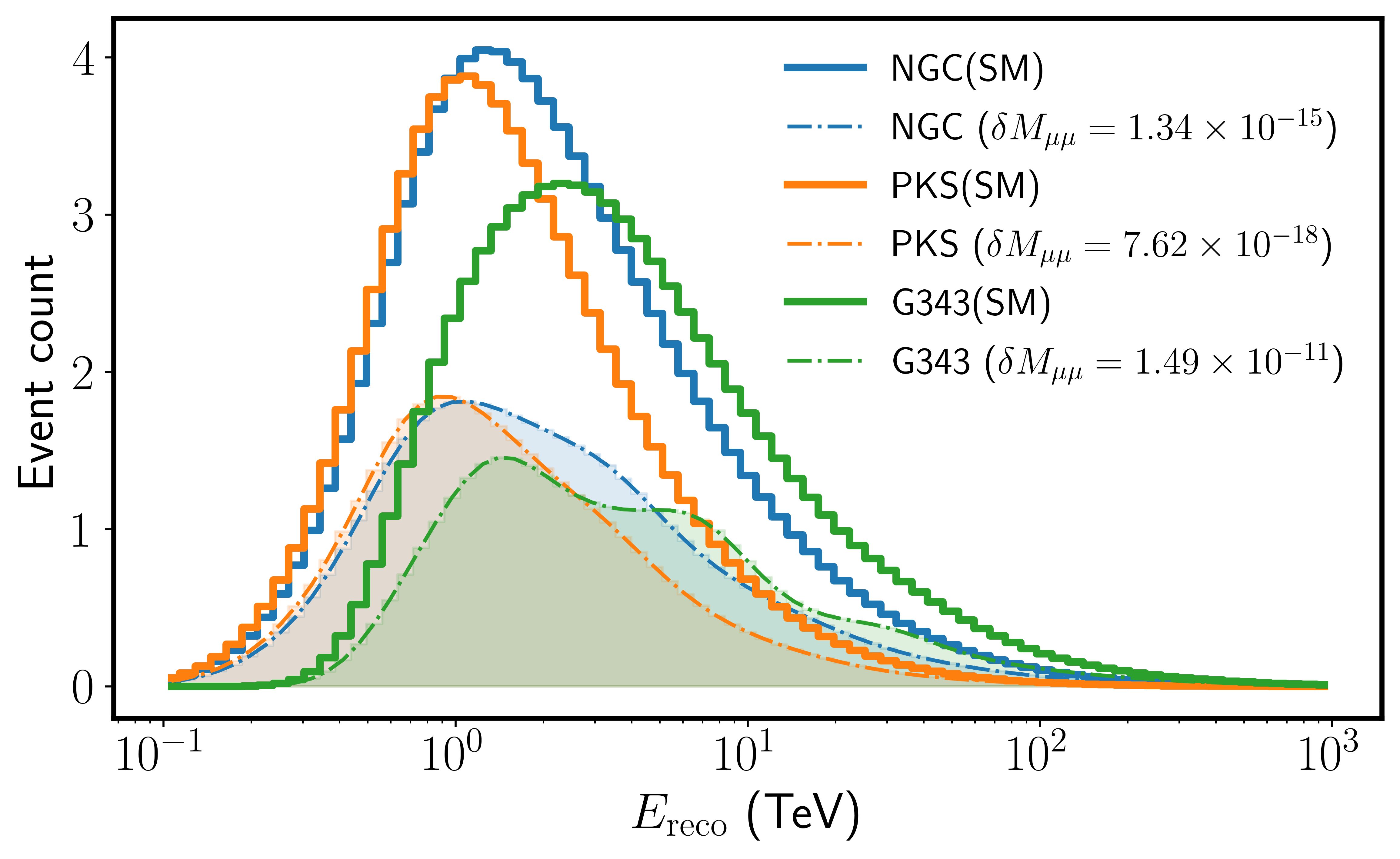}
\end{minipage}

\begin{minipage}[c]{0.48\textwidth}
  \centering
  \includegraphics[width=\textwidth]{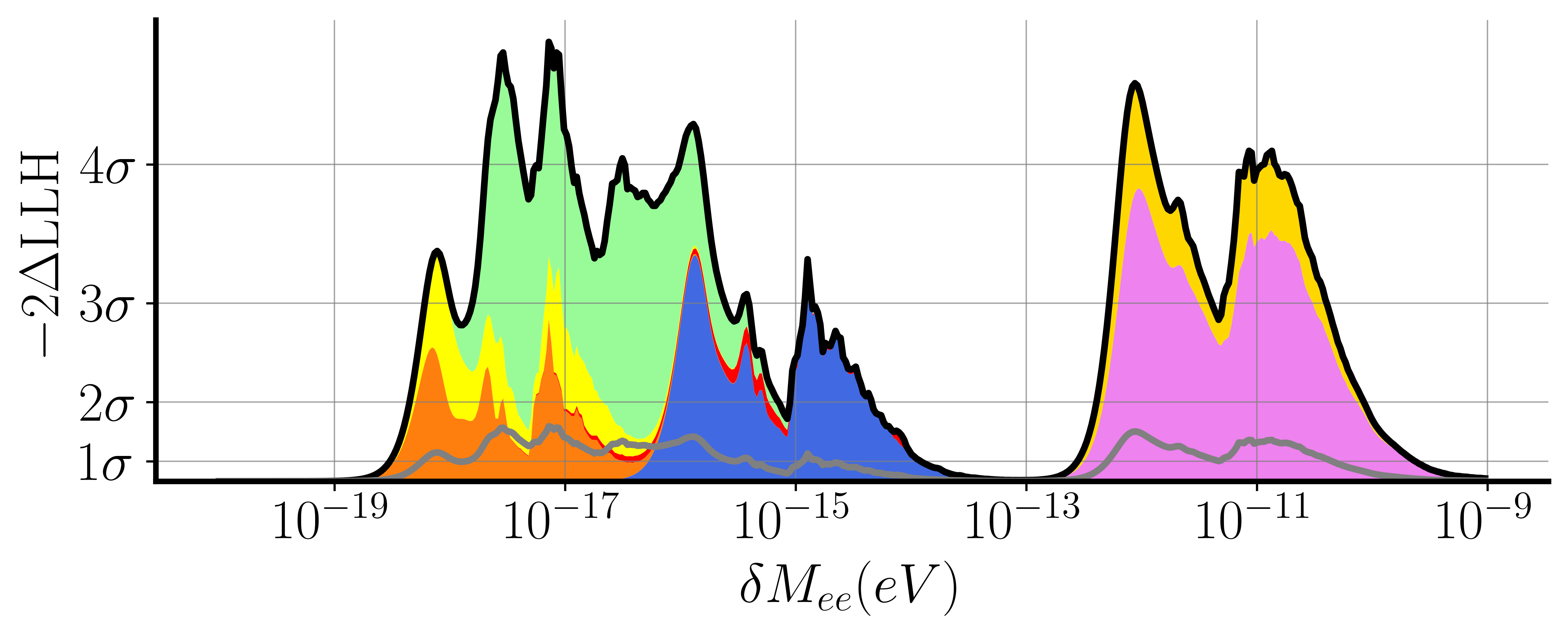}
\end{minipage}
\quad
\begin{minipage}[c]{0.48\textwidth}
  \centering
  \includegraphics[width=\textwidth]{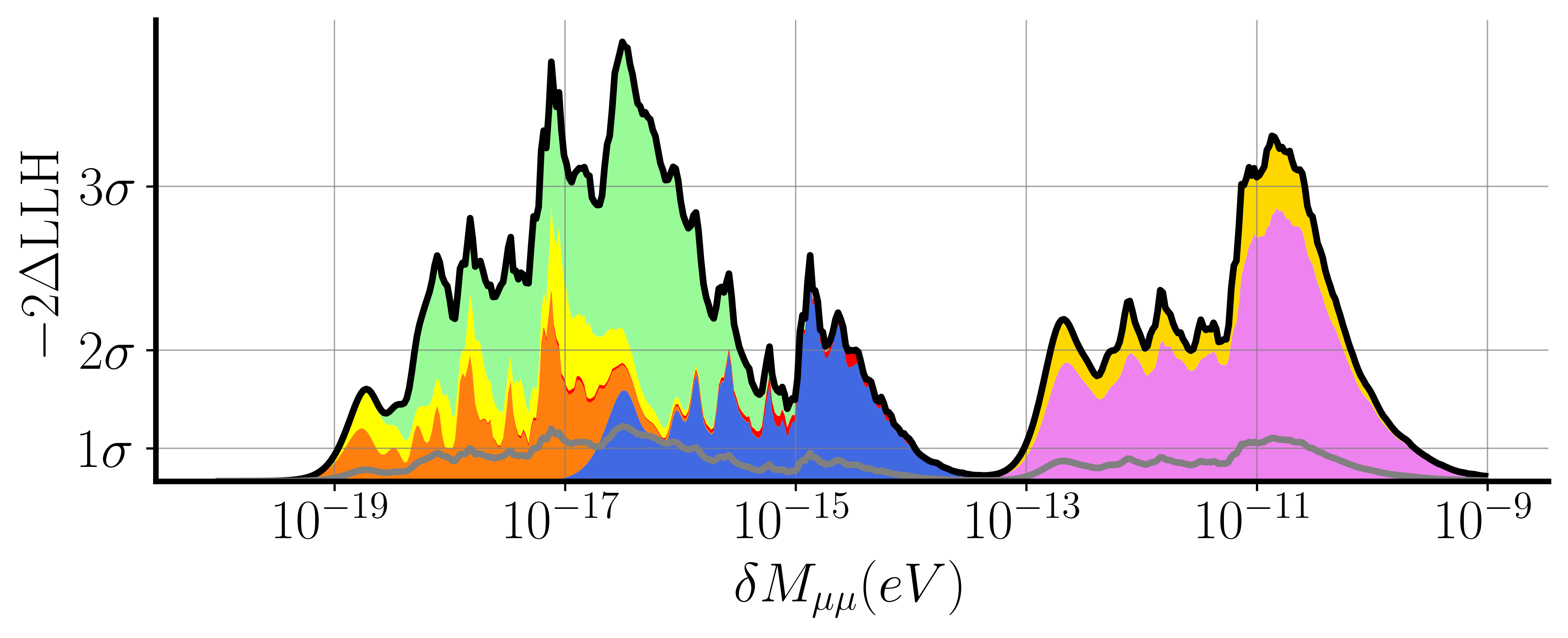}  
\end{minipage}

\begin{minipage}[c]{0.48\textwidth}
  \centering
  \includegraphics[width=\textwidth]{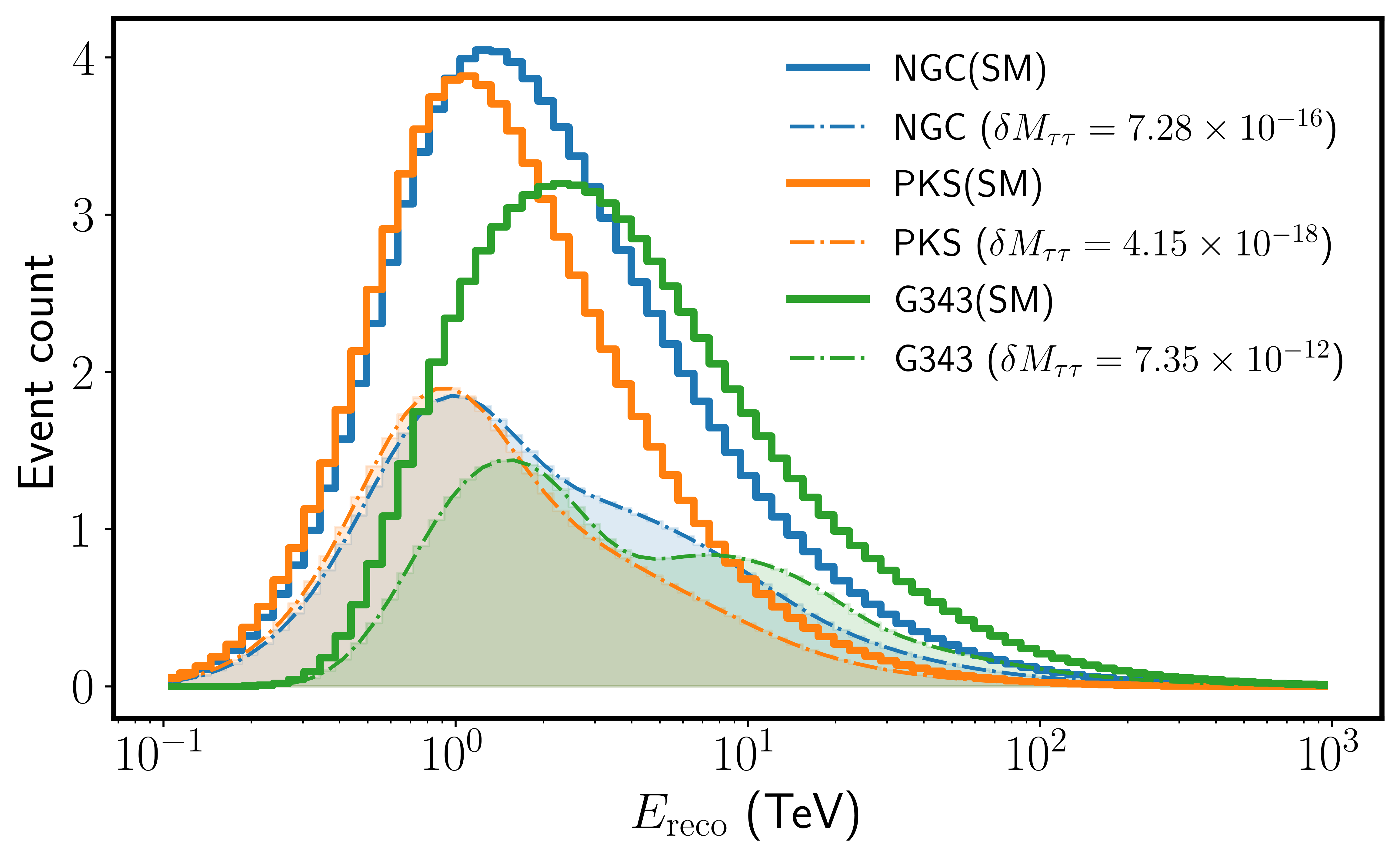}
\end{minipage}

\begin{minipage}[c]{0.60\textwidth}
  \centering
  \includegraphics[width=\textwidth]{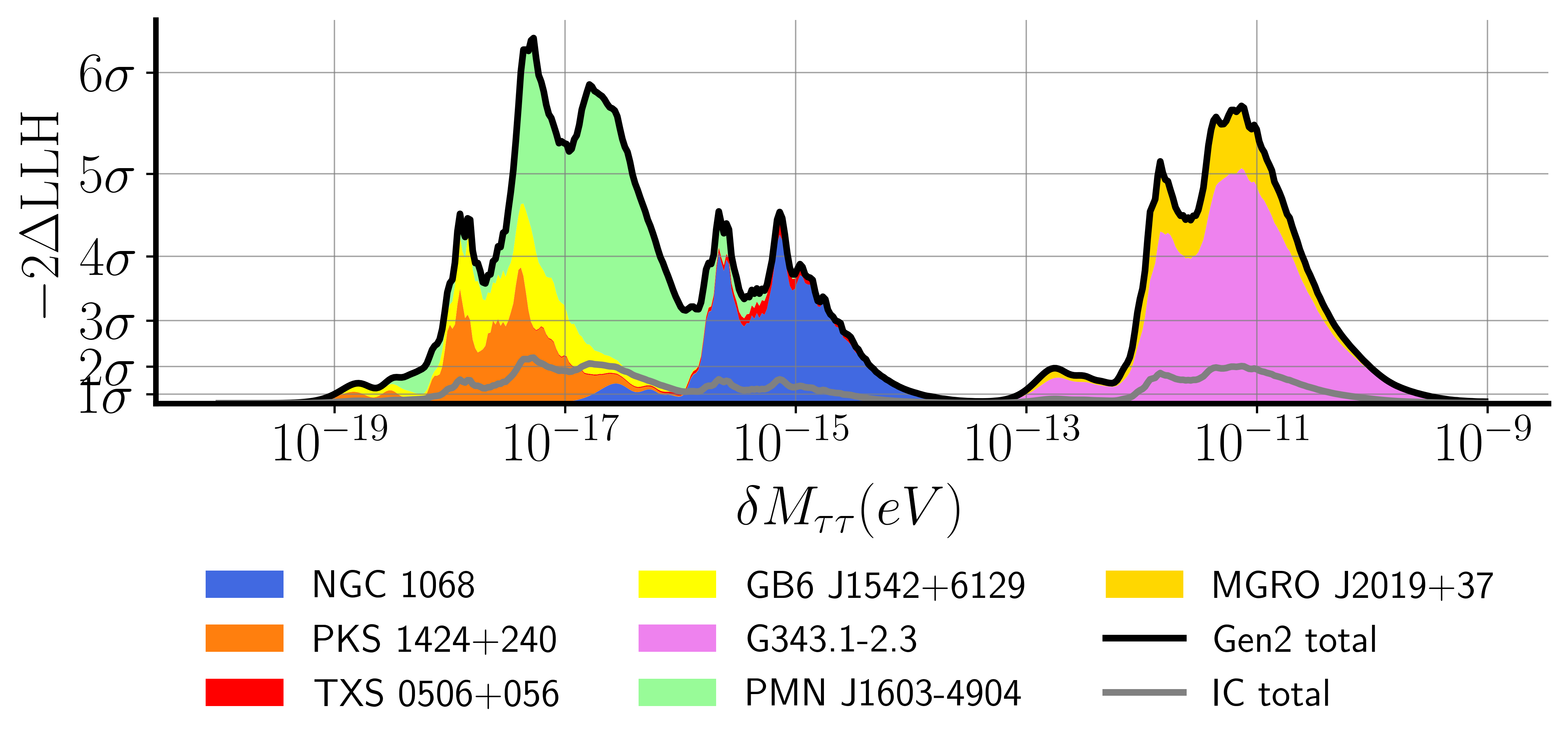}  
\end{minipage}

\caption{\label{fig:pt_shape} Point-source spectral-shape analysis for diagonal SNSI parameters. Top/middle/bottom rows correspond to $\delta M_{e e,} \delta M_{\mu \mu,}$ and $\delta M_{\tau \tau}$, respectively. Top panels show the reconstructed energy spectra for the IceCube point-source sample (tracks+cascades) in the SM (solid histograms) and with SNSI pseudo-Dirac splittings (dash-dotted curves) at the representative $\delta M_{\alpha \alpha}$ values indicated. Bottom panels show the profiled test statistic $-2 \Delta \mathrm{LLH}\left(\delta M_{\alpha \alpha}\right)$ relative to the SM. The thick gray (black) curve gives the current IceCube (Gen2) total, while colored filled components show the individual-source contributions.}
\end{figure}
\Cref{fig:pt_shape} collects the outcome of the spectral–shape fit performed on the expected event distribution from several high-significance neutrino point sources identified by IceCube.
The upper panels show the reconstructed-energy spectra \(E_{\rm reco}\) expected for each source in the SM (solid histograms) and in the presence of SNSI induced pseudo-Dirac splittings (dash-dotted curves). For each source, the dashed curve is shown at the representative value of \(\delta M_{\alpha\alpha}\) indicated in the legend, chosen to illustrate the characteristic distortion of the spectral shape.
Across the benchmarks displayed, pseudo-Dirac oscillations primarily reshape the spectrum but also produce a depletion of the peak due to oscillations to sterile states.
In addition, because the sources span different distances and have different spectral indices, each one would be most sensitive in the $\delta M_{\alpha \alpha}$ range where the pseudo-Dirac induced spectral modulation falls within the detector’s most populated $\mathcal{O}(\mathrm{TeV})$ energy region where the event statistics are largest.  Consequently, each source probes a distinct slice of the  \(\delta M_{\alpha\alpha}\) parameter space. 

The lower panels show the profiled test statistic, \(-2\Delta{\rm LLH}(\delta M_{\alpha\alpha})\), evaluated relative to the SM (\(\delta M_{\alpha\alpha}=0\)).
The thick gray line denotes the current total sensitivity from IceCube's track and cascades dataset, while the thick black line shows the corresponding forecast for IceCube-Gen2 assuming an eightfold increase in statistics with the same systematic treatment. The colored filled components indicate the individual-source contributions to the Gen2 total. Over most of the scanned parameter space, the present IceCube sensitivity remains below the nominal \(1\sigma\) level, reflecting the limited point-source statistics and the fact that the discrimination power comes mainly from subtle spectral-shape distortions rather than large rate changes.
In contrast, the Gen2 combination develops broad sensitivity bands and multiple localized maxima.
This multimodal structure in \(-2\Delta{\rm LLH}(\delta M_{\alpha\alpha})\) arises because $\delta M_{\alpha\alpha}$ does not map to a single oscillation frequency. Rather, for a given $\delta M_{\alpha\alpha}$, the probability receives three contributions--one for each mass eigenstate  $j=1,2,3$--through the effective splittings $\delta m_j^2\left(\delta M_{\alpha \alpha}\right)$ that enter the phase. Each $j$ therefore imprints a distinct characteristic $\delta M_{\alpha \alpha}$ window where the phase is near an extremum (modulo $2\pi$), generating multiple preferred regions rather than a single peak. 
When multiple sources are combined, these eigenstate-dependent windows further split and broaden in a source-dependent way primarily through different baselines (setting the preferred $\delta M_{\alpha \alpha}$ scale) and different spectral weights and detector response (setting the peak breadth). The superposition of the three eigenstate contributions across several sources naturally produces primary maxima accompanied by secondary shoulders.
The combined test statistic develops two separated sensitivity regions: a low-\(\delta M_{\alpha\alpha}\) band sourced primarily by extragalactic baselines and a high-\(\delta M_{\alpha\alpha}\) band sourced by Galactic baselines, reflecting the characteristic \(L/E\) scaling of the oscillation phase. 
The joint Gen2 analysis becomes sensitive at the \(\gtrsim 2\sigma\) level across wide intervals, roughly \(\delta M_{\alpha\alpha}\sim 10^{-19}\)--\(10^{-14}~{\rm eV}\) from extragalactic sources and \(\delta M_{\alpha\alpha}\sim 10^{-12}\)--\(10^{-10}~{\rm eV}\) from Galactic sources.  
NGC~1068, PKS~1424+240, and PMN~J1603$-$4904 provide the dominant leverage in the extragalactic regime, and among the Galactic sources G343.1-2.3 probes most significantly in substantially larger splittings toward the \(\delta M_{\alpha\alpha}\sim 10^{-12}\)--\(10^{-10}~{\rm eV}\) regime. 
In the combined likelihood, TXS~0506+056 contributes negligibly because only $\sim 9$ track+cascade events are attributed to that source in the current samples. Note that contributions from G343.1$-$2.3 and PMN~J1603$-$4904 are largely cascade-dominated, with no constraining power from track-like events in the present point-source sample.

\subsection{Combined sensitivities}
\begin{figure}[tbp]
    \centering
    \includegraphics[width=0.7\linewidth]{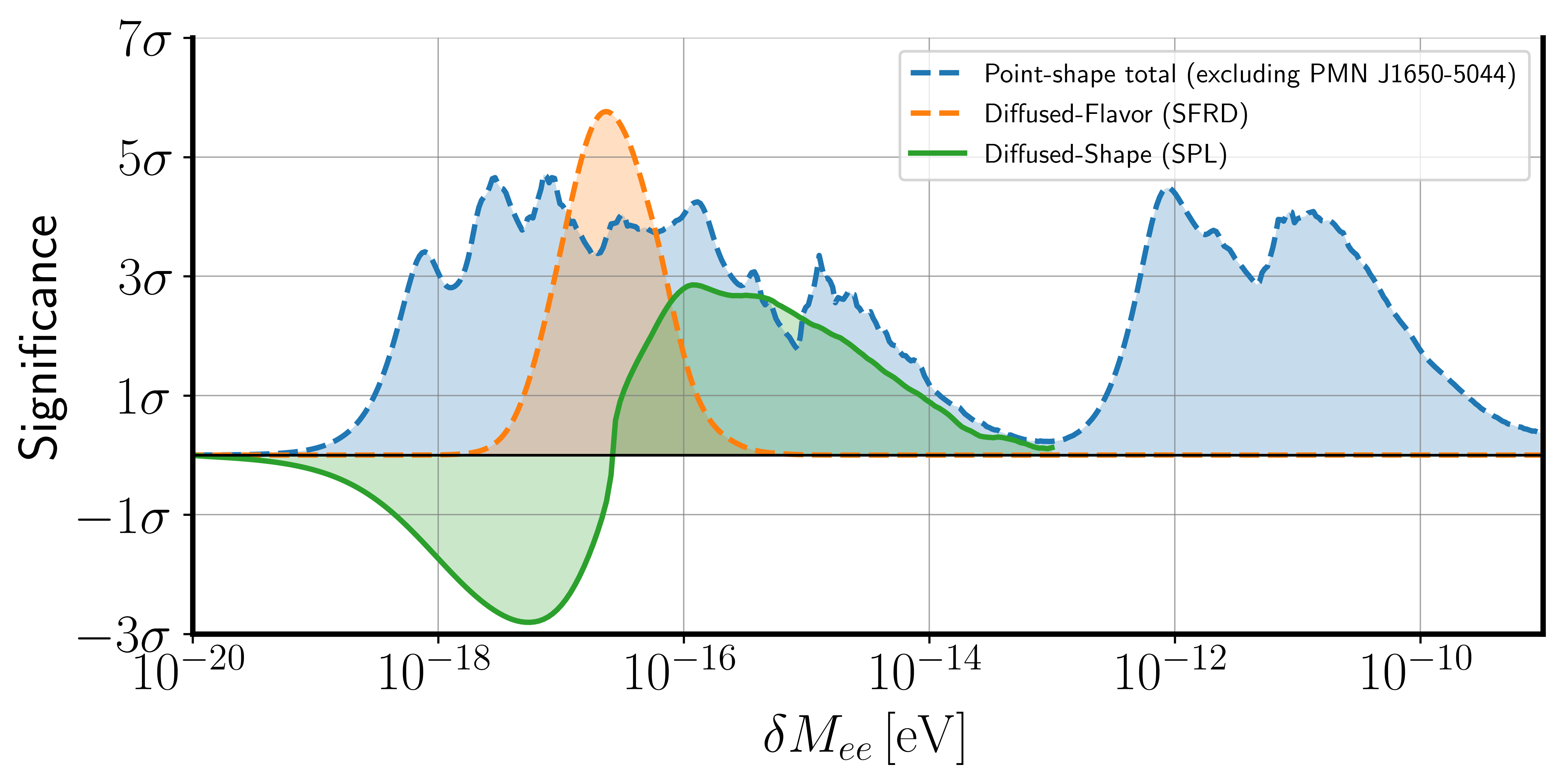}
           \includegraphics[width=0.7\linewidth]{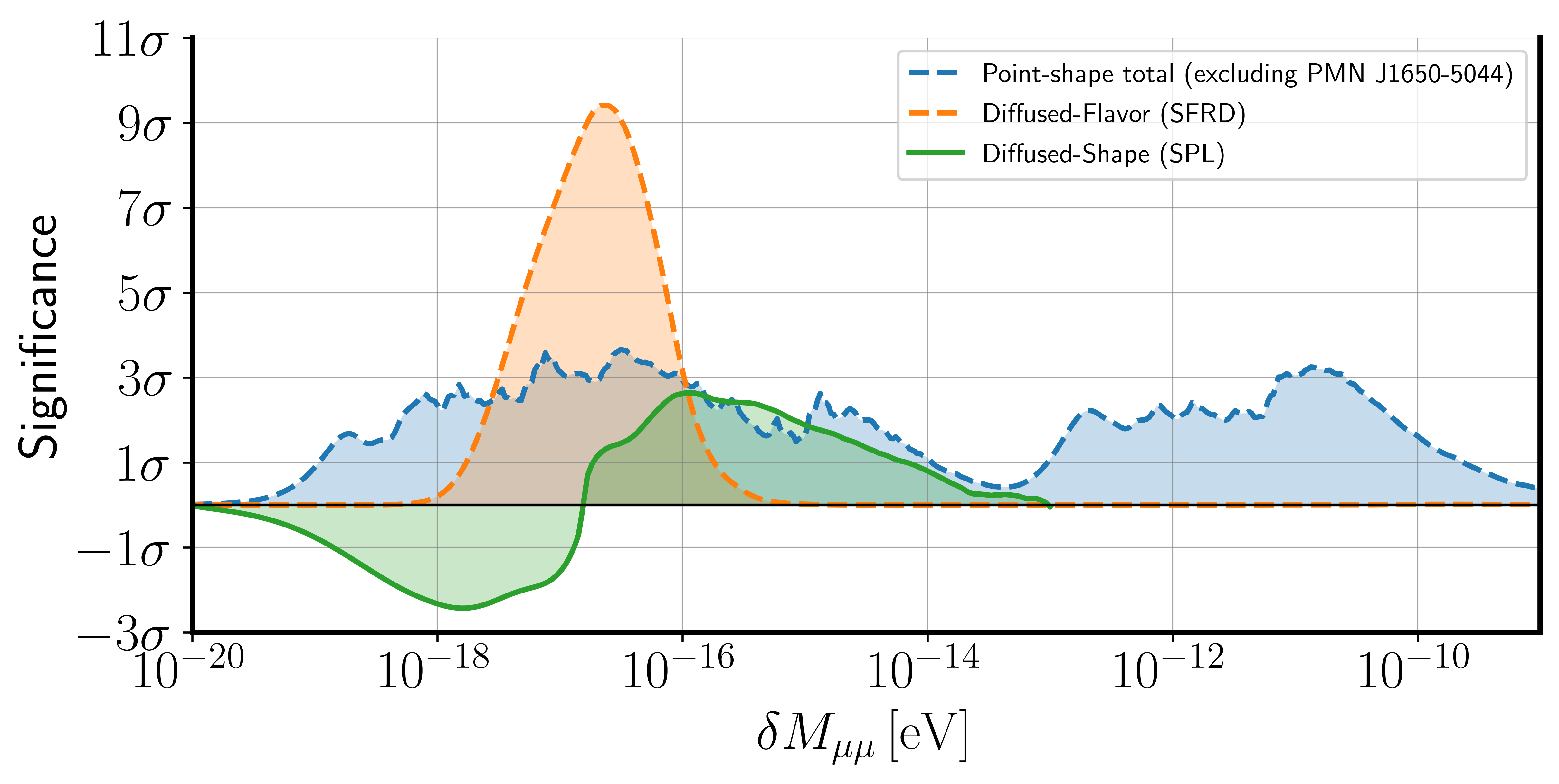}
     \includegraphics[width=0.7\linewidth]{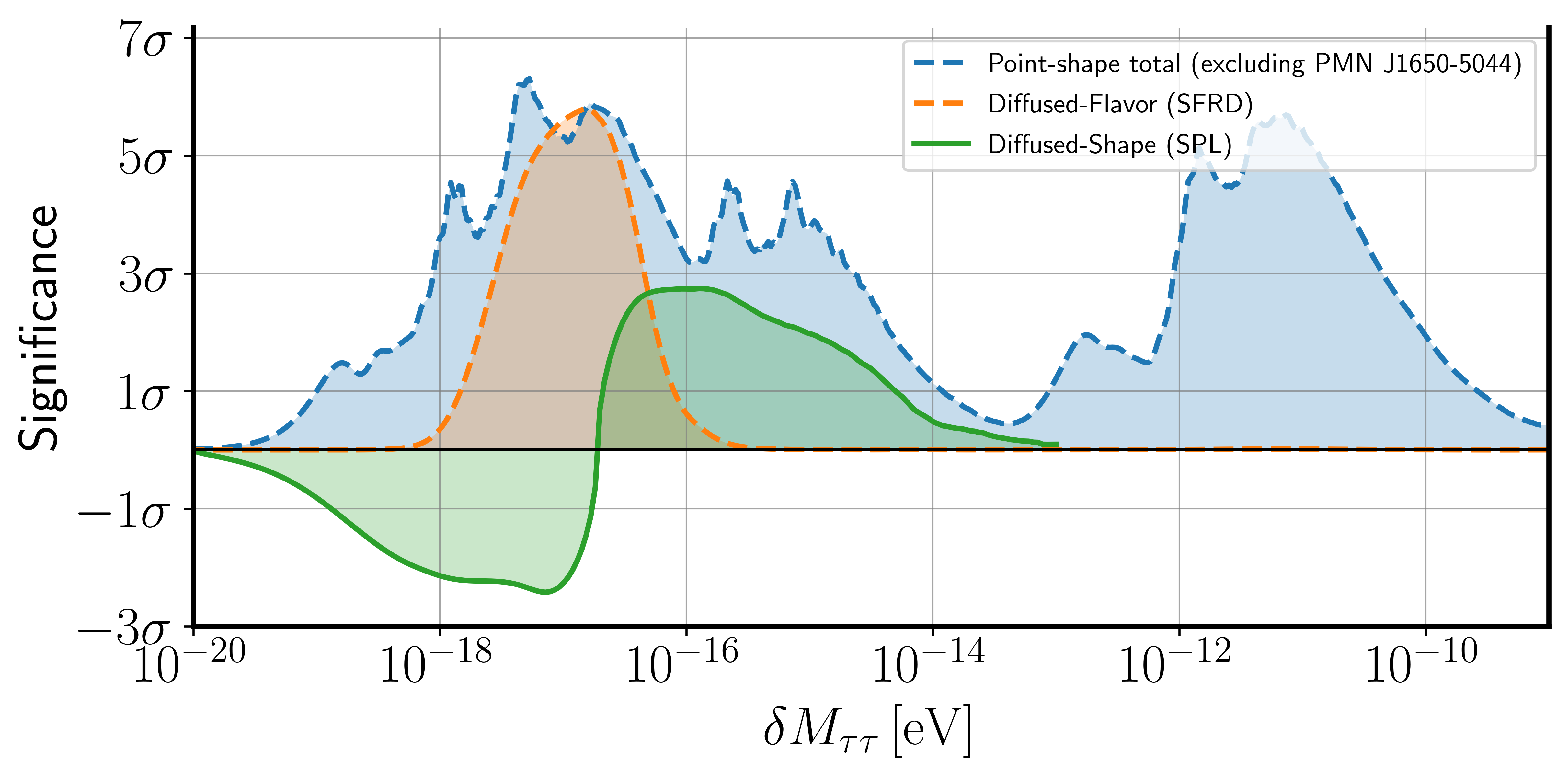}

    \caption{Overlay of signed significances for $\delta M_{ee}, \delta M_{\mu \mu}$ and $\delta M_{\tau\tau}$ obtained from three complementary channels: current diffuse-flux spectral-shape constraints (green, solid), projected diffuse-flux flavor-triangle likelihood (orange, dashed) and projected point-source spectral fits (blue, dashed). The diffuse curves assume SFRD source evolution and a single power law spectrum. Positive ordinates indicate exclusions relative to standard three-flavor oscillations, negative ordinates indicate a mild preference for the SNSI scenario. }
    \label{fig:sig_combined}
\end{figure}

\Cref{fig:sig_combined} finally overlays the significances obtained for each diagonal SNSI parameter $\delta M_{e e}$, $\delta M_{\mu \mu}$, $\delta M_{\tau \tau}$ from the three complementary IceCube/Gen2 strategies considered in this work:
\begin{enumerate}
\item \textbf{Point-source spectral fit}  (blue, dashed);
\item \textbf{Diffuse-flux flavor-triangle} likelihood (orange, dashed);
\item \textbf{ Current limits from Diffuse–flux spectral shape} analysis using the IceCube \emph{CombinedFit}: cascades $+$ tracks sample (green, solid).
\end{enumerate}
We use the line style to separate present constraints from future reach:  solid curves denote current IceCube results, and dashed curves denote IceCube-Gen2 projections. 
Throughout, positive values of the significance correspond to exclusions with respect to the standard oscillation picture, while negative values indicate a mild preference for the SNSI scenario over the SM expectation. 
The diffuse-flux channels exhibit their strongest sensitivities in the intermediate region $\delta M_{\alpha\alpha}\sim 10^{-18}$--$10^{-14}\,\mathrm{eV}$: the flavor likelihood features a pronounced peak around $\delta M_{\alpha\alpha}\sim 10^{-16}\,\mathrm{eV}$, while the current spectral shape data show a mild preference at $\delta M_{\alpha\alpha}\sim 10^{-18}$--$10^{-17}\,\mathrm{eV}$ and a modest exclusion band around $\delta M_{\alpha\alpha}\sim 10^{-16}$--$10^{-15}\,\mathrm{eV}$. 
At larger splittings, $\delta M_{\alpha\alpha}\gtrsim 10^{-14}\,\mathrm{eV}$, these diffuse observables lose resolving power as rapid oscillations wash out both flavor and energy information. 
By contrast, the point-source channel retains sensitivity and develops a second, high-$\delta M_{\alpha\alpha}$ band at $\delta M_{\alpha\alpha}\sim 10^{-12}$--$10^{-10}\,\mathrm{eV}$ (driven by shorter Galactic baselines), while continuing to probe the lower-$\delta M_{\alpha\alpha}$ region via extragalactic sources. 
For very small splittings, $\delta M_{\alpha\alpha}\lesssim 10^{-19}\,\mathrm{eV}$, all channels lose sensitivity because the induced phases remain too small over the relevant baselines and energies.
We also see that future IceCube-Gen2 diffuse-flux flavor measurements and point-source spectral analyses will be able to probe the small, non-zero SNSI splittings currently preferred by the diffuse-flux spectral data. 
Improved cascade energy resolution and the extended Gen2 radio array should sharpen the current dip in the significance curves and either confirm or rule out the tentative preference for small $\delta M$ values by diffuse-flux spectral data. 
At the same time, an expanding list of well-characterized extragalactic neutrino sources will enhance the reach of the point-source channel. 

\begin{figure}[tbp]
\centering 
\includegraphics[width=.8\textwidth]{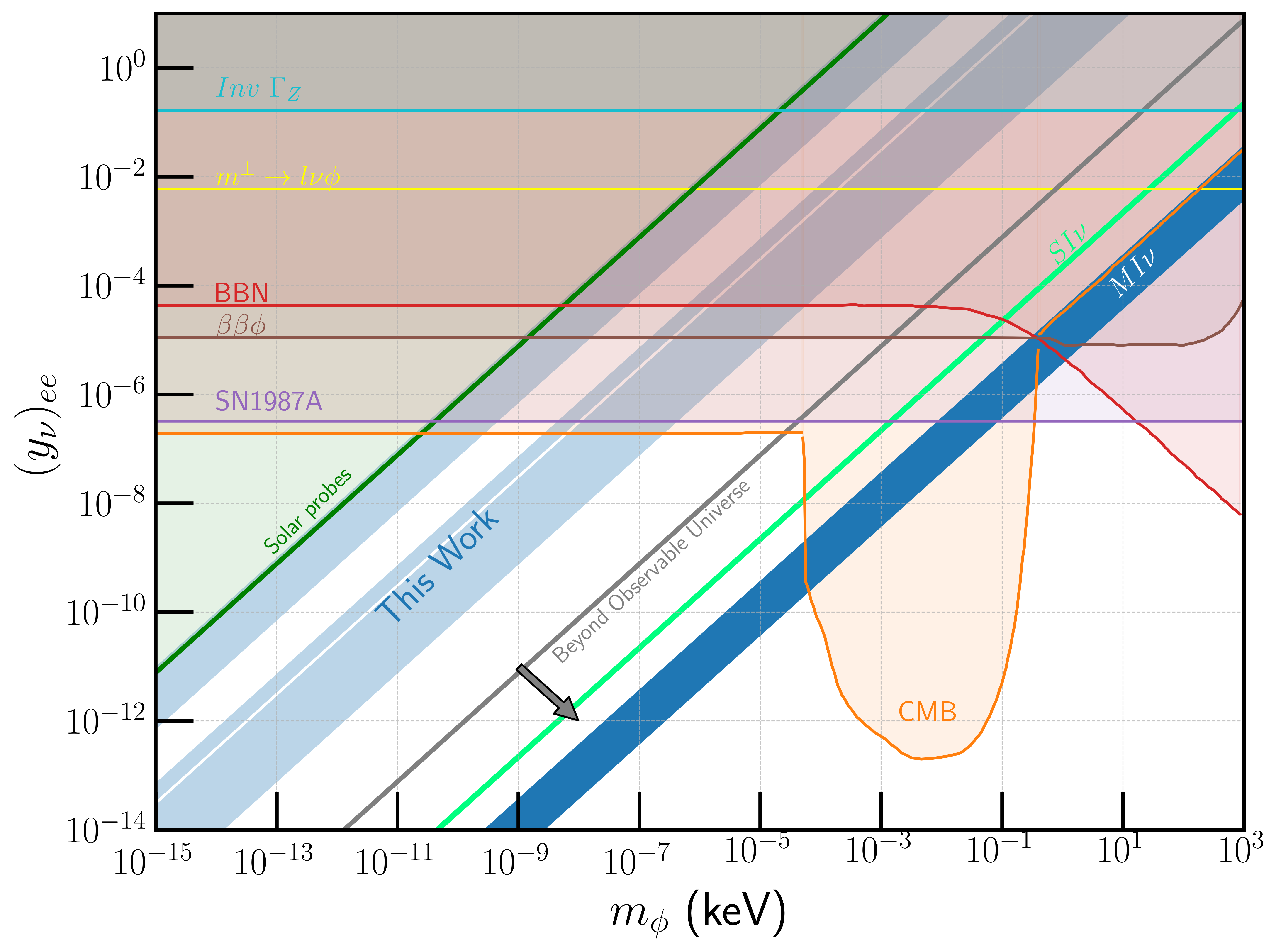}
\includegraphics[width=.48\textwidth]{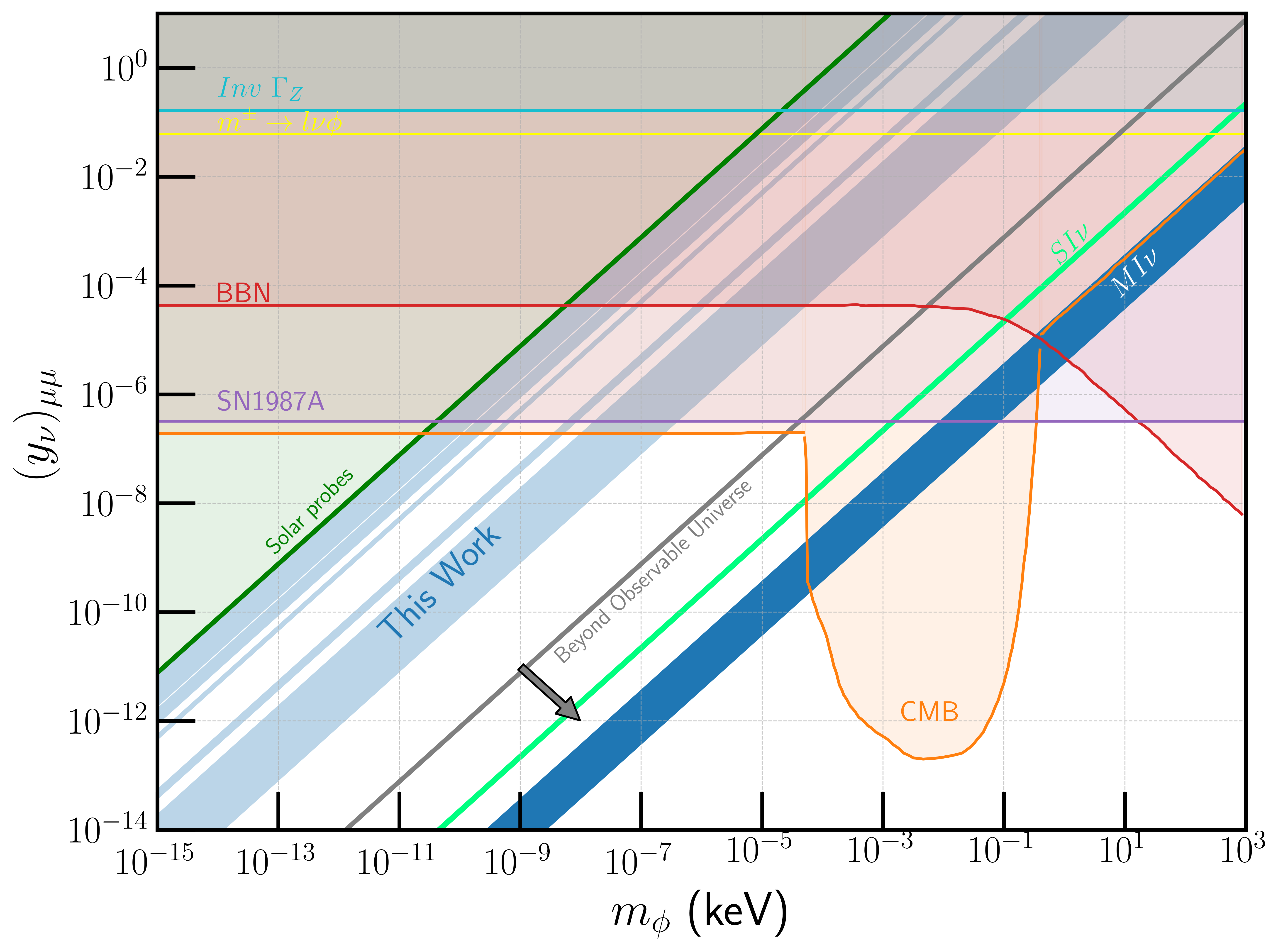}
\includegraphics[width=.48\textwidth]{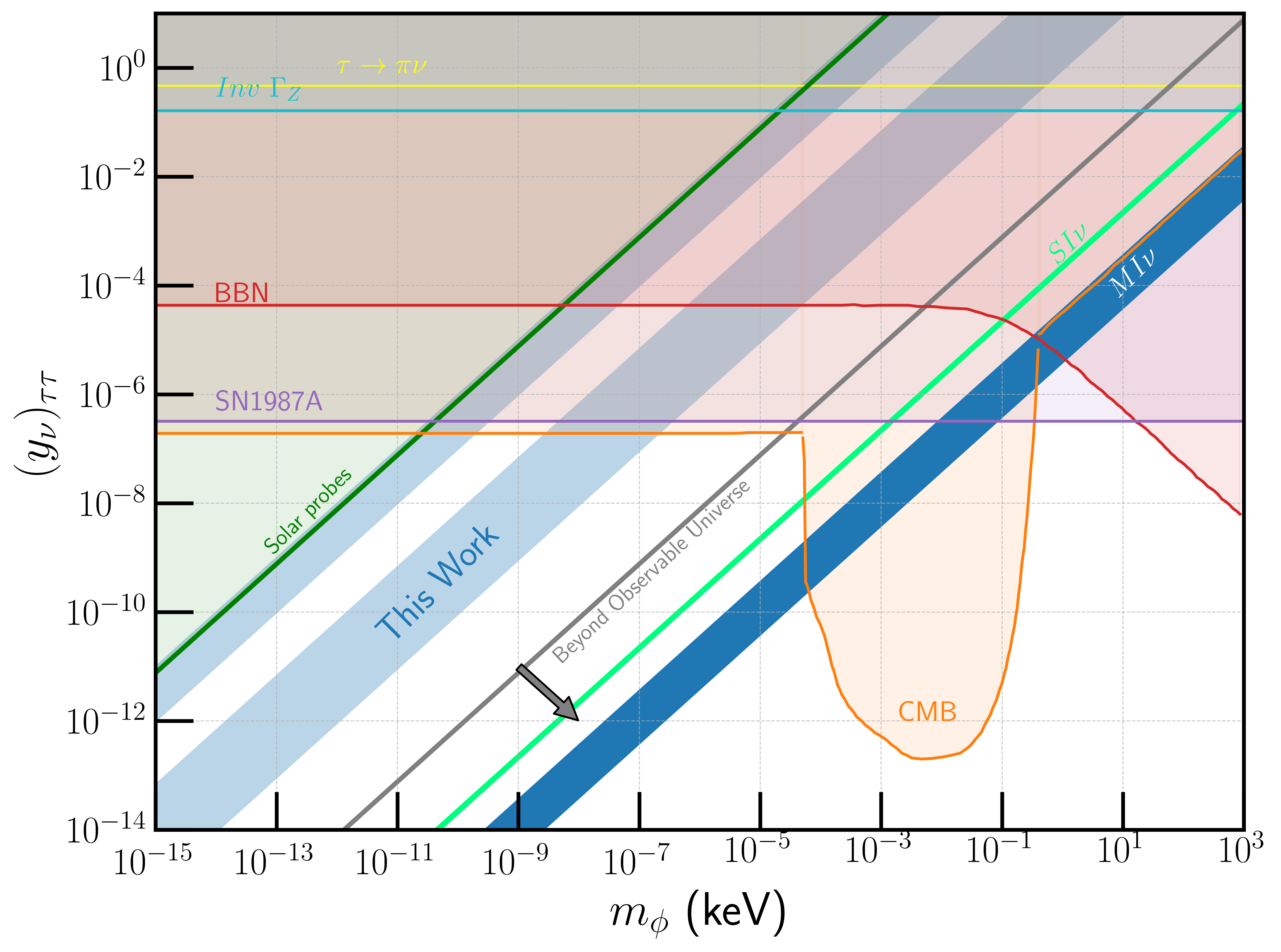}
\hfill

\caption{\label{fig:yukawaplane} 
Parameter space of a light scalar mediator with mass $m_\phi$ vs Yukawa coupling $y_\nu$ for neutrinos. The light-blue band shows the region that a combined IceCube analysis can probe at $95 \%$ CL, extending existing laboratory, cosmological and supernova constraints into the previously unexplored ultra-light regime $\mathrm{m}_\phi<10^{-4}~ \mathrm{eV}$ and demonstrating that high-energy astrophysical neutrinos already rival, and in this corner surpass, the reach of all other probes. }
\end{figure}

In \cref{fig:yukawaplane}, we finally translate these sensitivities and current constraints into the Yukawa coupling that controls the medium-induced pseudo-Dirac splitting in our model, $y_\nu\equiv \sqrt{|y^M y^D|}$, as a function of the light-scalar mass $m_{\phi}$. The region sensitive to the combined IceCube analysis at 95\% CL is presented in the figure as a light blue band. Notably, IceCube sensitivity extends into the ultra-light scalar mediator regime, where the induced pseudo-Dirac splitting can accumulate over astrophysical baselines. The green-shaded region on upper left corner corresponds to $\delta m^2\gtrsim 10^{-11}$ eV$^2$ which is excluded by solar neutrino data~\cite{Ansarifard:2022kvy}, while the gray region with an arrow indicates $\delta m^2\lesssim 10^{-23}$ eV$^2$ that corresponds to active-sterile neutrino oscillations over baselines exceeding the size of the observable Universe. The regions labeled SI$\nu$ and MI$\nu$ are the preferred regions to explain the Hubble tension with strong and medium neutrino self-interactions~\cite{Kreisch:2019yzn,RoyChoudhury:2020dmd}. 
%Notably, IceCube accesses an ultra-light mediator window extending to $m_\phi \lesssim 10^{-4}\,\mathrm{eV}$.
Also overlaid are the existing bounds from supernova SN1987A~\cite{Ivanez-Ballesteros:2024nws}, CMB~\cite{abenza_neutrino_nodate,Sandner:2023ptm,Poudou:2025qcx}, BBN~\cite{Babu:2019iml,Fong:2025xhh}, meson decay~\cite{Brdar:2020nbj,Dev:2024ygx}, $Z$ decay, and double beta decay ~\cite{Berryman:2022hds}.
%(the latter applies only to $(y_\nu)_{ee}$)
These constraints were generally derived in simplified models with a single scalar-neutrino coupling $y^M$ or $y^D$ and therefore should be interpreted as benchmark comparisons rather than direct exclusions in the $\sqrt{\left|y^M y^D\right|}$ plane, unless an additional model assumption, such as $y^M \simeq y^D \simeq y_{\nu}$ is imposed.
In particular, meson and $Z$-decay limits constrain scalar-emission vertices involving the individual couplings, double beta decay constrains only the lepton-number-violating electron-flavor coupling
$\left(y^M\right)_{ee}$, and the CMB, BBN, and SN1987A limits depend on thermalization, free streaming, or energy-loss rates controlled by the individual interactions. A dedicated recast would therefore be required for a fully model-independent comparison. For the BBN and supernova bounds based on in-medium neutrino masses ~\cite{Babu:2019iml,Venzor:2020ova}, we discuss this point further in \cref{app:inmedium_bounds}, where we show that these limits do not directly apply once the scalar propagator is consistently dressed.

With these qualifications, the figure illustrates that the IceCube flavor and spectral data probe an ultra-light-mediator region that is not directly covered by existing laboratory, cosmological, or supernova bounds.
%Strikingly, none of these laboratory, cosmological, or supernova probes currently provides comparable sensitivity in the ultra-light regime relevant for our scalar interaction model. 
IceCube therefore pushes directly into an unexplored region of the $(m_\phi,\,y_{\nu\alpha})$ parameter space. Thus by converting the flavor-triangle and spectral results into the ( $m_\phi, y_{\nu \alpha}$ ) plane, we show that astrophysical neutrino propagation provides a complementary and, in this ultra-light regime, uniquely sensitive probe of the mixed scalar interaction responsible for the induced pseudo-Dirac splitting.

%we show that high-energy astrophysical neutrinos already rival and, in the ultra-light regime, surpass the sensitivity of all other existing probes.

%%%%%%%%%%%%%%%%%%%%%%%%%%%%%%%%%%%%%%%%%%%%%%%%%%%%%%%%%%%%%%%%%%%%%%%
%%%%%%%%%%%%%%%%%%%%%%%%%%%%%%%%%%%%%%%%%%%%%%%%%%%%%%%%%%%%%%%%%%%%%%%
%%%%%%%%%%%%%%                 Conclusion                %%%%%%%%%%%%%%
%%%%%%%%%%%%%%%%%%%%%%%%%%%%%%%%%%%%%%%%%%%%%%%%%%%%%%%%%%%%%%%%%%%%%%%
%%%%%%%%%%%%%%%%%%%%%%%%%%%%%%%%%%%%%%%%%%%%%%%%%%%%%%%%%%%%%%%%%%%%%%%

\section{Conclusions} \label{sec:conclusion}
Majorana-type SNSIs may alter the neutrino–oscillation Hamiltonian, inducing pseudo-Dirac mass splittings for fundamentally Dirac neutrinos. This can leave two distinct imprints on high-energy astrophysical neutrinos:
1) a shift in the flavor composition that arrives at Earth, and
2) energy-dependent spectral ``wiggles" whose exact phase and amplitude depend on the neutrino baseline and the size of the pseudo-Dirac splitting. 
Here, we exploited both signatures separately with the current IceCube data set and with a forward projection for IceCube-Gen2, and also performed a combined analysis of both signatures to carve out the SNSI parameter space sensitivity. Present-day IceCube statistics already provide meaningful constraints in the “sweet spot” where oscillation phases are neither fully averaged nor too slow to matter. IceCube-Gen2 will enlarge this lever arm in energy, source identification, and overall statistics, pushing the sensitivity to SNSI parameters. 
These sensitivity regions are translated into limits on the underlying model parameters, resulting in new sensitivities/constraints on the parameter space of light SNSI. Over the coming decade, tighter constraints on key oscillation parameters will sharpen astrophysical flavor measurements as well, and improved statistics will boost the ability of next-generation detectors to distinguish new physics.
If in the future, IceCube and its successor IceCube-Gen2 collect a statistically significant sample of neutrinos from identifiable point sources, the collaboration could even perform a full flavor‐ratio analysis source by source; knowing the individual source flavor and spectral composition data would significantly improve the sensitivities to the accessible SNSI parameter space. As shown in the summary plot~\cref{fig:yukawaplane}, neutrino telescopes provide a unique opportunity to probe the ultra-light SNSI parameter space. 

\section*{Acknowledgments}

The work of PSBD was supported in part by the US Department of Energy under grant No. DE-SC0017987 and by a Humboldt Fellowship from the Alexander von Humboldt Foundation. The work of BD  is supported by the U.S. Department of Energy
Grant~DE-SC0010813.
I.M.S is supported by the STFC under Grant No. ST/T001011/1.
CAA are supported by the Faculty of Arts and Sciences of Harvard University, the National Science Foundation, the Canadian Institute for Advanced Research, the Research Corporation for Science Advancement, the John Templeton Foundation, and the David \& Lucile Packard Foundation.

\appendix
\section{In-medium neutrino mass bounds on SNSI from BBN and supernovae}
\label{app:inmedium_bounds}
The existing BBN and supernova limits on SNSI considered in Refs.~\cite{Babu:2019iml,Venzor:2020ova} are from the medium-induced neutrino mass from a tadpole self energy diagram in a background of fermions $f$. In the  relativistic limit $T\gg m_f$, this correction is usually written as
\begin{equation}
\Delta m_{\nu, \alpha \beta}=\frac{y_f y_{\alpha \beta} m_f}{3 m_\phi^2}\left(\frac{\pi^2}{12 \zeta(3)}\right)^{\frac{2}{3}}\left(n_f^{2 / 3}+n_{\bar{f}}^{2 / 3}\right).
\label{eqn:deltaMnu}
\end{equation}
This expression, however, assumes a vacuum scalar propagator. 
We emphasize that in a finite-density or finite-temperature medium, the scalar propagator must also be dressed by the in-medium self-energy
\begin{equation}
\frac{1}{q^2-m_\phi^2} \rightarrow\frac{1}{q^2-m_\phi^2-\Pi_\phi\left(q_0, \mathbf{q} ; T, \mu\right)}.
\end{equation}

In the static coherent forward-scattering limit, this can be approximated by
\begin{equation}
m_\phi^2 \rightarrow m_{\phi,\rm eff}^2
\simeq
m_\phi^2+\Delta m_\phi^2,
\qquad
\Delta m_\phi^2 \equiv \Pi_\phi(0, \mathbf{q}\to0;T,\mu).
\end{equation}

For relativistic background fermions  $m_f \ll T$, the scalar self-energy scales parametrically as 
\begin{equation}
\Delta m_\phi^2(T,N_f)=\frac{y_f^2}{3}\left(\frac{\pi^2}{12 \zeta(3)}\right)^{\frac{2}{3}}\left(n_f^{2 / 3}+n_{\bar{f}}^{2 / 3}\right).
\label{eq:self2}
\end{equation}
Thus the same background fermions that source the scalar-induced neutrino mass correction must also contribute to the in-medium self-energy of the scalar through a fermion loop as shown in Section 5 of Ref.~\cite{Babu:2019iml}. This dressing should be included even when the scalar mediator itself is not thermalized as long as the background fermions coupled to $\phi$ are populated in the medium.

This effect is particularly important in the regime $\Delta m_\phi^2\gg  m_\phi^2$. In this limit, the apparent density enhancement in \eqref{eqn:deltaMnu} , which follows from the use of the vacuum scalar propagator, is screened by the in-medium scalar mass. The relevant replacement in the coherent forward scattering limit is therefore 
$\frac{1}{m_\phi^2} \longrightarrow \frac{1}{m_{\phi, \text { eff }}^2} \simeq \frac{1}{\Delta m_\phi^2}$.
Consequently, the induced neutrino mass no longer grows parametrically as $(n_f^{2/3}+n_{\bar f}^{2/3})$; instead, this density dependence is largely canceled by the density dependence of the scalar self-energy given by Eq.~\eqref{eq:self2}.

For the scenario where background fermions are neutrinos themselves ($f=\nu$),the same coupling that appears in the numerator of the induced neutrino mass also generates the in-medium scalar self-energy.
Defining
\begin{equation}
\mathcal{N}_\nu \equiv\left(\frac{\pi^2}{12 \zeta(3)}\right)^{2 / 3}\left(n_\nu^{2 / 3}+n_{\bar{\nu}}^{2 / 3}\right),
\end{equation}
the in-medium neutrino mass correction from the tadpole self energy diagram  after including the dressed scalar propagator is given by
\begin{equation}
\Delta m_\nu=\frac{y_\nu^2 m_\nu}{3\left(m_\phi^2+\Delta m_\phi^2\right)} \mathcal{N}_\nu=
m_\nu\frac{\frac{y_\nu^2}{3}\mathcal{N}_\nu }{\left(m_\phi^2+\frac{y_\nu^2}{3} \mathcal{N}_\nu\right)} .
\end{equation}
Therefore,
\begin{equation}
\Delta m_\nu \leq m_\nu,
\end{equation}
and the induced correction no longer grows as $y_\nu^2 / m_\phi^2$ but instead saturates at a value of order the vacuum neutrino mass in a high density environment such as Supernova or at the time of BBN.
Consequently, the MeV-scale medium-induced mass corrections required for the BBN bound of Ref.~\cite{Venzor:2020ova} and the supernova SNSI bound of Ref.~\cite{Babu:2019iml} are not obtained in the neutrino-only coupling scenario once the scalar propagator is consistently dressed. These bounds therefore do not constrain the corresponding parameter space shown in \cref{fig:yukawaplane}.

%We find that, in the parameter region relevant for the neutrino-only coupling scenario(where the background fermions are neutrinos), the scalar thermal correction satisfies $\Delta m_\phi^2 \gg m_\phi^2$ for the regions probed by the BBN bound of Ref. ~\cite{Venzor:2020ova} and the supernova SNSI bound of Ref.~\cite{Babu:2019iml}. Thus, in this region, the scalar propagator is controlled by the in-medium mass, $m_{\phi,\mathrm{eff}}^2 \simeq \Delta m_\phi^2$, rather than by the vacuum mass $m_\phi^2$. In this case the neutrino in-medium mass correction remains small and consequently, the corresponding BBN and supernova limits do not constrain the parameter space.

\section{Finite-density exchange self-energy in the cosmic neutrino background}
\label{app:exchange-self-energy}

\begin{figure}[tbp]
\centering 
\includegraphics[width=.6\textwidth]{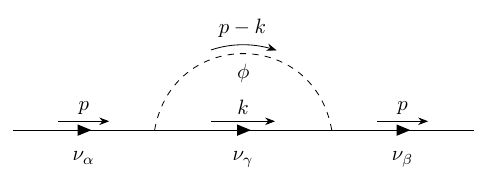}
\hfill
\caption{\label{fig:Exchange_self_energy}Exchange contribution to the finite-density neutrino self-energy. }
\end{figure}

In \cref{sec:Scalar NSI}, the medium-induced Majorana mass was obtained from the scalar
tadpole sourced by the Dirac bilinear $\langle\overline{\nu}\nu\rangle_{{\rm C}\nu{\rm B}}$. A distinct
finite-density contribution arises from the exchange (Fock) diagram as in \cref{fig:Exchange_self_energy} in which
the propagating neutrino exchanges a scalar with a background neutrino. In
this appendix, we derive this contribution and compare it with the tadpole-induced term relevant for the pseudo-Dirac splitting.

For simplicity, we consider a single Dirac neutrino mass eigenstate and take the Dirac-type scalar coupling to be real

\begin{equation}
\mathcal{L}_{\mathrm{int}}=y \phi \bar{\nu} \nu
\end{equation}
Let
$$
p^\mu=(E, \mathbf{p}), \quad k^\mu=\left(k^0, \mathbf{k}\right), \quad E_k=\sqrt{\mathbf{k}^2+m_\nu^2},
$$

where $p^\mu$ is the momentum of the propagating neutrino and $k^\mu$ is the momentum carried by the background-neutrino line.

The general medium-dependent one-loop neutrino self-energy from the exchange self energy diagram is given by
\begin{equation}
\Sigma^{\operatorname{med}}(p)=-y^2 \int \frac{d^4 k}{(2 \pi)^4}\left(\not k+m_\nu\right)\left[\frac{\Gamma_\phi(k-p)}{k^2-m_\nu^2}+\frac{\Gamma_\nu(k)}{(k-p)^2-m_\phi^2}\right]
\end{equation}

where, the medium-dependent part of the real-time neutrino propagator can be written as 
\begin{align}
    \Gamma_\nu(k)
    =2\pi\delta(k^2-m_\nu^2)
    \left[ f_\nu(|\mathbf{k}|)\theta(k^0)
       +f_{\bar\nu}(|\mathbf{k}|)\theta(-k^0)
    \right].
    \label{eq:ex-medium-insertion}
\end{align}

For the free streaming relic-neutrino background, the distribution functions retain their decoupled momentum dependence given by,
\begin{equation}
f_\nu(k)=\frac{1}{e^{(k-\mu) / T_{\nu0}}+1} \quad\text{and}
\quad
f_{\bar{\nu}}(k)=\frac{1}{e^{(k+\mu) / T_{\nu0}}+1}.
\end{equation}

Assuming no scalar $\phi$ population $\Gamma_\phi=0$, the self-energy in a neutrino background is
\begin{equation}
\Sigma^{\operatorname{med}}(p)=-y^2 \int \frac{d^4 k}{(2 \pi)^4}\left(\not k+m_\nu\right) \frac{\Gamma_\nu(k)}{(k-p)^2-m_\phi^2}.
\end{equation}
Using, 
\begin{equation}
\delta\left(k^2-m_\nu^2\right)=\frac{1}{2 E_k}\left[\delta\left(k^0-E_k\right)+\delta\left(k^0+E_k\right)\right],
\end{equation}
we can perform the $k^0$ integral. Relabeling $\mathbf{k} \rightarrow-\mathbf{k}$ in the negative-energy contribution gives
\begin{equation}
\Sigma_{\text {med }}(p)=-y^2 \int \frac{d^3 \mathbf{k}}{(2 \pi)^3 2 E_k}\left[f_\nu(k) \frac{\not k+m_\nu}{
(p-k)^2-m_\phi^2
}+f_{\bar{\nu}}(k) \frac{-\not k+m_\nu}{(p+k)^2-m_\phi^2}\right].
\end{equation}

The present relic-neutrino temperature being $T_{\nu, 0} \simeq 1.68 \times 10^{-4} \mathrm{eV}$, the populated massive C$\nu$B eigenstates are non-relativistic and we can consequently use
\begin{equation}
k^\mu \simeq m_\nu u^\mu, \quad u^\mu=(1, \mathbf{0}), \quad p \cdot k \simeq E m_\nu.
\end{equation}
Corrections to this approximation are of relative order $|\mathbf{k}| / m_\nu$.
Simplifying the scalar propagator denominators in the exchange self energy expression using the vacuum on-shell conditions for the external and background neutrinos we get

\begin{equation}
\begin{aligned}
(p-k)^2-m_\phi^2 & =p^2+k^2-m_\phi^2-2 p \cdot k \\
& \simeq 2 m_\nu^2-m_\phi^2-2 E m_\nu ,\\
(p+k)^2-m_\phi^2 & =p^2+k^2-m_\phi^2+2 p \cdot k \\
& \simeq 2 m_\nu^2-m_\phi^2+2 E m_\nu
\end{aligned}
\end{equation}

For the high-energy astrophysical neutrinos and ultralight scalar masses considered in this work, $2 E m_\nu \gg\left|2 m_\nu^2-m_\phi^2\right|$
and therefore the denominators then reduce to $(p-k)^2-m_\phi^2 \simeq-2 E m_\nu$ and $(p+k)^2-m_\phi^2 \simeq+2 E m_\nu$.

Using $E_k \simeq m_\nu$ and  $\not k \simeq m_\nu\not u$, the self-energy becomes

\begin{equation}
\begin{aligned}
& \Sigma_{\operatorname{med}}(p) \simeq-\frac{y^2}{2}\left[n_\nu \frac{\not u+1}{-2 E m_\nu}+ n_{\bar{\nu}} \frac{-\not u+1}{+2 E m_\nu}\right]
\end{aligned}
\end{equation}
It is useful to write this as 
\begin{equation}
\Sigma_{\mathrm{med}}^{\mathrm{UL}}(p)=\delta m_{\mathrm{ex}}^{\mathrm{UL}}+V_{\mathrm{ex}}^{\mathrm{UL}} \not u,
\end{equation}
where the mass correction is given by
\begin{equation}
\delta m_{\mathrm{ex}}^{\mathrm{UL}} \simeq \frac{y^2}{4 E m_\nu}\left(n_\nu-n_{\bar{\nu}}\right)
\end{equation}
and the contribution to the potential term is given by 
\begin{equation}
V_{\mathrm{ex}}^{\mathrm{UL}} \simeq \frac{y^2}{4 E m_\nu}\left(n_\nu+n_{\bar{\nu}}\right)
\end{equation}
The masslike contribution is thus proportional to the neutrino-antineutrino asymmetry and is strongly suppressed for nearly CP symmetric C$\nu$B. In contrast, the coefficient of $\not u$ , which modifies the neutrino dispersion relation as a refractive matter potential, is proportional to the total relic-neutrino density $n_\nu+n_{\bar{\nu}}$.
Nevertheless, in the ultralight-mediator regime, the exchange contribution is controlled by $2 E m_\nu$, rather than by $m_\phi^2$, and does not receive the $m_\phi^{-2}$ enhancement of the tadpole contribution.

%\clearpage
\bibliographystyle{JHEP}
\bibliography{scalarNSI}

@article{Groth:2025aan,
    author = "Groth, Kathrine M{\o}rch and Ahlers, Markus",
    title = "{Deciphering the sources of cosmic neutrinos}",
    eprint = "2503.07718",
    archivePrefix = "arXiv",
    primaryClass = "astro-ph.HE",
    doi = "10.1103/ntm4-lgbk",
    journal = "Phys. Rev. D",
    volume = "111",
    number = "10",
    pages = "103052",
    year = "2025"
}

@article{Coloma:2019mbs,
    author = "Coloma, Pilar and Esteban, Ivan and Gonzalez-Garcia, M. C. and Maltoni, Michele",
    title = "{Improved global fit to Non-Standard neutrino Interactions using COHERENT energy and timing data}",
    eprint = "1911.09109",
    archivePrefix = "arXiv",
    primaryClass = "hep-ph",
    reportNumber = "YITP-SB-19-38, IFT-UAM/CSIC-19-152, IFIC-19-49",
    doi = "10.1007/JHEP02(2020)023",
    journal = "JHEP",
    volume = "02",
    pages = "023",
    year = "2020",
    note = "[Addendum: JHEP 12, 071 (2020)]"
}

@article{Capel:2020txc,
    author = "Capel, Francesca and Mortlock, Daniel J. and Finley, Chad",
    title = "{Bayesian constraints on the astrophysical neutrino source population from IceCube data}",
    eprint = "2005.02395",
    archivePrefix = "arXiv",
    primaryClass = "astro-ph.HE",
    doi = "10.1103/PhysRevD.101.123017",
    journal = "Phys. Rev. D",
    volume = "101",
    number = "12",
    pages = "123017",
    year = "2020",
    note = "[Erratum: Phys.Rev.D 105, 129904 (2022)]"
}

@article{Dev:2024yrg,
    author = "Dev, P. S. Bhupal and Machado, Pedro A. N. and Martinez-Soler, Ivan",
    title = "{Pseudo-Dirac neutrinos and relic neutrino matter effect on the high-energy neutrino flavor composition}",
    eprint = "2406.18507",
    archivePrefix = "arXiv",
    primaryClass = "hep-ph",
    reportNumber = "CETUP-2023-022, FERMILAB-PUB-24-0317-T, IPPP/24/35",
    doi = "10.1016/j.physletb.2025.139306",
    journal = "Phys. Lett. B",
    volume = "862",
    pages = "139306",
    year = "2025"
}

@article{Rothstein:1992rh,
    author = "Rothstein, I. Z. and Babu, K. S. and Seckel, D.",
    title = "{Planck scale symmetry breaking and majoron physics}",
    eprint = "hep-ph/9301213",
    archivePrefix = "arXiv",
    reportNumber = "UM-TH-92-31, BA-92-77",
    doi = "10.1016/0550-3213(93)90368-Y",
    journal = "Nucl. Phys. B",
    volume = "403",
    pages = "725--748",
    year = "1993"
}

@article{Esmaili:2009fk,
    author = "Esmaili, Arman",
    title = "{Pseudo-Dirac Neutrino Scenario: Cosmic Neutrinos at Neutrino Telescopes}",
    eprint = "0909.5410",
    archivePrefix = "arXiv",
    primaryClass = "hep-ph",
    reportNumber = "IPM-P-2009-039",
    doi = "10.1103/PhysRevD.81.013006",
    journal = "Phys. Rev. D",
    volume = "81",
    pages = "013006",
    year = "2010"
}

@article{Gandhi:1995tf,
    author = "Gandhi, Raj and Quigg, Chris and Reno, Mary Hall and Sarcevic, Ina",
    title = "{Ultrahigh-energy neutrino interactions}",
    eprint = "hep-ph/9512364",
    archivePrefix = "arXiv",
    reportNumber = "FERMILAB-PUB-95-221-T, CLNS-95-1357, MRI-PHY-16-95, UIOWA-95-06, AZPH-TH-95-15",
    doi = "10.1016/0927-6505(96)00008-4",
    journal = "Astropart. Phys.",
    volume = "5",
    pages = "81--110",
    year = "1996"
}

@article{IceCube:2016zyt,
    author = "Aartsen, M. G. and others",
    collaboration = "IceCube",
    title = "{The IceCube Neutrino Observatory: Instrumentation and Online Systems}",
    eprint = "1612.05093",
    archivePrefix = "arXiv",
    primaryClass = "astro-ph.IM",
    doi = "10.1088/1748-0221/12/03/P03012",
    journal = "JINST",
    volume = "12",
    number = "03",
    pages = "P03012",
    year = "2017",
    note = "[Erratum: JINST 19, E05001 (2024)]"
}

@article{IceCube:2015rro,
    author = "Aartsen, M. G. and others",
    collaboration = "IceCube",
    title = "{Flavor Ratio of Astrophysical Neutrinos above 35 TeV in IceCube}",
    eprint = "1502.03376",
    archivePrefix = "arXiv",
    primaryClass = "astro-ph.HE",
    doi = "10.1103/PhysRevLett.114.171102",
    journal = "Phys. Rev. Lett.",
    volume = "114",
    number = "17",
    pages = "171102",
    year = "2015"
}

@article{IceCube:2020acn,
    author = "Aartsen, M. G. and others",
    collaboration = "IceCube",
    title = "{Characteristics of the diffuse astrophysical electron and tau neutrino flux with six years of IceCube high energy cascade data}",
    eprint = "2001.09520",
    archivePrefix = "arXiv",
    primaryClass = "astro-ph.HE",
    doi = "10.1103/PhysRevLett.125.121104",
    journal = "Phys. Rev. Lett.",
    volume = "125",
    number = "12",
    pages = "121104",
    year = "2020"
}

@article{IceCube:2024fxo,
    author = "Abbasi, R. and others",
    collaboration = "IceCube",
    title = "{Characterization of the astrophysical diffuse neutrino flux using starting track events in IceCube}",
    eprint = "2402.18026",
    archivePrefix = "arXiv",
    primaryClass = "astro-ph.HE",
    doi = "10.1103/PhysRevD.110.022001",
    journal = "Phys. Rev. D",
    volume = "110",
    number = "2",
    pages = "022001",
    year = "2024"
}

@article{IceCube-Gen2:2020qha,
    author = "Aartsen, M. G. and others",
    collaboration = "IceCube-Gen2",
    title = "{IceCube-Gen2: the window to the extreme Universe}",
    eprint = "2008.04323",
    archivePrefix = "arXiv",
    primaryClass = "astro-ph.HE",
    doi = "10.1088/1361-6471/abbd48",
    journal = "J. Phys. G",
    volume = "48",
    number = "6",
    pages = "060501",
    year = "2021"
}

@article{IceCube:2022der,
    author = "Abbasi, R. and others",
    collaboration = "IceCube",
    title = "{Evidence for neutrino emission from the nearby active galaxy NGC 1068}",
    eprint = "2211.09972",
    archivePrefix = "arXiv",
    primaryClass = "astro-ph.HE",
    doi = "10.1126/science.abg3395",
    journal = "Science",
    volume = "378",
    number = "6619",
    pages = "538--543",
    year = "2022"
}

@article{IceCube:2020wum,
    author = "Abbasi, R. and others",
    collaboration = "IceCube",
    title = "{The IceCube high-energy starting event sample: Description and flux characterization with 7.5 years of data}",
    eprint = "2011.03545",
    archivePrefix = "arXiv",
    primaryClass = "astro-ph.HE",
    doi = "10.1103/PhysRevD.104.022002",
    journal = "Phys. Rev. D",
    volume = "104",
    pages = "022002",
    year = "2021"
}

@article{Fong:2024msb,
    author = "Fong, Chee Sheng and Porto, Yago",
    title = "{Constraining the pseudo-Dirac nature of neutrinos using astrophysical neutrino flavor data}",
    eprint = "2406.15566",
    archivePrefix = "arXiv",
    primaryClass = "hep-ph",
    doi = "10.1103/yc27-61w5",
    journal = "Phys. Rev. D",
    volume = "112",
    number = "6",
    pages = "063001",
    year = "2025"
}

@article{Esteban:2024eli,
    author = "Esteban, Ivan and Gonzalez-Garcia, M. C. and Maltoni, Michele and Martinez-Soler, Ivan and Pinheiro, Jo\~ao Paulo and Schwetz, Thomas",
    title = "{NuFit-6.0: updated global analysis of three-flavor neutrino oscillations}",
    eprint = "2410.05380",
    archivePrefix = "arXiv",
    primaryClass = "hep-ph",
    reportNumber = "IFT-UAM/CSIC-24-140, YITP-SB-2024-24, IPPP/24/64, IPPP/24/64, IFT-UAM/CSIC-24-140, YITP-SB-2024-24",
    doi = "10.1007/JHEP12(2024)216",
    journal = "JHEP",
    volume = "12",
    pages = "216",
    year = "2024"
}

@article{Elias-Chavez:2018dru,
    author = "El\'\i{}as-Ch\'avez, M. and Mart\'\i{}nez, O. M.",
    title = "{Estimation of the Star Formation Rate using Long-Gamma Ray Burst observed by SWIFT}",
    eprint = "2006.03367",
    archivePrefix = "arXiv",
    primaryClass = "astro-ph.GA",
    journal = "Rev. Mex. Astron. Astrofis.",
    volume = "54",
    pages = "309--316",
    year = "2018"
}

@article{Carloni:2025dhv,
    author = {Carloni, Kiara and Porto, Yago and Arg{\"u}elles, Carlos A. and Dev, P. S. Bhupal and Jana, Sudip},
    title = "{Signatures of quasi-Dirac neutrinos in diffuse high-energy astrophysical neutrino data}",
    eprint = "2503.19960",
    archivePrefix = "arXiv",
    primaryClass = "hep-ph",
    doi = "10.1103/2hxm-qpsm",
    journal = "Phys. Rev. D",
    volume = "113",
    number = "11",
    pages = "115004",
    year = "2026"
}

@article{IceCube:2016tpw,
    author = "Aartsen, M. G. and others",
    collaboration = "IceCube",
    title = "{All-sky Search for Time-integrated Neutrino Emission from Astrophysical Sources with 7 yr of IceCube Data}",
    eprint = "1609.04981",
    archivePrefix = "arXiv",
    primaryClass = "astro-ph.HE",
    doi = "10.3847/1538-4357/835/2/151",
    journal = "Astrophys. J.",
    volume = "835",
    number = "2",
    pages = "151",
    year = "2017"
}

@article{IceCube-Gen2:2023rds,
    author = "Lad, Neha and others",
    collaboration = "IceCube-Gen2",
    title = "{Sensitivity of IceCube-Gen2 to measure flavor composition of Astrophysical neutrinos}",
    eprint = "2308.15220",
    archivePrefix = "arXiv",
    primaryClass = "astro-ph.HE",
    reportNumber = "PoS-ICRC2023-1123",
    doi = "10.22323/1.444.1123",
    journal = "PoS",
    volume = "ICRC2023",
    pages = "1123",
    year = "2023"
}

@article{Berryman:2022hds,
    author = "Berryman, Jeffrey M. and others",
    title = "{Neutrino self-interactions: A white paper}",
    eprint = "2203.01955",
    archivePrefix = "arXiv",
    primaryClass = "hep-ph",
    reportNumber = "CERN-TH-2022-024, DESY-22-035, FERMILAB-PUB-22-099-T",
    doi = "10.1016/j.dark.2023.101267",
    journal = "Phys. Dark Univ.",
    volume = "42",
    pages = "101267",
    year = "2023"
}

@misc{abenza_neutrino_nodate,
	title = {Neutrino (self)-{Interactions} in {Cosmology}},
	language = {en},
	author = {Escudero, Miguel},
	howpublished="\url{https://indico.cern.ch/event/1210319/contributions/5267731/attachments/2613472/4516764/Escudero_NeutrinoPlatform.pdf}",
}

@article{Poudou:2025qcx,
    author = "Poudou, Ad{\`e}le and Simon, Th{\'e}o and Montandon, Thomas and Teixeira, Elsa M. and Poulin, Vivian",
    title = "{Self-interacting neutrinos in light of recent CMB and LSS data}",
    eprint = "2503.10485",
    archivePrefix = "arXiv",
    primaryClass = "astro-ph.CO",
    doi = "10.1103/mljb-42fm",
    journal = "Phys. Rev. D",
    volume = "112",
    number = "10",
    pages = "103535",
    year = "2025"
}

@article{Brdar:2020nbj,
    author = "Brdar, Vedran and Lindner, Manfred and Vogl, Stefan and Xu, Xun-Jie",
    title = "{Revisiting neutrino self-interaction constraints from $Z$ and $\tau$ decays}",
    eprint = "2003.05339",
    archivePrefix = "arXiv",
    primaryClass = "hep-ph",
    doi = "10.1103/PhysRevD.101.115001",
    journal = "Phys. Rev. D",
    volume = "101",
    number = "11",
    pages = "115001",
    year = "2020"
}

@article{Ivanez-Ballesteros:2024nws,
    author = "Iv{\'a}{\~n}ez-Ballesteros, Pilar and Volpe, M. Cristina",
    title = "{Constraints on neutrino-Majoron coupling using SN1987A data}",
    eprint = "2410.11517",
    archivePrefix = "arXiv",
    primaryClass = "hep-ph",
    doi = "10.1103/d4vp-m261",
    journal = "Phys. Rev. D",
    volume = "112",
    number = "10",
    pages = "L101301",
    year = "2025"
}

@inproceedings{Naab:2023xcz,
    author = "Naab, Richard and Ganster, Erik and Zhang, Zelong",
    collaboration = "IceCube",
    title = "{Measurement of the astrophysical diffuse neutrino flux in a combined fit of IceCube's high energy neutrino data}",
    booktitle = "{38th International Cosmic Ray Conference}",
    eprint = "2308.00191",
    archivePrefix = "arXiv",
    primaryClass = "astro-ph.HE",
    reportNumber = "PoS-ICRC2023-1064",
    month = "7",
    year = "2023"
}

@article{Chang:1999pb,
    author = "Chang, Darwin and Kong, Otto C. W.",
    title = "{Pseudo-Dirac neutrinos}",
    eprint = "hep-ph/9912268",
    archivePrefix = "arXiv",
    reportNumber = "UR-1571",
    doi = "10.1016/S0370-2693(00)00228-8",
    journal = "Phys. Lett. B",
    volume = "477",
    pages = "416--423",
    year = "2000"
}

@article{deGouvea:2009fp,
    author = "de Gouvea, Andre and Huang, Wei-Chih and Jenkins, James",
    title = "{Pseudo-Dirac Neutrinos in the New Standard Model}",
    eprint = "0906.1611",
    archivePrefix = "arXiv",
    primaryClass = "hep-ph",
    reportNumber = "LA-UR-09-03593, NUHEP-TH-09-08",
    doi = "10.1103/PhysRevD.80.073007",
    journal = "Phys. Rev. D",
    volume = "80",
    pages = "073007",
    year = "2009"
}

@article{Nir:2000xn,
    author = "Nir, Yosef",
    title = "{PseudoDirac solar neutrinos}",
    eprint = "hep-ph/0002168",
    archivePrefix = "arXiv",
    reportNumber = "IASSNS-HEP-00-08",
    doi = "10.1088/1126-6708/2000/06/039",
    journal = "JHEP",
    volume = "06",
    pages = "039",
    year = "2000"
}

@article{Balaji:2001fi,
    author = "Balaji, K. R. S. and Kalliomaki, Anna and Maalampi, Jukka",
    title = "{Revisiting pseudoDirac neutrinos}",
    eprint = "hep-ph/0110314",
    archivePrefix = "arXiv",
    reportNumber = "DO-TH-01-15, HIP-2001-52-TH",
    doi = "10.1016/S0370-2693(01)01356-9",
    journal = "Phys. Lett. B",
    volume = "524",
    pages = "153--160",
    year = "2002"
}

@article{Kobayashi:2000md,
    author = "Kobayashi, Makoto and Lim, C. S.",
    title = "{Pseudo Dirac scenario for neutrino oscillations}",
    eprint = "hep-ph/0012266",
    archivePrefix = "arXiv",
    reportNumber = "KEK-TH-733, KOBE-TH-00-10",
    doi = "10.1103/PhysRevD.64.013003",
    journal = "Phys. Rev. D",
    volume = "64",
    pages = "013003",
    year = "2001"
}

@article{Dutta:2022fdt,
    author = "Dutta, Bhaskar and Ghosh, Sumit and Li, Tianjun and Thompson, Adrian and Verma, Ankur",
    title = "{Non-standard neutrino interactions in light mediator models at reactor experiments}",
    eprint = "2209.13566",
    archivePrefix = "arXiv",
    primaryClass = "hep-ph",
    reportNumber = "MI-HET-775, KIAS-P22043",
    doi = "10.1007/JHEP03(2023)163",
    journal = "JHEP",
    volume = "03",
    pages = "163",
    year = "2023"
}

@article{Babu:2019iml,
    author = "Babu, K. S. and Chauhan, Garv and Dev, P. S. Bhupal",
    title = "{Neutrino nonstandard interactions via light scalars in the Earth, Sun, supernovae, and the early Universe}",
    eprint = "1912.13488",
    archivePrefix = "arXiv",
    primaryClass = "hep-ph",
    reportNumber = "OSU-HEP-19-11",
    doi = "10.1103/PhysRevD.101.095029",
    journal = "Phys. Rev. D",
    volume = "101",
    number = "9",
    pages = "095029",
    year = "2020"
}

@article{IceCube:2025lev,
    author = "Abbasi, R. and others",
    collaboration = "IceCube",
    title = "{All-sky Neutrino Point-source Search with IceCube Combined Track and Cascade Data}",
    eprint = "2507.07275",
    archivePrefix = "arXiv",
    primaryClass = "astro-ph.HE",
    doi = "10.3847/1538-4357/ae113f",
    journal = "Astrophys. J.",
    volume = "995",
    number = "1",
    pages = "11",
    year = "2025"
}

@article{Carloni:2022cqz,
    author = "Carloni, Kiara and Mart{\'\i}nez-Soler, Ivan and Arguelles, Carlos A. and Babu, K. S. and Dev, P. S. Bhupal",
    title = "{Probing pseudo-Dirac neutrinos with astrophysical sources at IceCube}",
    eprint = "2212.00737",
    archivePrefix = "arXiv",
    primaryClass = "astro-ph.HE",
    doi = "10.1103/PhysRevD.109.L051702",
    journal = "Phys. Rev. D",
    volume = "109",
    pages = "L051702",
    year = "2024"
}

@article{IceCube:2025ewu,
    author = "Abbasi, R. and others",
    collaboration = "IceCube",
    title = "{Improved measurements of the TeV-PeV extragalactic neutrino spectrum from joint analyses of IceCube tracks and cascades}",
    eprint = "2507.22234",
    archivePrefix = "arXiv",
    primaryClass = "astro-ph.HE",
    doi = "10.1103/4n6v-r7n4",
    journal = "Phys. Rev. D",
    volume = "113",
    pages = "062002",
    year = "2026"
}

@article{Sandner:2023ptm,
    author = "Sandner, Stefan and Escudero, Miguel and Witte, Samuel J.",
    title = "{Precision CMB constraints on eV-scale bosons coupled to neutrinos}",
    eprint = "2305.01692",
    archivePrefix = "arXiv",
    primaryClass = "hep-ph",
    reportNumber = "IFIC/23-13, FTUV-23-0413.0599, CERN-TH-2023-073",
    doi = "10.1140/epjc/s10052-023-11864-6",
    journal = "Eur. Phys. J. C",
    volume = "83",
    number = "8",
    pages = "709",
    year = "2023"
}

@article{Meyer:2004hr,
    author = "Meyer, M. J. and others",
    title = "{The HIPASS Catalog. 1. Data presentation}",
    eprint = "astro-ph/0406384",
    archivePrefix = "arXiv",
    doi = "10.1111/j.1365-2966.2004.07710.x",
    journal = "Mon. Not. Roy. Astron. Soc.",
    volume = "350",
    pages = "1195",
    year = "2004"
}

@article{Rovero:2016igo,
    author = "Rovero, A. C. and Muriel, H. and Donzelli, C. and Pichel, A.",
    title = "{The BL-Lacertae gamma-ray blazar PKS 1424+240 associated with a group of galaxies at z = 0.6010}",
    eprint = "1602.08364",
    archivePrefix = "arXiv",
    primaryClass = "astro-ph.HE",
    doi = "10.1051/0004-6361/201527778",
    journal = "Astron. Astrophys.",
    volume = "589",
    pages = "A92",
    year = "2016"
}

@article{Paiano:2018qeq,
    author = "Paiano, Simona and Falomo, Renato and Treves, Aldo and Scarpa, Riccardo",
    title = "{The redshift of the BL Lac object TXS 0506+056}",
    eprint = "1802.01939",
    archivePrefix = "arXiv",
    primaryClass = "astro-ph.GA",
    doi = "10.3847/2041-8213/aaad5e",
    journal = "Astrophys. J. Lett.",
    volume = "854",
    number = "2",
    pages = "L32",
    year = "2018"
}

@article{Marcha:2013tt,
    author = "Marcha, M. J. M. and Caccianiga, A.",
    title = "{The CLASS BL Lac sample: The Radio Luminosity Function}",
    eprint = "1301.6550",
    archivePrefix = "arXiv",
    primaryClass = "astro-ph.CO",
    doi = "10.1093/mnras/stt065",
    journal = "Mon. Not. Roy. Astron. Soc.",
    volume = "430",
    pages = "2464",
    year = "2013"
}

@article{Dodson:2002pz,
    author = "Dodson, R. and Golap, K.",
    title = "{On the association of g343.1-2.3 and psr 1706-44}",
    eprint = "astro-ph/0201188",
    archivePrefix = "arXiv",
    reportNumber = "RGGD-02-01",
    doi = "10.1046/j.1365-8711.2002.05572.x",
    journal = "Mon. Not. Roy. Astron. Soc.",
    volume = "334",
    pages = "L1",
    year = "2002"
}

@article{HESS:2011zks,
    author = "Abramowski, A. and others",
    collaboration = "H.E.S.S.",
    title = "{Detection of very-high-energy gamma-ray emission from the vicinity of PSR B1706-44 and G343.1-2.3 with H.E.S.S}",
    eprint = "1102.0773",
    archivePrefix = "arXiv",
    primaryClass = "astro-ph.HE",
    doi = "10.1051/0004-6361/201015381",
    journal = "Astron. Astrophys.",
    volume = "528",
    pages = "A143",
    year = "2011"
}

@article{Goldoni:2015jua,
    author = {Goldoni, P. and Pita, S. and Boisson, C. and M{\"u}ller, C. and Dauser, T. and Jung, I. and Krau{\ss}, F. and Lenain, J. P. and Sol, H.},
    title = "{Optical-NIR spectroscopy of the puzzling $\gamma$-ray source 3FGL 1603.9-4903/PMN J1603-4904 with X-Shooter}",
    eprint = "1510.06234",
    archivePrefix = "arXiv",
    primaryClass = "astro-ph.HE",
    doi = "10.1051/0004-6361/201527582",
    journal = "Astron. Astrophys.",
    volume = "586",
    pages = "L2",
    year = "2016"
}

@article{Abdo:2006fq,
    author = "Abdo, A. A. and others",
    title = "{Discovery of TeV Gamma-Ray Emission from the Cygnus Region of the Galaxy}",
    eprint = "astro-ph/0611691",
    archivePrefix = "arXiv",
    reportNumber = "SLAC-PUB-12233",
    doi = "10.1086/513696",
    journal = "Astrophys. J. Lett.",
    volume = "658",
    pages = "L33--L36",
    year = "2007"
}

@article{Franklin:2024amy,
    author = "Franklin, Jack and Martinez-Soler, Ivan and Perez-Gonzalez, Yuber F. and Turner, Jessica",
    title = "{Probing the cosmic neutrino background and new physics with TeV-scale astrophysical neutrinos}",
    eprint = "2404.02202",
    archivePrefix = "arXiv",
    primaryClass = "hep-ph",
    reportNumber = "IPPP/24/12",
    doi = "10.1016/j.physletb.2025.139615",
    journal = "Phys. Lett. B",
    volume = "867",
    pages = "139615",
    year = "2025"
}

@article{IceCube:2010dpc,
    author = "Abbasi, R. and others",
    collaboration = "IceCube",
    title = "{Calibration and Characterization of the IceCube Photomultiplier Tube}",
    eprint = "1002.2442",
    archivePrefix = "arXiv",
    primaryClass = "astro-ph.IM",
    doi = "10.1016/j.nima.2010.03.102",
    journal = "Nucl. Instrum. Meth. A",
    volume = "618",
    pages = "139--152",
    year = "2010"
}

@article{Song:2020nfh,
    author = {Song, Ningqiang and Li, Shirley Weishi and Arg{\"u}elles, Carlos A. and Bustamante, Mauricio and Vincent, Aaron C.},
    title = "{The Future of High-Energy Astrophysical Neutrino Flavor Measurements}",
    eprint = "2012.12893",
    archivePrefix = "arXiv",
    primaryClass = "hep-ph",
    reportNumber = "FERMILAB-PUB-21-006-T",
    doi = "10.1088/1475-7516/2021/04/054",
    journal = "JCAP",
    volume = "04",
    pages = "054",
    year = "2021"
}

@article{Beacom:2003eu,
    author = "Beacom, John F. and Bell, Nicole F. and Hooper, Dan and Learned, John G. and Pakvasa, Sandip and Weiler, Thomas J.",
    title = "{PseudoDirac Neutrinos: A Challenge for Neutrino Telescopes}",
    eprint = "hep-ph/0307151",
    archivePrefix = "arXiv",
    reportNumber = "FERMILAB-PUB-03-201-A, MADPH-03-1337",
    doi = "10.1103/PhysRevLett.92.011101",
    journal = "Phys. Rev. Lett.",
    volume = "92",
    pages = "011101",
    year = "2004"
}

@article{Keranen:2003xd,
  author = "Keranen, P. and Maalampi, J. and Myyrylainen, M. and Riittinen, J.",
  title = "{Effects of sterile neutrinos on the ultrahigh-energy cosmic neutrino flux}",
  eprint = "hep-ph/0307041",
  archivePrefix = "arXiv",
  doi = "10.1016/j.physletb.2003.09.006",
  journal = "Phys. Lett. B",
  volume = "574",
  pages = "162--168",
  year = "2003"
}

@article{Chatterjee:2020kkm,
    author = "Chatterjee, Sabya Sachi and Palazzo, Antonio",
    title = "{Nonstandard Neutrino Interactions as a Solution to the $NO\nu A$ and T2K Discrepancy}",
    eprint = "2008.04161",
    archivePrefix = "arXiv",
    primaryClass = "hep-ph",
    reportNumber = "IPPP/20/35",
    doi = "10.1103/PhysRevLett.126.051802",
    journal = "Phys. Rev. Lett.",
    volume = "126",
    number = "5",
    pages = "051802",
    year = "2021"
}

@article{Denton:2020uda,
    author = "Denton, Peter B. and Gehrlein, Julia and Pestes, Rebekah",
    title = "{$CP$ -Violating Neutrino Nonstandard Interactions in Long-Baseline-Accelerator Data}",
    eprint = "2008.01110",
    archivePrefix = "arXiv",
    primaryClass = "hep-ph",
    doi = "10.1103/PhysRevLett.126.051801",
    journal = "Phys. Rev. Lett.",
    volume = "126",
    number = "5",
    pages = "051801",
    year = "2021"
}

@article{Chatterjee:2024kbn,
    author = "Chatterjee, Sabya Sachi and Palazzo, Antonio",
    title = "{Status of tension between NOvA and T2K after Neutrino 2024 and possible role of nonstandard neutrino interactions}",
    eprint = "2409.10599",
    archivePrefix = "arXiv",
    primaryClass = "hep-ph",
    doi = "10.1103/PhysRevD.110.113002",
    journal = "Phys. Rev. D",
    volume = "110",
    number = "11",
    pages = "113002",
    year = "2024"
}

@article{Abbasi:2025fjc,
    author = "Abbasi, R. and others",
    title = "{Characterization of the Three-Flavor Composition of Cosmic Neutrinos with IceCube}",
    eprint = "2510.24957",
    archivePrefix = "arXiv",
    primaryClass = "astro-ph.HE",
    month = "10",
    year = "2025"
}

@article{Notzold:1987ik,
    author = {N{\"o}tzold, Dirk and Raffelt, Georg},
    title = "{Neutrino dispersion at finite temperature and density}",
    reportNumber = "MPI-PAE/PTh-87/87",
    doi = "10.1016/0550-3213(88)90113-7",
    journal = "Nucl. Phys. B",
    volume = "307",
    pages = "924--936",
    year = "1988"
}

@article{Proceedings:2019qno,
    author = "Dev, P. S. Bhupal and others",
    title = "{Neutrino Non-Standard Interactions: A Status Report}",
    eprint = "1907.00991",
    archivePrefix = "arXiv",
    primaryClass = "hep-ph",
    reportNumber = "FERMILAB-CONF-19-299-T",
    doi = "10.21468/SciPostPhysProc.2.001",
    journal = "SciPost. Phys. Proc.", 
    volume = "2",
    pages = "001",
    year = "2019"
}

@article{Farzan:2017xzy,
    author = "Farzan, Y. and Tortola, M.",
    title = "{Neutrino oscillations and Non-Standard Interactions}",
    eprint = "1710.09360",
    archivePrefix = "arXiv",
    primaryClass = "hep-ph",
    doi = "10.3389/fphy.2018.00010",
    journal = "Front. in Phys.",
    volume = "6",
    pages = "10",
    year = "2018"
}

@article{Wolfenstein:1977ue,
    author = "Wolfenstein, L.",
    title = "{Neutrino Oscillations in Matter}",
    reportNumber = "COO-3066-102",
    doi = "10.1103/PhysRevD.17.2369",
    journal = "Phys. Rev. D",
    volume = "17",
    pages = "2369--2374",
    year = "1978"
}

@article{Cherchiglia:2023ojf,
    author = "Cherchiglia, Adriano and Pasquini, Pedro and Peres, O. L. G. and Rodrigues, F. F. and Rossi, R. R. and Souza, E. S.",
    title = "{Alleviating the present tension between T2K and NO{\ensuremath{\nu}}A with nonstandard neutrino interactions}",
    eprint = "2310.18401",
    archivePrefix = "arXiv",
    primaryClass = "hep-ph",
    doi = "10.1103/55qm-zbhv",
    journal = "Phys. Rev. D",
    volume = "112",
    number = "9",
    pages = "093004",
    year = "2025"
}

@article{Gehrlein:2024vwz,
    author = "Gehrlein, Julia and Machado, Pedro A. N. and Pinheiro, Jo{\~a}o Paulo",
    title = "{Constraining non-standard neutrino interactions with neutral current events at long-baseline oscillation experiments}",
    eprint = "2412.08712",
    archivePrefix = "arXiv",
    primaryClass = "hep-ph",
    reportNumber = "FERMILAB-PUB-24-0776-T, FERMILAB-PUB-24-0776-T",
    doi = "10.1007/JHEP05(2025)065",
    journal = "JHEP",
    volume = "05",
    pages = "065",
    year = "2025"
}

@article{Coloma:2024ict,
    author = "Coloma, Pilar and Fern{\'a}ndez-Mart{\'\i}nez, Enrique and L{\'o}pez-Pav{\'o}n, Jacobo and Marcano, Xabier and Naredo-Tuero, Daniel and Urrea, Salvador",
    title = "{Improving the global SMEFT picture with bounds on neutrino NSI}",
    eprint = "2411.00090",
    archivePrefix = "arXiv",
    primaryClass = "hep-ph",
    reportNumber = "IFT-UAM/CSIC-24-151, FTUV-24-1025.8856",
    doi = "10.1007/JHEP02(2025)137",
    journal = "JHEP",
    volume = "02",
    pages = "137",
    year = "2025"
}

@article{BOREXINO:2026owb,
    author = "Antonelli, V. and others",
    collaboration = "BOREXINO",
    title = "{Constraints on non-standard neutrino interactions from Borexino extended data-set}",
    eprint = "2602.08685",
    archivePrefix = "arXiv",
    primaryClass = "hep-ex",
    month = "2",
    year = "2026"
}

@article{RES-NOVA:2026fii,
    author = "Alloni, D. and others",
    collaboration = "RES-NOVA",
    title = "{Neutrino NSI in archaeological Pb}",
    eprint = "2602.23419",
    archivePrefix = "arXiv",
    primaryClass = "hep-ph",
    month = "2",
    year = "2026"
}

@article{IceCube:2022ubv,
    author = "Abbasi, R. and others",
    collaboration = "IceCube",
    title = "{Strong Constraints on Neutrino Nonstandard Interactions from TeV-Scale $\nu_u$ Disappearance at IceCube}",
    eprint = "2201.03566",
    archivePrefix = "arXiv",
    primaryClass = "hep-ex",
    doi = "10.1103/PhysRevLett.129.011804",
    journal = "Phys. Rev. Lett.",
    volume = "129",
    number = "1",
    pages = "011804",
    year = "2022"
}

@article{Dutta:2020che,
    author = "Dutta, Bhaskar and Lang, Rafael F. and Liao, Shu and Sinha, Samiran and Strigari, Louis and Thompson, Adrian",
    title = "{A global analysis strategy to resolve neutrino NSI degeneracies with scattering and oscillation data}",
    eprint = "2002.03066",
    archivePrefix = "arXiv",
    primaryClass = "hep-ph",
    reportNumber = "MI-TH-204",
    doi = "10.1007/JHEP09(2020)106",
    journal = "JHEP",
    volume = "09",
    pages = "106",
    year = "2020"
}

@article{Coloma:2023ixt,
    author = "Coloma, Pilar and Gonzalez-Garcia, M. C. and Maltoni, Michele and Pinheiro, Jo{\~a}o Paulo and Urrea, Salvador",
    title = "{Global constraints on non-standard neutrino interactions with quarks and electrons}",
    eprint = "2305.07698",
    archivePrefix = "arXiv",
    primaryClass = "hep-ph",
    reportNumber = "IFT-UAM/CSIC-23-47, IFIC/23-15, FTUV-23-0427.3710, YITP-SB-2023-05",
    doi = "10.1007/JHEP08(2023)032",
    journal = "JHEP",
    volume = "08",
    pages = "032",
    year = "2023"
}

@article{DayaBay:2024hya,
    author = "An, F. P. and others",
    collaboration = "Daya Bay",
    title = "{Charged-current non-standard neutrino interactions at Daya Bay}",
    eprint = "2401.02901",
    archivePrefix = "arXiv",
    primaryClass = "hep-ph",
    doi = "10.1007/JHEP05(2024)204",
    journal = "JHEP",
    volume = "05",
    pages = "204",
    year = "2024"
}

@article{NOvA:2024lti,
    author = "Acero, M. A. and others",
    collaboration = "NOvA",
    title = "{Search for CP-Violating Neutrino Nonstandard Interactions with the NOvA Experiment}",
    eprint = "2403.07266",
    archivePrefix = "arXiv",
    primaryClass = "hep-ex",
    reportNumber = "FERMILAB-PUB-24-0108-PPD",
    doi = "10.1103/PhysRevLett.133.201802",
    journal = "Phys. Rev. Lett.",
    volume = "133",
    number = "20",
    pages = "201802",
    year = "2024"
}

@article{KM3NeT:2024pte,
    author = "Aiello, S. and others",
    collaboration = "KM3NeT",
    title = "{Search for non-standard neutrino interactions with the first six detection units of KM3NeT/ORCA}",
    eprint = "2411.19078",
    archivePrefix = "arXiv",
    primaryClass = "hep-ex",
    doi = "10.1088/1475-7516/2025/02/073",
    journal = "JCAP",
    volume = "02",
    pages = "073",
    year = "2025"
}

@article{MINOS:2016sbv,
    author = "Adamson, P. and others",
    collaboration = "MINOS",
    title = "{Search for flavor-changing nonstandard neutrino interactions using $\nu_{e}$ appearance in MINOS}",
    eprint = "1605.06169",
    archivePrefix = "arXiv",
    primaryClass = "hep-ex",
    reportNumber = "FERMILAB-PUB-16-171-ND",
    doi = "10.1103/PhysRevD.95.012005",
    journal = "Phys. Rev. D",
    volume = "95",
    number = "1",
    pages = "012005",
    year = "2017"
}

@article{Super-Kamiokande:2022lyl,
    author = "Weatherly, P. and others",
    collaboration = "Super-Kamiokande",
    title = "{Testing Non-Standard Interactions Between Solar Neutrinos and Quarks with Super-Kamiokande}",
    eprint = "2203.11772",
    archivePrefix = "arXiv",
    primaryClass = "hep-ex",
    month = "3",
    year = "2022"
}

@article{ANTARES:2021crm,
    author = "Albert, A. and others",
    collaboration = "ANTARES",
    title = "{Search for non-standard neutrino interactions with 10 years of ANTARES data}",
    eprint = "2112.14517",
    archivePrefix = "arXiv",
    primaryClass = "hep-ex",
    doi = "10.1007/JHEP07(2022)048",
    journal = "JHEP",
    volume = "07",
    pages = "048",
    year = "2022"
}

@article{Super-Kamiokande:2017yvm,
    author = "Abe, K. and others",
    collaboration = "Super-Kamiokande",
    title = "{Atmospheric neutrino oscillation analysis with external constraints in Super-Kamiokande I-IV}",
    eprint = "1710.09126",
    archivePrefix = "arXiv",
    primaryClass = "hep-ex",
    doi = "10.1103/PhysRevD.97.072001",
    journal = "Phys. Rev. D",
    volume = "97",
    number = "7",
    pages = "072001",
    year = "2018"
}

@article{Babu:2019mfe,
    author = "Babu, K. S. and Dev, P. S. Bhupal and Jana, Sudip and Thapa, Anil",
    title = "{Non-Standard Interactions in Radiative Neutrino Mass Models}",
    eprint = "1907.09498",
    archivePrefix = "arXiv",
    primaryClass = "hep-ph",
    reportNumber = "FERMILAB-PUB-19-304-T, OSU-HEP-19-04",
    doi = "10.1007/JHEP03(2020)006",
    journal = "JHEP",
    volume = "03",
    pages = "006",
    year = "2020"
}

@article{Grossman:1995wx,
    author = "Grossman, Yuval",
    title = "{Nonstandard neutrino interactions and neutrino oscillation experiments}",
    eprint = "hep-ph/9507344",
    archivePrefix = "arXiv",
    reportNumber = "WIS-95-30-PH",
    doi = "10.1016/0370-2693(95)01069-3",
    journal = "Phys. Lett. B",
    volume = "359",
    pages = "141--147",
    year = "1995"
}

@article{Ge:2018uhz,
    author = "Ge, Shao-Feng and Parke, Stephen J.",
    title = "{Scalar Nonstandard Interactions in Neutrino Oscillation}",
    eprint = "1812.08376",
    archivePrefix = "arXiv",
    primaryClass = "hep-ph",
    reportNumber = "IPMU18-0206, FERMILAB-PUB-18-487-T",
    doi = "10.1103/PhysRevLett.122.211801",
    journal = "Phys. Rev. Lett.",
    volume = "122",
    number = "21",
    pages = "211801",
    year = "2019"
}

@article{Flores:2026vbx,
    author = "Flores, L. J. and Pacheco-Ak{\'e}, R. and Peinado, Eduardo and Sanchez Garcia, G. and V{\'a}zquez-J{\'a}uregui, E.",
    title = "{Constraints on light sterile neutrinos and scalar nonstandard interactions using the first reactor antineutrino oscillation results at JUNO}",
    eprint = "2603.24677",
    archivePrefix = "arXiv",
    primaryClass = "hep-ph",
    doi = "10.1103/5zws-wpcc",
    journal = "Phys. Rev. D",
    volume = "113",
    number = "11",
    pages = "115030",
    year = "2026"
}

@article{Alves:2026ydc,
    author = "Alves, Gustavo F. S. and Nunokawa, Hiroshi and Zukanovich Funchal, Renata",
    title = "{Impact of New Physics on the JUNO-Long-Baseline Synergy in Neutrino Mass Ordering Determination}",
    eprint = "2603.17181",
    archivePrefix = "arXiv",
    primaryClass = "hep-ph",
    reportNumber = "FERMILAB-PUB-26-0111-T",
    month = "3",
    year = "2026"
}

@article{Choubey:2026jiq,
    author = "Choubey, Sandhya and Lund, Andreas",
    title = "{Neutrino mass ordering in JUNO at risk from scalar NSI induced resonance}",
    eprint = "2602.05564",
    archivePrefix = "arXiv",
    primaryClass = "hep-ph",
    month = "2",
    year = "2026"
}

@article{ESSnuSB:2023lbg,
    author = "Aguilar, J. and others",
    collaboration = "ESSnuSB",
    title = "{Study of nonstandard interactions mediated by a scalar field at the ESSnuSB experiment}",
    eprint = "2310.10749",
    archivePrefix = "arXiv",
    primaryClass = "hep-ex",
    doi = "10.1103/PhysRevD.109.115010",
    journal = "Phys. Rev. D",
    volume = "109",
    number = "11",
    pages = "115010",
    year = "2024"
}

@article{Dutta:2025rxh,
    author = "Dutta, Bhaskar and Karthikeyan, Aparajitha and Mishra, Nityasa and Porto, Yago and Strigari, Louis E.",
    title = "{Scalar non-standard neutrino interactions in Galactic supernovae}",
    eprint = "2508.16558",
    archivePrefix = "arXiv",
    primaryClass = "hep-ph",
    month = "8",
    year = "2025"
}

@article{Das:2025zts,
    author = "Das, Sudipta and Reno, Mary Hall",
    title = "{Probing scalar non-standard interaction of supernova neutrinos in next-generation neutrino experiments}",
    eprint = "2508.16510",
    archivePrefix = "arXiv",
    primaryClass = "hep-ph",
    doi = "10.1088/1475-7516/2026/02/076",
    journal = "JCAP",
    volume = "02",
    pages = "076",
    year = "2026"
}

@article{Denton:2024upc,
    author = "Denton, Peter B. and Giarnetti, Alessio and Meloni, Davide",
    title = "{Solar neutrinos and the strongest oscillation constraints on scalar NSI}",
    eprint = "2409.15411",
    archivePrefix = "arXiv",
    primaryClass = "hep-ph",
    doi = "10.1007/JHEP01(2025)097",
    journal = "JHEP",
    volume = "01",
    pages = "097",
    year = "2025"
}

@article{Venzor:2020ova,
    author = "Venzor, Jorge and P{\'e}rez-Lorenzana, Abdel and De-Santiago, Josue",
    title = "{Bounds on neutrino-scalar nonstandard interactions from big bang nucleosynthesis}",
    eprint = "2009.08104",
    archivePrefix = "arXiv",
    primaryClass = "hep-ph",
    doi = "10.1103/PhysRevD.103.043534",
    journal = "Phys. Rev. D",
    volume = "103",
    number = "4",
    pages = "043534",
    year = "2021"
}

@article{Yadav:2024qav,
    author = "Yadav, Bhavna and Alok, Ashutosh Kumar",
    title = "{Impact of scalar NSI on spatial and temporal correlations in neutrino oscillations}",
    eprint = "2411.17503",
    archivePrefix = "arXiv",
    primaryClass = "hep-ph",
    doi = "10.1088/1361-6471/ae263c",
    journal = "J. Phys. G",
    volume = "52",
    number = "12",
    pages = "125004",
    year = "2025"
}

@article{DeRomeri:2024iaw,
    author = "De Romeri, Valentina and Papoulias, Dimitrios K. and Ternes, Christoph A.",
    title = "{Bounds on new neutrino interactions from the first CE{\ensuremath{\nu}}NS data at direct detection experiments}",
    eprint = "2411.11749",
    archivePrefix = "arXiv",
    primaryClass = "hep-ph",
    doi = "10.1088/1475-7516/2025/05/012",
    journal = "JCAP",
    volume = "05",
    pages = "012",
    year = "2025"
}

@article{Pusty:2024hcn,
    author = "Pusty, Sambit Kumar and Majhi, Rudra and Singha, Dinesh Kumar and Ghosh, Monojit and Mohanta, Rukmani",
    title = "{Impact of scalar NSI with off-diagonal parameters at DUNE and P2SO}",
    eprint = "2410.23014",
    archivePrefix = "arXiv",
    primaryClass = "hep-ph",
    doi = "10.1140/epjc/s10052-025-15014-y",
    journal = "Eur. Phys. J. C",
    volume = "85",
    number = "11",
    pages = "1294",
    year = "2025"
}

@article{Medhi:2024rsi,
    author = "Medhi, Abinash and Sarker, Arnab and Devi, Moon Moon",
    title = "{Scalar NSI: a unique tool for constraining absolute neutrino masses via neutrino oscillations}",
    eprint = "2307.05348",
    archivePrefix = "arXiv",
    primaryClass = "hep-ph",
    doi = "10.1140/epjc/s10052-025-14089-x",
    journal = "Eur. Phys. J. C",
    volume = "85",
    number = "4",
    pages = "380",
    year = "2025"
}

@article{Bezboruah:2024yhk,
    author = "Bezboruah, Dharitree and Chattopadhyay, Dibya S. and Medhi, Abinash and Sarker, Arnab and Devi, Moon Moon",
    title = "{Neutrino oscillations in presence of diagonal elements of scalar NSI: an analytic approach}",
    eprint = "2410.05250",
    archivePrefix = "arXiv",
    primaryClass = "hep-ph",
    doi = "10.1007/JHEP12(2024)222",
    journal = "JHEP",
    volume = "12",
    pages = "222",
    year = "2025"
}

@article{Sarker:2024ytu,
    author = "Sarker, Arnab and Bezboruah, Dharitree and Medhi, Abinash and Devi, Moon Moon",
    title = "{Sensitivity of DUNE in the presence of off-diagonal scalar NSI parameters}",
    eprint = "2406.15307",
    archivePrefix = "arXiv",
    primaryClass = "hep-ph",
    doi = "10.1103/dj83-rw9c",
    journal = "Phys. Rev. D",
    volume = "112",
    number = "3",
    pages = "035042",
    year = "2025"
}

@article{Dutta:2024hqq,
    author = "Dutta, Bhaskar and Ghosh, Sumit and Kelly, Kevin J. and Li, Tianjun and Thompson, Adrian and Verma, Ankur",
    title = "{Non-standard neutrino interactions mediated by a light scalar at DUNE}",
    eprint = "2401.02107",
    archivePrefix = "arXiv",
    primaryClass = "hep-ph",
    doi = "10.1007/JHEP07(2024)213",
    journal = "JHEP",
    volume = "07",
    pages = "213",
    year = "2024"
}

@article{Sarker:2023qzp,
    author = "Sarker, Arnab and Medhi, Abinash and Bezboruah, Dharitree and Devi, Moon Moon and Dutta, Debajyoti",
    title = "{Impact of scalar NSI on the neutrino mass ordering sensitivity at DUNE, HK and KNO}",
    eprint = "2309.12249",
    archivePrefix = "arXiv",
    primaryClass = "hep-ph",
    doi = "10.1007/JHEP06(2024)128",
    journal = "JHEP",
    volume = "06",
    pages = "128",
    year = "2024"
}

@article{Singha:2023set,
    author = "Singha, Dinesh Kumar and Majhi, Rudra and Panda, Lipsarani and Ghosh, Monojit and Mohanta, Rukmani",
    title = "{Study of scalar nonstandard interaction at the Protvino to super-ORCA experiment}",
    eprint = "2308.10789",
    archivePrefix = "arXiv",
    primaryClass = "hep-ph",
    doi = "10.1103/PhysRevD.109.095038",
    journal = "Phys. Rev. D",
    volume = "109",
    number = "9",
    pages = "095038",
    year = "2024"
}

@article{Schwemberger:2023hee,
    author = "Schwemberger, Thomas and Takhistov, Volodymyr and Yu, Tien-Tien",
    title = "{Hunting nonstandard neutrino interactions and leptoquarks in dark matter experiments}",
    eprint = "2307.15736",
    archivePrefix = "arXiv",
    primaryClass = "hep-ph",
    reportNumber = "KEK-QUP-2023-0007, KEK-TH-2516, KEK-Cosmo-0309, IPMU23-0009",
    doi = "10.1088/1475-7516/2024/11/068",
    journal = "JCAP",
    volume = "11",
    pages = "068",
    year = "2024"
}

@article{Gupta:2023wct,
    author = "Gupta, Aman and Majumdar, Debasish and Prakash, Suprabh",
    title = "{Neutrino oscillation measurements with KamLAND and JUNO in the presence of scalar NSI}",
    eprint = "2306.07343",
    archivePrefix = "arXiv",
    primaryClass = "hep-ph",
    doi = "10.1016/j.dark.2025.102011",
    journal = "Phys. Dark Univ.",
    volume = "49",
    pages = "102011",
    year = "2025"
}

@article{Denton:2022pxt,
    author = "Denton, Peter B. and Giarnetti, Alessio and Meloni, Davide",
    title = "{How to identify different new neutrino oscillation physics scenarios at DUNE}",
    eprint = "2210.00109",
    archivePrefix = "arXiv",
    primaryClass = "hep-ph",
    doi = "10.1007/JHEP02(2023)210",
    journal = "JHEP",
    volume = "02",
    pages = "210",
    year = "2023"
}

@article{Medhi:2022qmu,
    author = "Medhi, Abinash and Devi, Moon Moon and Dutta, Debajyoti",
    title = "{Imprints of scalar NSI on the CP-violation sensitivity using synergy among DUNE, T2HK and T2HKK}",
    eprint = "2209.05287",
    archivePrefix = "arXiv",
    primaryClass = "hep-ph",
    doi = "10.1007/JHEP01(2023)079",
    journal = "JHEP",
    volume = "01",
    pages = "079",
    year = "2023"
}

@article{Bergmann:1999rz,
    author = "Bergmann, Sven and Grossman, Yuval and Nardi, Enrico",
    title = "{Neutrino propagation in matter with general interactions}",
    eprint = "hep-ph/9903517",
    archivePrefix = "arXiv",
    reportNumber = "SLAC-PUB-8083, WIS-99-11-DPP, UDEA-PE-99-001, WIS-99-11-MAR.-DPP",
    doi = "10.1103/PhysRevD.60.093008",
    journal = "Phys. Rev. D",
    volume = "60",
    pages = "093008",
    year = "1999"
}

@article{Schwemberger:2022fjl,
    author = "Schwemberger, Thomas and Yu, Tien-Tien",
    title = "{Detecting beyond the standard model interactions of solar neutrinos in low-threshold dark matter detectors}",
    eprint = "2202.01254",
    archivePrefix = "arXiv",
    primaryClass = "hep-ph",
    doi = "10.1103/PhysRevD.106.015002",
    journal = "Phys. Rev. D",
    volume = "106",
    number = "1",
    pages = "015002",
    year = "2022"
}

@article{Medhi:2021wxj,
    author = "Medhi, Abinash and Dutta, Debajyoti and Devi, Moon Moon",
    title = "{Exploring the effects of scalar non standard interactions on the CP violation sensitivity at DUNE}",
    eprint = "2111.12943",
    archivePrefix = "arXiv",
    primaryClass = "hep-ph",
    doi = "10.1007/JHEP06(2022)129",
    journal = "JHEP",
    volume = "06",
    pages = "129",
    year = "2022"
}

@article{Chaves:2021kxe,
    author = "Chaves, Mariano E. and de Holanda, Pedro Cunha and Peres, Orlando Luis Goulart",
    title = "{Testing non-standard neutrino interactions in (anti)-electron neutrino disappearance experiments}",
    eprint = "2106.15725",
    archivePrefix = "arXiv",
    primaryClass = "hep-ph",
    doi = "10.1007/JHEP03(2023)180",
    journal = "JHEP",
    volume = "03",
    pages = "180",
    year = "2023"
}

@article{Escrihuela:2021mud,
    author = "Escrihuela, F. J. and Flores, L. J. and Miranda, O. G. and Rend{\'o}n, Javier",
    title = "{Global constraints on neutral-current generalized neutrino interactions}",
    eprint = "2105.06484",
    archivePrefix = "arXiv",
    primaryClass = "hep-ph",
    doi = "10.1007/JHEP07(2021)061",
    journal = "JHEP",
    volume = "07",
    pages = "061",
    year = "2021"
}

@article{Smirnov:2019cae,
    author = "Smirnov, Alexei Yu and Xu, Xun-Jie",
    title = "{Wolfenstein potentials for neutrinos induced by ultra-light mediators}",
    eprint = "1909.07505",
    archivePrefix = "arXiv",
    primaryClass = "hep-ph",
    doi = "10.1007/JHEP12(2019)046",
    journal = "JHEP",
    volume = "12",
    pages = "046",
    year = "2019"
}

@article{Khan:2019jvr,
    author = "Khan, Amir N. and Rodejohann, Werner and Xu, Xun-Jie",
    title = "{Borexino and general neutrino interactions}",
    eprint = "1906.12102",
    archivePrefix = "arXiv",
    primaryClass = "hep-ph",
    reportNumber = "FERMILAB-PUB-19-348-T",
    doi = "10.1103/PhysRevD.101.055047",
    journal = "Phys. Rev. D",
    volume = "101",
    number = "5",
    pages = "055047",
    year = "2020"
}

@article{deGouvea:2019qaz,
    author = "de Gouv{\^e}a, Andr{\'e} and Dev, P. S. Bhupal and Dutta, Bhaskar and Ghosh, Tathagata and Han, Tao and Zhang, Yongchao",
    title = "{Leptonic Scalars at the LHC}",
    eprint = "1910.01132",
    archivePrefix = "arXiv",
    primaryClass = "hep-ph",
    reportNumber = "PITT-PACC 1909, MI-TH-1936",
    doi = "10.1007/JHEP07(2020)142",
    journal = "JHEP",
    volume = "07",
    pages = "142",
    year = "2020"
}

@article{Berryman:2018ogk,
    author = "Berryman, Jeffrey M. and De Gouv{\^e}a, Andr{\'e} and Kelly, Kevin J. and Zhang, Yue",
    title = "{Lepton-Number-Charged Scalars and Neutrino Beamstrahlung}",
    eprint = "1802.00009",
    archivePrefix = "arXiv",
    primaryClass = "hep-ph",
    reportNumber = "NUHEP-TH-18-03, FERMILAB-PUB-18-020-T",
    doi = "10.1103/PhysRevD.97.075030",
    journal = "Phys. Rev. D",
    volume = "97",
    number = "7",
    pages = "075030",
    year = "2018"
}

@article{Pasquini:2015fjv,
    author = "Pasquini, P. S. and Peres, O. L. G.",
    title = "{Bounds on Neutrino-Scalar Yukawa Coupling}",
    eprint = "1511.01811",
    archivePrefix = "arXiv",
    primaryClass = "hep-ph",
    doi = "10.1103/PhysRevD.93.053007",
    journal = "Phys. Rev. D",
    volume = "93",
    number = "5",
    pages = "053007",
    year = "2016",
    note = "[Erratum: Phys.Rev.D 93, 079902 (2016)]"
}

@article{Barger:1981vd,
    author = "Barger, Vernon D. and Keung, Wai-Yee and Pakvasa, S.",
    title = "{Majoron Emission by Neutrinos}",
    reportNumber = "MAD/PH/15",
    doi = "10.1103/PhysRevD.25.907",
    journal = "Phys. Rev. D",
    volume = "25",
    pages = "907",
    year = "1982"
}

@article{Lessa:2007up,
    author = "Lessa, A. P. and Peres, O. L. G.",
    title = "{Revising limits on neutrino-Majoron couplings}",
    eprint = "hep-ph/0701068",
    archivePrefix = "arXiv",
    doi = "10.1103/PhysRevD.75.094001",
    journal = "Phys. Rev. D",
    volume = "75",
    pages = "094001",
    year = "2007"
}

@article{Dev:2024ygx,
    author = "Dev, P. S. Bhupal and Kim, Doojin and Sathyan, Deepak and Sinha, Kuver and Zhang, Yongchao",
    title = "{New laboratory constraints on neutrinophilic mediators}",
    eprint = "2407.12738",
    archivePrefix = "arXiv",
    primaryClass = "hep-ph",
    reportNumber = "CETUP-2024-005",
    doi = "10.1016/j.physletb.2025.139765",
    journal = "Phys. Lett. B",
    volume = "868",
    pages = "139765",
    year = "2025"
}

@article{DeGouvea:2019wpf,
    author = "De Gouv{\^e}a, Andr{\'e} and Sen, Manibrata and Tangarife, Walter and Zhang, Yue",
    title = "{Dodelson-Widrow Mechanism in the Presence of Self-Interacting Neutrinos}",
    eprint = "1910.04901",
    archivePrefix = "arXiv",
    primaryClass = "hep-ph",
    reportNumber = "NUHEP-TH/19-15, FERMILAB-PUB-19-522-T",
    doi = "10.1103/PhysRevLett.124.081802",
    journal = "Phys. Rev. Lett.",
    volume = "124",
    number = "8",
    pages = "081802",
    year = "2020"
}

@article{Foroughi-Abari:2025mhj,
    author = "Foroughi-Abari, Saeid and Kelly, Kevin J. and Zhang, Yue",
    title = "{Radiative correction from secret neutrino interactions and implications for neutrino-scattering experiments}",
    eprint = "2510.15023",
    archivePrefix = "arXiv",
    primaryClass = "hep-ph",
    reportNumber = "MI-HET-868",
    doi = "10.1103/9lrf-pw2b",
    journal = "Phys. Rev. D",
    volume = "113",
    number = "7",
    pages = "075030",
    year = "2026"
}

@article{Dev:2025tdv,
    author = "Dev, P. S. Bhupal and Kim, Doojin and Sathyan, Deepak and Sinha, Kuver and Zhang, Yongchao",
    title = "{New Constraints on Neutrino-Dark Matter Interactions: A Comprehensive Analysis}",
    eprint = "2507.01000",
    archivePrefix = "arXiv",
    primaryClass = "hep-ph",
    reportNumber = "CETUP-2023-022",
    month = "7",
    year = "2025"
}

@article{Noriega:2025ulc,
    author = "Noriega, Hern{\'a}n E. and De-Santiago, Josue and Garcia-Arroyo, Gabriela and Venzor, Jorge and P{\'e}rez-Lorenzana, Abdel",
    title = "{Resonant neutrino self-interactions: Insights from the full shape galaxy power spectrum}",
    eprint = "2506.07994",
    archivePrefix = "arXiv",
    primaryClass = "astro-ph.CO",
    doi = "10.1103/b9x4-hnqn",
    journal = "Phys. Rev. D",
    volume = "112",
    number = "6",
    pages = "063509",
    year = "2025"
}

@article{Das:2025asx,
    author = "Das, Anirban and Dev, P. S. Bhupal and Gao, Christina and Ghosh, Subhajit and Kim, Taegyun",
    title = "{Impostor among Neutrinos: Dark Radiation Masquerading as Self-Interacting Neutrinos}",
    eprint = "2506.08085",
    archivePrefix = "arXiv",
    primaryClass = "hep-ph",
    reportNumber = "MITP-25-044",
    doi = "10.1103/jprg-jll6",
    journal = "Phys. Rev. Lett.",
    volume = "136",
    number = "13",
    pages = "131003",
    year = "2026"
}

@article{He:2025bex,
    author = "He, Yuxuan and Liu, Jia and Wang, Xiao-Ping and Zhong, Yi-Ming",
    title = "{Implications of the KM3NeT ultrahigh-energy event on neutrino self-interactions}",
    eprint = "2504.20163",
    archivePrefix = "arXiv",
    primaryClass = "hep-ph",
    doi = "10.1103/bb85-7kyf",
    journal = "Phys. Rev. D",
    volume = "113",
    number = "4",
    pages = "043022",
    year = "2026"
}

@article{Wang:2025qap,
    author = "Wang, Isaac R. and Xu, Xun-Jie and Zhou, Bei",
    title = "{Widen the Resonance: Probing a New Regime of Neutrino Self-Interactions with Astrophysical Neutrinos}",
    eprint = "2501.07624",
    archivePrefix = "arXiv",
    primaryClass = "hep-ph",
    reportNumber = "FERMILAB-PUB-25-0016-T",
    doi = "10.1103/9ddp-j1z9",
    journal = "Phys. Rev. Lett.",
    volume = "135",
    number = "18",
    pages = "181002",
    year = "2025"
}

@article{Foroughi-Abari:2025upe,
    author = "Foroughi-Abari, Saeid and Kelly, Kevin J. and Rai, Mudit and Zhang, Yue",
    title = "{Enabling Strong Neutrino Self-Interaction with an Unparticle Mediator}",
    eprint = "2501.02049",
    archivePrefix = "arXiv",
    primaryClass = "hep-ph",
    reportNumber = "MI-HET-847",
    doi = "10.1103/PhysRevLett.134.181001",
    journal = "Phys. Rev. Lett.",
    volume = "134",
    number = "18",
    pages = "181001",
    year = "2025"
}

@article{Liu:2024ywd,
    author = "Liu, Hongkai and Ueda, Daiki",
    title = "{Searching for neutrino self-interactions at future muon colliders}",
    eprint = "2412.11910",
    archivePrefix = "arXiv",
    primaryClass = "hep-ph",
    doi = "10.1016/j.physletb.2025.139748",
    journal = "Phys. Lett. B",
    volume = "868",
    pages = "139748",
    year = "2025"
}

@article{deLima:2024ohf,
    author = "de Lima, Carlos Henrique and McKeen, David and Ng, John N. and Shamma, Michael and Tuckler, Douglas",
    title = "{Probing lepton number violation at same-sign lepton colliders}",
    eprint = "2411.15303",
    archivePrefix = "arXiv",
    primaryClass = "hep-ph",
    doi = "10.1103/PhysRevD.111.075002",
    journal = "Phys. Rev. D",
    volume = "111",
    number = "7",
    pages = "075002",
    year = "2025"
}

@article{Zhang:2024meg,
    author = "Zhang, Yue",
    title = "{Neutrino self-interaction and weak mixing-angle measurements}",
    eprint = "2411.05070",
    archivePrefix = "arXiv",
    primaryClass = "hep-ph",
    doi = "10.1103/7y8g-56fz",
    journal = "Phys. Rev. D",
    volume = "112",
    number = "3",
    pages = "035027",
    year = "2025"
}

@article{Benso:2024qrg,
    author = "Benso, Cristina and Schwetz, Thomas and Vatsyayan, Drona",
    title = "{Large neutrino mass in cosmology and keV sterile neutrino dark matter from a dark sector}",
    eprint = "2410.23926",
    archivePrefix = "arXiv",
    primaryClass = "hep-ph",
    doi = "10.1088/1475-7516/2025/04/054",
    journal = "JCAP",
    volume = "04",
    pages = "054",
    year = "2025"
}

@article{Suliga:2024nng,
    author = "Suliga, Anna M. and Cheong, Patrick Chi-Kit and Froustey, Julien and Fuller, George M. and Gr{\'a}f, Luk{\'a}{\v{s}} and Kehrer, Kyle and Scholer, Oliver and Shalgar, Shashank",
    title = "{Nonconservation of Lepton Numbers in the Neutrino Sector Could Change the Prospects for Core Collapse Supernova Explosions}",
    eprint = "2410.01080",
    archivePrefix = "arXiv",
    primaryClass = "hep-ph",
    reportNumber = "CETUP2024-009, FERMILAB-PUB-24-1033-T, INT-PUB-24-050, N3AS-24-033",
    doi = "10.1103/gnp5-4y8k",
    journal = "Phys. Rev. Lett.",
    volume = "134",
    number = "24",
    pages = "241002",
    year = "2025"
}

@article{Chikashige:1980qk,
    author = "Chikashige, Y. and Mohapatra, Rabindra N. and Peccei, R. D.",
    title = "{Spontaneously Broken Lepton Number and Cosmological Constraints on the Neutrino Mass Spectrum}",
    reportNumber = "MPI-PAE/PTh 40/80",
    doi = "10.1103/PhysRevLett.45.1926",
    journal = "Phys. Rev. Lett.",
    volume = "45",
    pages = "1926",
    year = "1980"
}

@article{Bardin:1970wq,
    author = "Bardin, D. Yu. and Bilenky, Samoil M. and Pontecorvo, B.",
    title = "{On the nu - nu interaction}",
    doi = "10.1016/0370-2693(70)90602-7",
    journal = "Phys. Lett. B",
    volume = "32",
    pages = "121--124",
    year = "1970"
}

@article{Beacom:2002cb,
    author = "Beacom, John F. and Bell, Nicole F.",
    title = "{Do Solar Neutrinos Decay?}",
    eprint = "hep-ph/0204111",
    archivePrefix = "arXiv",
    reportNumber = "FERMILAB-PUB-02-061-A",
    doi = "10.1103/PhysRevD.65.113009",
    journal = "Phys. Rev. D",
    volume = "65",
    pages = "113009",
    year = "2002"
}

@article{Laha:2013xua,
    author = "Laha, Ranjan and Dasgupta, Basudeb and Beacom, John F.",
    title = "{Constraints on New Neutrino Interactions via Light Abelian Vector Bosons}",
    eprint = "1304.3460",
    archivePrefix = "arXiv",
    primaryClass = "hep-ph",
    doi = "10.1103/PhysRevD.89.093025",
    journal = "Phys. Rev. D",
    volume = "89",
    number = "9",
    pages = "093025",
    year = "2014"
}

@article{Ioka:2014kca,
    author = "Ioka, Kunihto and Murase, Kohta",
    title = "{IceCube PeV{\textendash}EeV neutrinos and secret interactions of neutrinos}",
    eprint = "1404.2279",
    archivePrefix = "arXiv",
    primaryClass = "astro-ph.HE",
    reportNumber = "KEK-TH-1723, KEK-COSMO-141",
    doi = "10.1093/ptep/ptu090",
    journal = "PTEP",
    volume = "2014",
    number = "6",
    pages = "061E01",
    year = "2014"
}

@article{Ng:2014pca,
    author = "Ng, Kenny C. Y. and Beacom, John F.",
    title = "{Cosmic neutrino cascades from secret neutrino interactions}",
    eprint = "1404.2288",
    archivePrefix = "arXiv",
    primaryClass = "astro-ph.HE",
    doi = "10.1103/PhysRevD.90.065035",
    journal = "Phys. Rev. D",
    volume = "90",
    number = "6",
    pages = "065035",
    year = "2014",
    note = "[Erratum: Phys.Rev.D 90, 089904 (2014)]"
}

@article{Forastieri:2015paa,
    author = "Forastieri, Francesco and Lattanzi, Massimiliano and Natoli, Paolo",
    title = "{Constraints on secret neutrino interactions after Planck}",
    eprint = "1504.04999",
    archivePrefix = "arXiv",
    primaryClass = "astro-ph.CO",
    doi = "10.1088/1475-7516/2015/07/014",
    journal = "JCAP",
    volume = "07",
    pages = "014",
    year = "2015"
}

@article{Bialynicka-Birula:1964ddi,
    author = "Bialynicka-Birula, Z.",
    title = "{Do Neutrinos Interact between Themselves?}",
    doi = "10.1007/BF02749481",
    journal = "Nuovo Cim.",
    volume = "33",
    pages = "1484--1487",
    year = "1964"
}

@article{Shoemaker:2015qul,
    author = "Shoemaker, Ian M. and Murase, Kohta",
    title = "{Probing BSM Neutrino Physics with Flavor and Spectral Distortions: Prospects for Future High-Energy Neutrino Telescopes}",
    eprint = "1512.07228",
    archivePrefix = "arXiv",
    primaryClass = "astro-ph.HE",
    doi = "10.1103/PhysRevD.93.085004",
    journal = "Phys. Rev. D",
    volume = "93",
    number = "8",
    pages = "085004",
    year = "2016"
}

@article{Das:2017iuj,
    author = "Das, Anirban and Dighe, Amol and Sen, Manibrata",
    title = "{New effects of non-standard self-interactions of neutrinos in a supernova}",
    eprint = "1705.00468",
    archivePrefix = "arXiv",
    primaryClass = "hep-ph",
    reportNumber = "TIFR-TH-17-18",
    doi = "10.1088/1475-7516/2017/05/051",
    journal = "JCAP",
    volume = "05",
    pages = "051",
    year = "2017"
}

@article{Huang:2017egl,
    author = "Huang, Guo-yuan and Ohlsson, Tommy and Zhou, Shun",
    title = "{Observational Constraints on Secret Neutrino Interactions from Big Bang Nucleosynthesis}",
    eprint = "1712.04792",
    archivePrefix = "arXiv",
    primaryClass = "hep-ph",
    doi = "10.1103/PhysRevD.97.075009",
    journal = "Phys. Rev. D",
    volume = "97",
    number = "7",
    pages = "075009",
    year = "2018"
}

@article{Blum:2018ljv,
    author = "Blum, Kfir and Nir, Yosef and Shavit, Michal",
    title = "{Neutrinoless double-beta decay with massive scalar emission}",
    eprint = "1802.08019",
    archivePrefix = "arXiv",
    primaryClass = "hep-ph",
    doi = "10.1016/j.physletb.2018.08.022",
    journal = "Phys. Lett. B",
    volume = "785",
    pages = "354--361",
    year = "2018"
}

@article{Brune:2018sab,
    author = {Brune, Tim and P{\"a}s, Heinrich},
    title = "{Massive Majorons and constraints on the Majoron-neutrino coupling}",
    eprint = "1808.08158",
    archivePrefix = "arXiv",
    primaryClass = "hep-ph",
    reportNumber = "DO-TH 18/23",
    doi = "10.1103/PhysRevD.99.096005",
    journal = "Phys. Rev. D",
    volume = "99",
    number = "9",
    pages = "096005",
    year = "2019"
}

@article{Forastieri:2019cuf,
    author = "Forastieri, Francesco and Lattanzi, Massimiliano and Natoli, Paolo",
    title = "{Cosmological constraints on neutrino self-interactions with a light mediator}",
    eprint = "1904.07810",
    archivePrefix = "arXiv",
    primaryClass = "astro-ph.CO",
    doi = "10.1103/PhysRevD.100.103526",
    journal = "Phys. Rev. D",
    volume = "100",
    number = "10",
    pages = "103526",
    year = "2019"
}

@article{Blinov:2019gcj,
    author = "Blinov, Nikita and Kelly, Kevin James and Krnjaic, Gordan Z and McDermott, Samuel D",
    title = "{Constraining the Self-Interacting Neutrino Interpretation of the Hubble Tension}",
    eprint = "1905.02727",
    archivePrefix = "arXiv",
    primaryClass = "astro-ph.CO",
    reportNumber = "FERMILAB-PUB-19-175-A-T",
    doi = "10.1103/PhysRevLett.123.191102",
    journal = "Phys. Rev. Lett.",
    volume = "123",
    number = "19",
    pages = "191102",
    year = "2019"
}

@article{Shalgar:2019rqe,
    author = "Shalgar, Shashank and Tamborra, Irene and Bustamante, Mauricio",
    title = "{Core-collapse supernovae stymie secret neutrino interactions}",
    eprint = "1912.09115",
    archivePrefix = "arXiv",
    primaryClass = "astro-ph.HE",
    doi = "10.1103/PhysRevD.103.123008",
    journal = "Phys. Rev. D",
    volume = "103",
    number = "12",
    pages = "123008",
    year = "2021"
}

@article{EscuderoAbenza:2020cmq,
    author = "Escudero Abenza, Miguel",
    title = "{Precision early universe thermodynamics made simple: $N_{\rm eff}$ and neutrino decoupling in the Standard Model and beyond}",
    eprint = "2001.04466",
    archivePrefix = "arXiv",
    primaryClass = "hep-ph",
    reportNumber = "KCL-2019-85",
    doi = "10.1088/1475-7516/2020/05/048",
    journal = "JCAP",
    volume = "05",
    pages = "048",
    year = "2020"
}

@article{Bustamante:2020mep,
    author = "Bustamante, Mauricio and Rosenstr{\o}m, Charlotte and Shalgar, Shashank and Tamborra, Irene",
    title = "{Bounds on secret neutrino interactions from high-energy astrophysical neutrinos}",
    eprint = "2001.04994",
    archivePrefix = "arXiv",
    primaryClass = "astro-ph.HE",
    doi = "10.1103/PhysRevD.101.123024",
    journal = "Phys. Rev. D",
    volume = "101",
    number = "12",
    pages = "123024",
    year = "2020"
}

@article{Grohs:2020xxd,
    author = "Grohs, E. and Fuller, George M. and Sen, Manibrata",
    title = "{Consequences of neutrino self interactions for weak decoupling and big bang nucleosynthesis}",
    eprint = "2002.08557",
    archivePrefix = "arXiv",
    primaryClass = "astro-ph.CO",
    doi = "10.1088/1475-7516/2020/07/001",
    journal = "JCAP",
    volume = "07",
    pages = "001",
    year = "2020"
}

@article{Lyu:2020lps,
    author = "Lyu, Kun-Feng and Stamou, Emmanuel and Wang, Lian-Tao",
    title = "{Self-interacting neutrinos: Solution to Hubble tension versus experimental constraints}",
    eprint = "2004.10868",
    archivePrefix = "arXiv",
    primaryClass = "hep-ph",
    doi = "10.1103/PhysRevD.103.015004",
    journal = "Phys. Rev. D",
    volume = "103",
    number = "1",
    pages = "015004",
    year = "2021"
}

@article{Whitford:2025dmq,
    author = "Whitford, Abb{\'e} M. and Howlett, Cullan and Davis, Tamara M. and Camarena, David and Cyr-Racine, Francis-Yan",
    title = "{Limits on self-interacting neutrinos from the BAO and CMB phase shift}",
    eprint = "2511.00800",
    archivePrefix = "arXiv",
    primaryClass = "astro-ph.CO",
    doi = "10.1088/1475-7516/2026/03/064",
    journal = "JCAP",
    volume = "03",
    pages = "064",
    year = "2026"
}

@article{Camarena:2024daj,
    author = "Camarena, David and Cyr-Racine, Francis-Yan",
    title = "{Strong constraints on a simple self-interacting neutrino cosmology}",
    eprint = "2403.05496",
    archivePrefix = "arXiv",
    primaryClass = "astro-ph.CO",
    doi = "10.1103/PhysRevD.111.023504",
    journal = "Phys. Rev. D",
    volume = "111",
    number = "2",
    pages = "023504",
    year = "2025"
}

@article{Camarena:2023cku,
    author = "Camarena, David and Cyr-Racine, Francis-Yan and Houghteling, John",
    title = "{Confronting self-interacting neutrinos with the full shape of the galaxy power spectrum}",
    eprint = "2309.03941",
    archivePrefix = "arXiv",
    primaryClass = "astro-ph.CO",
    doi = "10.1103/PhysRevD.108.103535",
    journal = "Phys. Rev. D",
    volume = "108",
    number = "10",
    pages = "103535",
    year = "2023"
}

@article{Kreisch:2022zxp,
    author = "Kreisch, Christina D. and others",
    title = "{Atacama Cosmology Telescope: The persistence of neutrino self-interaction in cosmological measurements}",
    eprint = "2207.03164",
    archivePrefix = "arXiv",
    primaryClass = "astro-ph.CO",
    doi = "10.1103/PhysRevD.109.043501",
    journal = "Phys. Rev. D",
    volume = "109",
    number = "4",
    pages = "043501",
    year = "2024"
}

@article{Park:2019ibn,
    author = "Park, Minsu and Kreisch, Christina D. and Dunkley, Jo and Hadzhiyska, Boryana and Cyr-Racine, Francis-Yan",
    title = "{$\Lambda$CDM or self-interacting neutrinos: How CMB data can tell the two models apart}",
    eprint = "1904.02625",
    archivePrefix = "arXiv",
    primaryClass = "astro-ph.CO",
    doi = "10.1103/PhysRevD.100.063524",
    journal = "Phys. Rev. D",
    volume = "100",
    number = "6",
    pages = "063524",
    year = "2019"
}

@article{Kreisch:2019yzn,
    author = "Kreisch, Christina D. and Cyr-Racine, Francis-Yan and Dor{\'e}, Olivier",
    title = "{Neutrino puzzle: Anomalies, interactions, and cosmological tensions}",
    eprint = "1902.00534",
    archivePrefix = "arXiv",
    primaryClass = "astro-ph.CO",
    doi = "10.1103/PhysRevD.101.123505",
    journal = "Phys. Rev. D",
    volume = "101",
    number = "12",
    pages = "123505",
    year = "2020"
}

@article{Deppisch:2020sqh,
    author = "Deppisch, Frank F. and Graf, Lukas and Rodejohann, Werner and Xu, Xun-Jie",
    title = "{Neutrino Self-Interactions and Double Beta Decay}",
    eprint = "2004.11919",
    archivePrefix = "arXiv",
    primaryClass = "hep-ph",
    doi = "10.1103/PhysRevD.102.051701",
    journal = "Phys. Rev. D",
    volume = "102",
    number = "5",
    pages = "051701",
    year = "2020"
}

@article{Esteban:2021tub,
    author = "Esteban, Ivan and Pandey, Sujata and Brdar, Vedran and Beacom, John F.",
    title = "{Probing secret interactions of astrophysical neutrinos in the high-statistics era}",
    eprint = "2107.13568",
    archivePrefix = "arXiv",
    primaryClass = "hep-ph",
    reportNumber = "FERMILAB-PUB-21-328-T, nuhep-th/21-06",
    doi = "10.1103/PhysRevD.104.123014",
    journal = "Phys. Rev. D",
    volume = "104",
    number = "12",
    pages = "123014",
    year = "2021"
}

@article{Ge:2021lur,
    author = "Ge, Shao-Feng and Pasquini, Pedro",
    title = "{Probing light mediators in the radiative emission of neutrino pair}",
    eprint = "2110.03510",
    archivePrefix = "arXiv",
    primaryClass = "hep-ph",
    doi = "10.1140/epjc/s10052-022-10131-4",
    journal = "Eur. Phys. J. C",
    volume = "82",
    number = "3",
    pages = "208",
    year = "2022"
}

@article{Chichiri:2021wvw,
    author = "Chichiri, Carlos and Gelmini, Graciela B. and Lu, Philip and Takhistov, Volodymyr",
    title = "{Cosmological dependence of sterile neutrino dark matter with self-interacting neutrinos}",
    eprint = "2111.04087",
    archivePrefix = "arXiv",
    primaryClass = "hep-ph",
    reportNumber = "IPMU21-0076",
    doi = "10.1088/1475-7516/2022/09/036",
    journal = "JCAP",
    volume = "09",
    pages = "036",
    year = "2022"
}

@article{Smirnov:2022sfo,
    author = "Smirnov, Alexei Yu. and Xu, Xun-Jie",
    title = "{Neutrino bound states and bound systems}",
    eprint = "2201.00939",
    archivePrefix = "arXiv",
    primaryClass = "hep-ph",
    doi = "10.1007/JHEP08(2022)170",
    journal = "JHEP",
    volume = "08",
    pages = "170",
    year = "2022"
}

@article{deGouvea:2022cmo,
    author = "de Gouv{\^e}a, Andr{\'e} and Sen, Manibrata and Weill, Jean",
    title = "{Visible neutrino decays and the impact of the daughter-neutrino mass}",
    eprint = "2203.14976",
    archivePrefix = "arXiv",
    primaryClass = "hep-ph",
    doi = "10.1103/PhysRevD.106.013005",
    journal = "Phys. Rev. D",
    volume = "106",
    number = "1",
    pages = "013005",
    year = "2022"
}

@article{Das:2022xsz,
    author = "Das, Anirban and Perez-Gonzalez, Yuber F. and Sen, Manibrata",
    title = "{Neutrino secret self-interactions: A booster shot for the cosmic neutrino background}",
    eprint = "2204.11885",
    archivePrefix = "arXiv",
    primaryClass = "hep-ph",
    reportNumber = "SLAC-PUB-17673, IPPP/22/24",
    doi = "10.1103/PhysRevD.106.095042",
    journal = "Phys. Rev. D",
    volume = "106",
    number = "9",
    pages = "095042",
    year = "2022"
}

@article{Das:2023npl,
    author = "Das, Anirban and Ghosh, Subhajit",
    title = "{The magnificent ACT of flavor-specific neutrino self-interaction}",
    eprint = "2303.08843",
    archivePrefix = "arXiv",
    primaryClass = "astro-ph.CO",
    reportNumber = "SLAC-PUB-17708",
    doi = "10.1088/1475-7516/2023/09/042",
    journal = "JCAP",
    volume = "09",
    pages = "042",
    year = "2023"
}

@article{Akita:2022etk,
    author = "Akita, Kensuke and Im, Sang Hui and Masud, Mehedi",
    title = "{Probing non-standard neutrino interactions with a light boson from next galactic and diffuse supernova neutrinos}",
    eprint = "2206.06852",
    archivePrefix = "arXiv",
    primaryClass = "hep-ph",
    reportNumber = "CTPU-PTC-22-13",
    doi = "10.1007/JHEP12(2022)050",
    journal = "JHEP",
    volume = "12",
    pages = "050",
    year = "2022"
}

@article{Chang:2022aas,
    author = "Chang, Po-Wen and Esteban, Ivan and Beacom, John F. and Thompson, Todd A. and Hirata, Christopher M.",
    title = "{Toward Powerful Probes of Neutrino Self-Interactions in Supernovae}",
    eprint = "2206.12426",
    archivePrefix = "arXiv",
    primaryClass = "hep-ph",
    doi = "10.1103/PhysRevLett.131.071002",
    journal = "Phys. Rev. Lett.",
    volume = "131",
    number = "7",
    pages = "071002",
    year = "2023"
}

@article{Chen:2022kal,
    author = "Chen, Yu-Ming and Sen, Manibrata and Tangarife, Walter and Tuckler, Douglas and Zhang, Yue",
    title = "{Core-collapse supernova constraint on the origin of sterile neutrino dark matter via neutrino self-interactions}",
    eprint = "2207.14300",
    archivePrefix = "arXiv",
    primaryClass = "hep-ph",
    doi = "10.1088/1475-7516/2022/11/014",
    journal = "JCAP",
    volume = "11",
    pages = "014",
    year = "2022"
}

@article{Farzan:2002wx,
    author = "Farzan, Yasaman",
    title = "{Bounds on the coupling of the Majoron to light neutrinos from supernova cooling}",
    eprint = "hep-ph/0211375",
    archivePrefix = "arXiv",
    reportNumber = "SLAC-PUB-9543, SISSA-69-2002-EP",
    doi = "10.1103/PhysRevD.67.073015",
    journal = "Phys. Rev. D",
    volume = "67",
    pages = "073015",
    year = "2003"
}

@article{Chikashige:1980ui,
    author = "Chikashige, Y. and Mohapatra, Rabindra N. and Peccei, R. D.",
    title = "{Are There Real Goldstone Bosons Associated with Broken Lepton Number?}",
    reportNumber = "MPI-PAE-PTH-36-80",
    doi = "10.1016/0370-2693(81)90011-3",
    journal = "Phys. Lett. B",
    volume = "98",
    pages = "265--268",
    year = "1981"
}

@article{Gelmini:1980re,
    author = "Gelmini, G. B. and Roncadelli, M.",
    title = "{Left-Handed Neutrino Mass Scale and Spontaneously Broken Lepton Number}",
    reportNumber = "MPI-PAE-PTH-50-80",
    doi = "10.1016/0370-2693(81)90559-1",
    journal = "Phys. Lett. B",
    volume = "99",
    pages = "411--415",
    year = "1981"
}

@article{Schechter:1981cv,
    author = "Schechter, J. and Valle, J. W. F.",
    title = "{Neutrino Decay and Spontaneous Violation of Lepton Number}",
    reportNumber = "SU-4217-203, COO-3533-203",
    doi = "10.1103/PhysRevD.25.774",
    journal = "Phys. Rev. D",
    volume = "25",
    pages = "774",
    year = "1982"
}

@article{Fiorillo:2022cdq,
    author = "Fiorillo, Damiano F. G. and Raffelt, Georg G. and Vitagliano, Edoardo",
    title = "{Strong Supernova 1987A Constraints on Bosons Decaying to Neutrinos}",
    eprint = "2209.11773",
    archivePrefix = "arXiv",
    primaryClass = "hep-ph",
    doi = "10.1103/PhysRevLett.131.021001",
    journal = "Phys. Rev. Lett.",
    volume = "131",
    number = "2",
    pages = "021001",
    year = "2023"
}

@article{Coyle:2022bwa,
    author = "Coyle, Nina M. and Li, Shirley Weishi and Machado, Pedro A. N.",
    title = "{The impact of neutrino-nucleus interaction modeling on new physics searches}",
    eprint = "2210.03753",
    archivePrefix = "arXiv",
    primaryClass = "hep-ph",
    reportNumber = "FERMILAB-PUB-22-726-T",
    doi = "10.1007/JHEP12(2022)166",
    journal = "JHEP",
    volume = "12",
    pages = "166",
    year = "2022"
}

@article{Doring:2023vmk,
    author = {D{\"o}ring, Christian and Vogl, Stefan},
    title = "{Testing secret interaction with astrophysical neutrino point sources}",
    eprint = "2304.08533",
    archivePrefix = "arXiv",
    primaryClass = "hep-ph",
    reportNumber = "ULB-TH/23-05",
    doi = "10.1088/1475-7516/2024/07/015",
    journal = "JCAP",
    volume = "07",
    pages = "015",
    year = "2024"
}

@article{PandaX:2025tls,
    author = "Li, Tao and others",
    collaboration = "PandaX",
    title = "{Probing Scalar-Neutrino and Scalar-Dark-Matter Interactions with PandaX-4T}",
    eprint = "2511.13515",
    archivePrefix = "arXiv",
    primaryClass = "hep-ex",
    doi = "10.1103/pn9y-g5k7",
    journal = "Phys. Rev. Lett.",
    volume = "136",
    number = "24",
    pages = "241802",
    year = "2026"
}

@article{deVries:2025hqa,
    author = "de Vries, Jordy and Gr{\'a}f, Luk{\'a}{\v{s}} and Plakkot, Vaisakh and Star{\'y}, Dominik",
    title = "{Scalarful double beta decay}",
    eprint = "2511.19173",
    archivePrefix = "arXiv",
    primaryClass = "hep-ph",
    doi = "10.1007/JHEP03(2026)102",
    journal = "JHEP",
    volume = "03",
    pages = "102",
    year = "2026"
}

@article{Boudjema:2025okq,
    author = "Boudjema, Noor-Ines and Deppisch, Frank F. and Herrero-Brocal, Antonio and Majumdar, Chayan and Senapati, Supriya",
    title = "{Probing dark sector particles coupling to neutrinos with double beta decay}",
    eprint = "2511.13606",
    archivePrefix = "arXiv",
    primaryClass = "hep-ph",
    doi = "10.1103/lxh2-wsrz",
    journal = "Phys. Rev. D",
    volume = "113",
    number = "7",
    pages = "075039",
    year = "2026"
}

@article{Li:2023puz,
    author = "Li, Shao-Ping and Xu, Xun-Jie",
    title = "{N$_{eff}$ constraints on light mediators coupled to neutrinos: the dilution-resistant effect}",
    eprint = "2307.13967",
    archivePrefix = "arXiv",
    primaryClass = "hep-ph",
    doi = "10.1007/JHEP10(2023)012",
    journal = "JHEP",
    volume = "10",
    pages = "012",
    year = "2023"
}

@article{Wu:2023twu,
    author = "Wu, Quan-feng and Xu, Xun-Jie",
    title = "{Shedding light on neutrino self-interactions with solar antineutrino searches}",
    eprint = "2308.15849",
    archivePrefix = "arXiv",
    primaryClass = "hep-ph",
    doi = "10.1088/1475-7516/2024/02/037",
    journal = "JCAP",
    volume = "02",
    pages = "037",
    year = "2024"
}

@article{Telalovic:2024cot,
    author = "Telalovic, Bernanda and Fiorillo, Damiano F. G. and Mart{\'\i}nez-Mirav{\'e}, Pablo and Vitagliano, Edoardo and Bustamante, Mauricio",
    title = "{The next galactic supernova can uncover mass and couplings of particles decaying to neutrinos}",
    eprint = "2406.15506",
    archivePrefix = "arXiv",
    primaryClass = "hep-ph",
    doi = "10.1088/1475-7516/2024/11/011",
    journal = "JCAP",
    volume = "11",
    pages = "011",
    year = "2024"
}

@article{Huang:2024tbo,
    author = "Huang, Jihong and Zhou, Shun",
    title = "{Helicity-changing decays of cosmological relic neutrinos}",
    eprint = "2407.04932",
    archivePrefix = "arXiv",
    primaryClass = "hep-ph",
    doi = "10.1088/1475-7516/2024/09/067",
    journal = "JCAP",
    volume = "09",
    pages = "067",
    year = "2024"
}

@article{Bai:2024kmt,
    author = "Bai, Weidong and Liao, Jiajun and Liu, Hongkai",
    title = "{Constraining neutrinophilic mediators at FASER{\ensuremath{\nu}}, FLArE, and FASER{\ensuremath{\nu}}2}",
    eprint = "2409.01826",
    archivePrefix = "arXiv",
    primaryClass = "hep-ph",
    doi = "10.1103/PhysRevD.110.115031",
    journal = "Phys. Rev. D",
    volume = "110",
    number = "11",
    pages = "115031",
    year = "2024"
}

@article{Machado:2025ltu,
    author = "Machado, Pedro A. N. and Wang, Isaac R. and Xu, Xun-Jie and Zhou, Bei",
    title = "{Widen the Resonance at Ultra-High Energies: Novel Probes of Neutrino Self-interactions in the High-Mass Regime}",
    eprint = "2512.00165",
    archivePrefix = "arXiv",
    primaryClass = "hep-ph",
    reportNumber = "FERMILAB-PUB-25-0853-T",
    month = "11",
    year = "2025"
}

@article{RoyChoudhury:2022rva,
    author = "Roy Choudhury, Shouvik and Hannestad, Steen and Tram, Thomas",
    title = "{Massive neutrino self-interactions and inflation}",
    eprint = "2207.07142",
    archivePrefix = "arXiv",
    primaryClass = "astro-ph.CO",
    doi = "10.1088/1475-7516/2022/10/018",
    journal = "JCAP",
    volume = "10",
    pages = "018",
    year = "2022"
}

@article{Craig:2024tky,
    author = "Craig, Nathaniel and Green, Daniel and Meyers, Joel and Rajendran, Surjeet",
    title = "{No {\ensuremath{\nu}}s is Good News}",
    eprint = "2405.00836",
    archivePrefix = "arXiv",
    primaryClass = "astro-ph.CO",
    reportNumber = "FERMILAB-PUB-24-0492-SQMS-V",
    doi = "10.1007/JHEP09(2024)097",
    journal = "JHEP",
    volume = "09",
    pages = "097",
    year = "2024"
}

@article{He:2023oke,
    author = "He, Adam and An, Rui and Ivanov, Mikhail M. and Gluscevic, Vera",
    title = "{Self-interacting neutrinos in light of large-scale structure data}",
    eprint = "2309.03956",
    archivePrefix = "arXiv",
    primaryClass = "astro-ph.CO",
    reportNumber = "MIT-CTP/5608",
    doi = "10.1103/PhysRevD.109.103527",
    journal = "Phys. Rev. D",
    volume = "109",
    number = "10",
    pages = "103527",
    year = "2024"
}

@article{Pal:2026cgj,
    author = "Pal, Sourav and Pal, Supratik",
    title = "{Neutrino self-interactions in post-reionization era: Lyman-$\alpha$, 21-cm and cross-spectra}",
    eprint = "2604.15287",
    archivePrefix = "arXiv",
    primaryClass = "astro-ph.CO",
    month = "4",
    year = "2026"
}

@article{Parashari:2026dxo,
    author = "Parashari, Priyank and Gluscevic, Vera and Zhang, Yue and Bird, Simeon and Ivanov, Mikhail M. and He, Adam",
    title = "{Ly{\ensuremath{\alpha}} forest bounds on sterile neutrino production via neutrino self-interactions}",
    eprint = "2602.17821",
    archivePrefix = "arXiv",
    primaryClass = "astro-ph.CO",
    reportNumber = "MIT-CTP/6004",
    month = "2",
    year = "2026"
}

@article{Perez-Castro:2026muj,
    author = "P{\'e}rez-Castro, Ivan and De-Santiago, Josue and Garcia-Arroyo, Gabriela and Venzor, Jorge and P{\'e}rez-Lorenzana, Abdel",
    title = "{Towards a complete scheme of cosmological neutrino self-interactions: Collision term for a wide range of mediator masses}",
    eprint = "2602.12477",
    archivePrefix = "arXiv",
    primaryClass = "hep-ph",
    month = "2",
    year = "2026"
}

@article{Montefalcone:2025ibh,
    author = "Montefalcone, Gabriele and Ghosh, Subhajit and Boddy, Kimberly K. and Ho, Daven Wei Ren and Tsai, Yuhsin",
    title = "{Directly probing neutrino interactions through CMB phase shift measurements}",
    eprint = "2509.20363",
    archivePrefix = "arXiv",
    primaryClass = "astro-ph.CO",
    reportNumber = "UTWI-23-2025",
    doi = "10.1103/lylf-3xyq",
    journal = "Phys. Rev. D",
    volume = "113",
    number = "2",
    pages = "023540",
    year = "2026"
}

@article{Libanore:2025ack,
    author = "Libanore, Sarah and Ghosh, Subhajit and Kovetz, Ely D. and Boddy, Kimberly K. and Raccanelli, Alvise",
    title = "{Joint 21-cm and CMB forecasts for constraining self-interacting massive neutrinos}",
    eprint = "2504.15348",
    archivePrefix = "arXiv",
    primaryClass = "astro-ph.CO",
    doi = "10.1103/tdms-6n76",
    journal = "Phys. Rev. D",
    volume = "112",
    number = "6",
    pages = "063502",
    year = "2025"
}

@article{He:2025jwp,
    author = "He, Adam and Ivanov, Mikhail M. and Bird, Simeon and An, Rui and Gluscevic, Vera",
    title = "{Fresh look at neutrino self-interactions with the Lyman-{\ensuremath{\alpha}} forest: Constraints from EFT and PRIYA simulations}",
    eprint = "2503.15592",
    archivePrefix = "arXiv",
    primaryClass = "astro-ph.CO",
    reportNumber = "MIT-CTP/5854",
    doi = "10.1103/wzpy-p7w8",
    journal = "Phys. Rev. D",
    volume = "112",
    number = "6",
    pages = "063540",
    year = "2025"
}

@article{Kaplan:2024ydw,
    author = "Kaplan, David E. and Luo, Xuheng and Rajendran, Surjeet",
    title = "{Probing long-range forces between neutrinos with cosmic structures}",
    eprint = "2412.20766",
    archivePrefix = "arXiv",
    primaryClass = "hep-ph",
    reportNumber = "FERMILAB-PUB-24-0977-SQMS-V",
    doi = "10.1103/PhysRevD.111.055019",
    journal = "Phys. Rev. D",
    volume = "111",
    number = "5",
    pages = "055019",
    year = "2025"
}

@article{Beacom:2004yd,
    author = "Beacom, John F. and Bell, Nicole F. and Dodelson, Scott",
    title = "{Neutrinoless universe}",
    eprint = "astro-ph/0404585",
    archivePrefix = "arXiv",
    reportNumber = "FERMILAB-PUB-04-050-A",
    doi = "10.1103/PhysRevLett.93.121302",
    journal = "Phys. Rev. Lett.",
    volume = "93",
    pages = "121302",
    year = "2004"
}

@article{Leal:2025eou,
    author = "Leal, Luighi P. S. and Naredo-Tuero, Daniel and Funchal, Renata Zukanovich",
    title = "{Cosmogenic neutrinos as probes of new physics}",
    eprint = "2504.10576",
    archivePrefix = "arXiv",
    primaryClass = "hep-ph",
    reportNumber = "IFT-UAM/CSIC-25-38",
    doi = "10.1007/JHEP08(2025)057",
    journal = "JHEP",
    volume = "08",
    pages = "057",
    year = "2025"
}

@article{Creque-Sarbinowski:2020qhz,
    author = "Creque-Sarbinowski, Cyril and Hyde, Jeffrey and Kamionkowski, Marc",
    title = "{Resonant neutrino self-interactions}",
    eprint = "2005.05332",
    archivePrefix = "arXiv",
    primaryClass = "hep-ph",
    doi = "10.1103/PhysRevD.103.023527",
    journal = "Phys. Rev. D",
    volume = "103",
    number = "2",
    pages = "023527",
    year = "2021"
}

@article{Ehring:2024mjx,
    author = "Ehring, Jakob and Abbar, Sajad and Janka, Hans-Thomas and Raffelt, Georg and Nakamura, Ko and Kotake, Kei",
    title = "{Gravitational-Wave Signatures of Nonstandard Neutrino Properties in Collapsing Stellar Cores}",
    eprint = "2412.02750",
    archivePrefix = "arXiv",
    primaryClass = "astro-ph.HE",
    doi = "10.1103/rv17-jm6g",
    journal = "Phys. Rev. Lett.",
    volume = "136",
    number = "2",
    pages = "021201",
    year = "2026"
}

@article{Brinckmann:2020bcn,
    author = "Brinckmann, Thejs and Chang, Jae Hyeok and LoVerde, Marilena",
    title = "{Self-interacting neutrinos, the Hubble parameter tension, and the cosmic microwave background}",
    eprint = "2012.11830",
    archivePrefix = "arXiv",
    primaryClass = "astro-ph.CO",
    reportNumber = "YITP-SB-2020-40",
    doi = "10.1103/PhysRevD.104.063523",
    journal = "Phys. Rev. D",
    volume = "104",
    number = "6",
    pages = "063523",
    year = "2021"
}

@article{Cyr-Racine:2013jua,
    author = "Cyr-Racine, Francis-Yan and Sigurdson, Kris",
    title = "{Limits on Neutrino-Neutrino Scattering in the Early Universe}",
    eprint = "1306.1536",
    archivePrefix = "arXiv",
    primaryClass = "astro-ph.CO",
    doi = "10.1103/PhysRevD.90.123533",
    journal = "Phys. Rev. D",
    volume = "90",
    number = "12",
    pages = "123533",
    year = "2014"
}

@article{Barenboim:2019tux,
    author = "Barenboim, Gabriela and Denton, Peter B. and Oldengott, Isabel M.",
    title = "{Constraints on inflation with an extended neutrino sector}",
    eprint = "1903.02036",
    archivePrefix = "arXiv",
    primaryClass = "astro-ph.CO",
    doi = "10.1103/PhysRevD.99.083515",
    journal = "Phys. Rev. D",
    volume = "99",
    number = "8",
    pages = "083515",
    year = "2019"
}

@article{Taule:2022jrz,
    author = "Taule, Petter and Escudero, Miguel and Garny, Mathias",
    title = "{Global view of neutrino interactions in cosmology: The free streaming window as seen by Planck}",
    eprint = "2207.04062",
    archivePrefix = "arXiv",
    primaryClass = "astro-ph.CO",
    reportNumber = "TUM-HEP-1406/22",
    doi = "10.1103/PhysRevD.106.063539",
    journal = "Phys. Rev. D",
    volume = "106",
    number = "6",
    pages = "063539",
    year = "2022"
}

@article{Das:2020xke,
    author = "Das, Anirban and Ghosh, Subhajit",
    title = "{Flavor-specific interaction favors strong neutrino self-coupling in the early universe}",
    eprint = "2011.12315",
    archivePrefix = "arXiv",
    primaryClass = "astro-ph.CO",
    reportNumber = "SLAC-PUB-17547",
    doi = "10.1088/1475-7516/2021/07/038",
    journal = "JCAP",
    volume = "07",
    pages = "038",
    year = "2021"
}

@article{Venzor:2023aka,
    author = "Venzor, Jorge and Garcia-Arroyo, Gabriela and De-Santiago, Josue and P{\'e}rez-Lorenzana, Abdel",
    title = "{Resonant neutrino self-interactions and the H0 tension}",
    eprint = "2303.12792",
    archivePrefix = "arXiv",
    primaryClass = "astro-ph.CO",
    doi = "10.1103/PhysRevD.108.043536",
    journal = "Phys. Rev. D",
    volume = "108",
    number = "4",
    pages = "043536",
    year = "2023"
}

@article{Abbar:2022jdm,
    author = "Abbar, Sajad",
    title = "{Nonstandard neutrino self-interactions can cause neutrino flavor equipartition inside the supernova core}",
    eprint = "2208.06023",
    archivePrefix = "arXiv",
    primaryClass = "astro-ph.HE",
    reportNumber = "MPP-2022-108",
    doi = "10.1103/PhysRevD.107.103002",
    journal = "Phys. Rev. D",
    volume = "107",
    number = "10",
    pages = "103002",
    year = "2023"
}

@article{Lei:2019nma,
    author = "Lei, Minjie and Steinberg, Noah and Wells, James D.",
    title = "{Probing Non-Standard Neutrino Interactions with Supernova Neutrinos at Hyper-K}",
    eprint = "1907.01059",
    archivePrefix = "arXiv",
    primaryClass = "hep-ph",
    doi = "10.1007/JHEP01(2020)179",
    journal = "JHEP",
    volume = "01",
    pages = "179",
    year = "2020"
}

@article{Murase:2019xqi,
    author = "Murase, Kohta and Shoemaker, Ian M.",
    title = "{Neutrino Echoes from Multimessenger Transient Sources}",
    eprint = "1903.08607",
    archivePrefix = "arXiv",
    primaryClass = "hep-ph",
    doi = "10.1103/PhysRevLett.123.241102",
    journal = "Phys. Rev. Lett.",
    volume = "123",
    number = "24",
    pages = "241102",
    year = "2019"
}

@article{Araki:2015mya,
    author = "Araki, Takeshi and Kaneko, Fumihiro and Ota, Toshihiko and Sato, Joe and Shimomura, Takashi",
    title = "{MeV scale leptonic force for cosmic neutrino spectrum and muon anomalous magnetic moment}",
    eprint = "1508.07471",
    archivePrefix = "arXiv",
    primaryClass = "hep-ph",
    reportNumber = "UME-PP-002, STUPP-15-223",
    doi = "10.1103/PhysRevD.93.013014",
    journal = "Phys. Rev. D",
    volume = "93",
    number = "1",
    pages = "013014",
    year = "2016"
}

@article{Bilenky:1994ma,
    author = "Bilenky, Mikhail S. and Santamaria, Arcadi",
    title = "{Bounding effective operators at the one loop level: The Case of four fermion neutrino interactions}",
    eprint = "hep-ph/9405427",
    archivePrefix = "arXiv",
    reportNumber = "CERN-TH-7230-94",
    doi = "10.1016/0370-2693(94)00961-9",
    journal = "Phys. Lett. B",
    volume = "336",
    pages = "91--99",
    year = "1994"
}

@article{Bilenky:1992xn,
    author = "Bilenky, Mikhail S. and Bilenky, Samoil M. and Santamaria, A.",
    title = "{Invisible width of the Z boson and 'secret' neutrino-neutrino interactions}",
    reportNumber = "FTUV-92-40, BI-TP-92-36",
    doi = "10.1016/0370-2693(93)90703-K",
    journal = "Phys. Lett. B",
    volume = "301",
    pages = "287--291",
    year = "1993"
}

@article{Fong:2025xhh,
    author = "Fong, Chee Sheng and Porto, Yago",
    title = "{$N_{\textrm{eff}}$ Constraint on Pseudo-Dirac Neutrinos}",
    eprint = "2512.19782",
    archivePrefix = "arXiv",
    primaryClass = "hep-ph",
    month = "12",
    year = "2025"
}

@article{MacDonald:2025jbm,
    author = {MacDonald, Miller and Carloni, Kiara and Arg{\"u}elles, Carlos A. and Mart{\'\i}nez-Soler, Ivan and Alves Batista, Rafael},
    title = "{Exploring New Propagation Scales With Galactic Neutrinos}",
    eprint = "2512.10744",
    archivePrefix = "arXiv",
    primaryClass = "hep-ph",
    month = "12",
    year = "2025"
}

@article{Dev:2021axj,
    author = "Dev, P. S. Bhupal and Dutta, Bhaskar and Ghosh, Tathagata and Han, Tao and Qin, Han and Zhang, Yongchao",
    title = "{Leptonic scalars and collider signatures in a UV-complete model}",
    eprint = "2109.04490",
    archivePrefix = "arXiv",
    primaryClass = "hep-ph",
    reportNumber = "PITT-PACC-2108, MI-TH-2112, HRI-RECAPP-2021-007",
    doi = "10.1007/JHEP03(2022)068",
    journal = "JHEP",
    volume = "03",
    pages = "068",
    year = "2022"
}

@article{Agashe:2024owh,
    author = "Agashe, Kaustubh and Airen, Sagar and Franceschini, Roberto and Kim, Doojin and Kotwal, Ashutosh V. and Ricci, Lorenzo and Sathyan, Deepak",
    title = "{{\textquotedblleft}Unification{\textquotedblright} of BSM searches and SM measurements: the case of lepton+[inline-graphic not available: see fulltext] and m$_{W}$}",
    eprint = "2404.17574",
    archivePrefix = "arXiv",
    primaryClass = "hep-ph",
    doi = "10.1007/JHEP02(2025)139",
    journal = "JHEP",
    volume = "02",
    pages = "139",
    year = "2025"
}

@article{Kolb:1987qy,
    author = "Kolb, Edward W. and Turner, Michael S.",
    title = "{Supernova 1987A and the Secret Interactions of Neutrinos}",
    reportNumber = "FERMILAB-PUB-87-110-A",
    doi = "10.1103/PhysRevD.36.2895",
    journal = "Phys. Rev. D",
    volume = "36",
    pages = "2895",
    year = "1987"
}

@article{Heurtier:2016otg,
    author = "Heurtier, Lucien and Zhang, Yongchao",
    title = "{Supernova Constraints on Massive (Pseudo)Scalar Coupling to Neutrinos}",
    eprint = "1609.05882",
    archivePrefix = "arXiv",
    primaryClass = "hep-ph",
    reportNumber = "ULB-TH-16-16",
    doi = "10.1088/1475-7516/2017/02/042",
    journal = "JCAP",
    volume = "02",
    pages = "042",
    year = "2017"
}

@article{Fiorillo:2023ytr,
    author = "Fiorillo, Damiano F. G. and Raffelt, Georg G. and Vitagliano, Edoardo",
    title = "{Large Neutrino Secret Interactions Have a Small Impact on Supernovae}",
    eprint = "2307.15115",
    archivePrefix = "arXiv",
    primaryClass = "hep-ph",
    doi = "10.1103/PhysRevLett.132.021002",
    journal = "Phys. Rev. Lett.",
    volume = "132",
    number = "2",
    pages = "021002",
    year = "2024"
}

@article{Fiorillo:2023cas,
    author = "Fiorillo, Damiano F. G. and Raffelt, Georg G. and Vitagliano, Edoardo",
    title = "{Supernova emission of secretly interacting neutrino fluid: Theoretical foundations}",
    eprint = "2307.15122",
    archivePrefix = "arXiv",
    primaryClass = "hep-ph",
    doi = "10.1103/PhysRevD.109.023017",
    journal = "Phys. Rev. D",
    volume = "109",
    number = "2",
    pages = "023017",
    year = "2024"
}

@article{Hufnagel:2021pso,
    author = "Hufnagel, Marco and Xu, Xun-Jie",
    title = "{Dark matter produced from neutrinos}",
    eprint = "2110.09883",
    archivePrefix = "arXiv",
    primaryClass = "hep-ph",
    reportNumber = "ULB-TH/21-14",
    doi = "10.1088/1475-7516/2022/01/043",
    journal = "JCAP",
    volume = "01",
    number = "01",
    pages = "043",
    year = "2022"
}

@article{Jana:2025vyb,
    author = "Jana, Sudip and Manna, Sudip and K, Vishnu P.",
    title = "{Gravitational Wave Signature and the Nature of Neutrino Masses: Majorana, Dirac, or Pseudo-Dirac?}",
    eprint = "2509.10456",
    archivePrefix = "arXiv",
    primaryClass = "hep-ph",
    reportNumber = "HRI-RECAPP-2025-09, MS-TP-25-19",
    doi = "10.1016/j.physletb.2026.140476",
    journal = "Phys. Lett. B",
    volume = "877",
    pages = "140476",
    year = "2026"
}

@article{Dixit:2024ldv,
    author = "Dixit, Khushboo and Miranda, Luis Salvador and Razzaque, Soebur",
    title = "{Searching for pseudo-dirac neutrinos from astrophysical sources in IceCube data}",
    eprint = "2406.06476",
    archivePrefix = "arXiv",
    primaryClass = "astro-ph.HE",
    doi = "10.1140/epjc/s10052-025-15207-5",
    journal = "Eur. Phys. J. C",
    volume = "85",
    number = "12",
    pages = "1481",
    year = "2025"
}

@article{Rink:2022nvw,
    author = "Rink, Thomas and Sen, Manibrata",
    title = "{Constraints on pseudo-Dirac neutrinos using high-energy neutrinos from NGC 1068}",
    eprint = "2211.16520",
    archivePrefix = "arXiv",
    primaryClass = "hep-ph",
    doi = "10.1016/j.physletb.2024.138558",
    journal = "Phys. Lett. B",
    volume = "851",
    pages = "138558",
    year = "2024"
}

@article{Wolfenstein:1981kw,
    author = "Wolfenstein, Lincoln",
    title = "{Different Varieties of Massive Dirac Neutrinos}",
    reportNumber = "COO-3066-164",
    doi = "10.1016/0550-3213(81)90096-1",
    journal = "Nucl. Phys. B",
    volume = "186",
    pages = "147--152",
    year = "1981"
}

@article{Petcov:1982ya,
    author = "Petcov, S. T.",
    title = "{On Pseudodirac Neutrinos, Neutrino Oscillations and Neutrinoless Double beta Decay}",
    doi = "10.1016/0370-2693(82)91246-1",
    journal = "Phys. Lett. B",
    volume = "110",
    pages = "245--249",
    year = "1982"
}

@article{Valle:1983dk,
    author = "Valle, J. W. F. and Singer, M.",
    title = "{Lepton Number Violation With Quasi Dirac Neutrinos}",
    reportNumber = "RL-83-018",
    doi = "10.1103/PhysRevD.28.540",
    journal = "Phys. Rev. D",
    volume = "28",
    pages = "540",
    year = "1983"
}

@article{Martinez-Soler:2021unz,
    author = "Martinez-Soler, Ivan and Perez-Gonzalez, Yuber F. and Sen, Manibrata",
    title = "{Signs of pseudo-Dirac neutrinos in SN1987A data}",
    eprint = "2105.12736",
    archivePrefix = "arXiv",
    primaryClass = "hep-ph",
    reportNumber = "FERMILAB-PUB-21-225-T, NUHEP-TH/21-05, N3AS-21-009",
    doi = "10.1103/PhysRevD.105.095019",
    journal = "Phys. Rev. D",
    volume = "105",
    number = "9",
    pages = "095019",
    year = "2022"
}

@article{Franklin:2023diy,
    author = "Franklin, Jack and Perez-Gonzalez, Yuber F. and Turner, Jessica",
    title = "{JUNO as a probe of the pseudo-Dirac nature using solar neutrinos}",
    eprint = "2304.05418",
    archivePrefix = "arXiv",
    primaryClass = "hep-ph",
    reportNumber = "IPPP/23/20",
    doi = "10.1103/PhysRevD.108.035010",
    journal = "Phys. Rev. D",
    volume = "108",
    number = "3",
    pages = "035010",
    year = "2023"
}

@article{Doi:1983wu,
    author = "Doi, Masaru and Kenmoku, Masakatsu and Kotani, Tsuneyuki and Nishiura, Hiroyuki and Takasugi, Eiichi",
    title = "{PSEUDODIRAC NEUTRINO}",
    reportNumber = "OS-GE-83-48",
    doi = "10.1143/PTP.70.1331",
    journal = "Prog. Theor. Phys.",
    volume = "70",
    pages = "1331",
    year = "1983"
}

@article{Giunti:1992hk,
    author = "Giunti, C. and Kim, C. W. and Lee, U. W.",
    title = "{Oscillations of pseudoDirac neutrinos and the solar neutrino problem}",
    eprint = "hep-ph/9205214",
    archivePrefix = "arXiv",
    reportNumber = "JHU-TIPAC-920011, DFTT-12-92",
    doi = "10.1103/PhysRevD.46.3034",
    journal = "Phys. Rev. D",
    volume = "46",
    pages = "3034--3039",
    year = "1992"
}

@article{Cirelli:2004cz,
    author = "Cirelli, Marco and Marandella, Guido and Strumia, Alessandro and Vissani, Francesco",
    title = "{Probing oscillations into sterile neutrinos with cosmology, astrophysics and experiments}",
    eprint = "hep-ph/0403158",
    archivePrefix = "arXiv",
    reportNumber = "IFUP-TH-2004-2",
    doi = "10.1016/j.nuclphysb.2004.11.056",
    journal = "Nucl. Phys. B",
    volume = "708",
    pages = "215--267",
    year = "2005"
}

@article{Anamiati:2017rxw,
    author = "Anamiati, G. and Fonseca, R. M. and Hirsch, M.",
    title = "{Quasi Dirac neutrino oscillations}",
    eprint = "1710.06249",
    archivePrefix = "arXiv",
    primaryClass = "hep-ph",
    reportNumber = "IFIC-17-45",
    doi = "10.1103/PhysRevD.97.095008",
    journal = "Phys. Rev. D",
    volume = "97",
    number = "9",
    pages = "095008",
    year = "2018"
}

@article{deGouvea:2021ymm,
    author = "de Gouv{\^e}a, Andr{\'e} and McGinness, Emma and Martinez-Soler, Ivan and Perez-Gonzalez, Yuber F.",
    title = "{pp solar neutrinos at DARWIN}",
    eprint = "2111.02421",
    archivePrefix = "arXiv",
    primaryClass = "hep-ph",
    reportNumber = "NUHEP-TH/21-17, FERMILAB-PUB-21-560-T, IPPP/21/46",
    doi = "10.1103/PhysRevD.106.096017",
    journal = "Phys. Rev. D",
    volume = "106",
    number = "9",
    pages = "096017",
    year = "2022"
}

@article{DeGouvea:2020ang,
    author = "De Gouv{\^e}a, Andr{\'e} and Martinez-Soler, Ivan and Perez-Gonzalez, Yuber F. and Sen, Manibrata",
    title = "{Fundamental physics with the diffuse supernova background neutrinos}",
    eprint = "2007.13748",
    archivePrefix = "arXiv",
    primaryClass = "hep-ph",
    reportNumber = "NUHEP-TH/20-08, FERMILAB-PUB-20-353-T",
    doi = "10.1103/PhysRevD.102.123012",
    journal = "Phys. Rev. D",
    volume = "102",
    pages = "123012",
    year = "2020"
}

@article{Crocker:1999yw,
    author = "Crocker, R. M. and Melia, F. and Volkas, R. R.",
    title = "{Oscillating neutrinos from the galactic center}",
    eprint = "astro-ph/9911292",
    archivePrefix = "arXiv",
    reportNumber = "UM-P-99-40, RCHEP-99-08",
    doi = "10.1086/317350",
    journal = "Astrophys. J. Suppl.",
    volume = "130",
    pages = "339--350",
    year = "2000"
}

@article{Crocker:2001zs,
    author = "Crocker, Roland M. and Melia, Fulvio and Volkas, Raymond R.",
    title = "{Searching for long wavelength neutrino oscillations in the distorted neutrino spectrum of galactic supernova remnants}",
    eprint = "astro-ph/0106090",
    archivePrefix = "arXiv",
    doi = "10.1086/340278",
    journal = "Astrophys. J. Suppl.",
    volume = "141",
    pages = "147--155",
    year = "2002"
}

@article{Esmaili:2012ac,
    author = "Esmaili, Arman and Farzan, Yasaman",
    title = "{Implications of the Pseudo-Dirac Scenario for Ultra High Energy Neutrinos from GRBs}",
    eprint = "1208.6012",
    archivePrefix = "arXiv",
    primaryClass = "hep-ph",
    doi = "10.1088/1475-7516/2012/12/014",
    journal = "JCAP",
    volume = "12",
    pages = "014",
    year = "2012"
}

@article{Ansarifard:2022kvy,
    author = "Ansarifard, Saeed and Farzan, Yasaman",
    title = "{Revisiting pseudo-Dirac neutrino scenario after recent solar neutrino data}",
    eprint = "2211.09105",
    archivePrefix = "arXiv",
    primaryClass = "hep-ph",
    doi = "10.1103/PhysRevD.107.075029",
    journal = "Phys. Rev. D",
    volume = "107",
    number = "7",
    pages = "075029",
    year = "2023"
}

@article{Joshipura:2013yba,
    author = "Joshipura, Anjan S. and Mohanty, Subhendra and Pakvasa, Sandip",
    title = "{Pseudo-Dirac neutrinos via a mirror world and depletion of ultrahigh energy neutrinos}",
    eprint = "1307.5712",
    archivePrefix = "arXiv",
    primaryClass = "hep-ph",
    doi = "10.1103/PhysRevD.89.033003",
    journal = "Phys. Rev. D",
    volume = "89",
    number = "3",
    pages = "033003",
    year = "2014"
}

@article{Brdar:2018tce,
    author = "Brdar, Vedran and Hansen, Rasmus S. L.",
    title = "{IceCube Flavor Ratios with Identified Astrophysical Sources: Towards Improving New Physics Testability}",
    eprint = "1812.05541",
    archivePrefix = "arXiv",
    primaryClass = "hep-ph",
    doi = "10.1088/1475-7516/2019/02/023",
    journal = "JCAP",
    volume = "02",
    pages = "023",
    year = "2019"
}

@article{Perez-Gonzalez:2023llw,
    author = "Perez-Gonzalez, Yuber F. and Sen, Manibrata",
    title = "{From Dirac to Majorana: The cosmic neutrino background capture rate in the minimally extended Standard Model}",
    eprint = "2308.05147",
    archivePrefix = "arXiv",
    primaryClass = "hep-ph",
    doi = "10.1103/PhysRevD.109.023022",
    journal = "Phys. Rev. D",
    volume = "109",
    number = "2",
    pages = "023022",
    year = "2024"
}

@article{Anamiati:2019maf,
    author = "Anamiati, G. and De Romeri, V. and Hirsch, M. and Ternes, C. A. and T{\'o}rtola, M.",
    title = "{Quasi-Dirac neutrino oscillations at DUNE and JUNO}",
    eprint = "1907.00980",
    archivePrefix = "arXiv",
    primaryClass = "hep-ph",
    doi = "10.1103/PhysRevD.100.035032",
    journal = "Phys. Rev. D",
    volume = "100",
    number = "3",
    pages = "035032",
    year = "2019"
}

@article{Babu:2022ikf,
    author = "Babu, K. S. and He, Xiao-Gang and Su, Mingxian and Thapa, Anil",
    title = "{Naturally light Dirac and pseudo-Dirac neutrinos from left-right symmetry}",
    eprint = "2205.09127",
    archivePrefix = "arXiv",
    primaryClass = "hep-ph",
    doi = "10.1007/JHEP08(2022)140",
    journal = "JHEP",
    volume = "08",
    pages = "140",
    year = "2022"
}

@article{IceCube:2025uyt,
    author = "Balagopal V., Aswathi and others",
    collaboration = "IceCube",
    title = "{Measurement of the Three-Flavor Composition of Astrophysical Neutrinos with Contained IceCube Events}",
    eprint = "2507.07212",
    archivePrefix = "arXiv",
    primaryClass = "astro-ph.HE",
    reportNumber = "PoS-ICRC2025-983",
    doi = "10.22323/1.501.0983",
    journal = "PoS",
    volume = "ICRC2025",
    pages = "983",
    year = "2025"
}

@article{IceCube:2025dlr,
    author = "Balagopal V., Aswathi and others",
    collaboration = "IceCube",
    title = "{Measurement of the Diffuse Astrophysical Neutrino Spectrum above a TeV with All Flavor Starting Events in IceCube}",
    eprint = "2507.06002",
    archivePrefix = "arXiv",
    primaryClass = "astro-ph.HE",
    reportNumber = "PoS-ICRC2025-985",
    doi = "10.22323/1.501.0985",
    journal = "PoS",
    volume = "ICRC2025",
    pages = "985",
    year = "2025"
}

@article{IceCube:2023fgt,
    author = "Abbasi, Rasha and others",
    collaboration = "IceCube",
    title = "{Summary of IceCube tau neutrino searches and flavor composition measurements of the diffuse astrophysical neutrino flux}",
    eprint = "2308.15213",
    archivePrefix = "arXiv",
    primaryClass = "astro-ph.HE",
    reportNumber = "PoS-ICRC2023-1122",
    doi = "10.22323/1.444.1122",
    journal = "PoS",
    volume = "ICRC2023",
    pages = "1122",
    year = "2023"
}

@article{IceCube:2024nhk,
    author = "Abbasi, R. and others",
    collaboration = "IceCube",
    title = "{Observation of Seven Astrophysical Tau Neutrino Candidates with IceCube}",
    eprint = "2403.02516",
    archivePrefix = "arXiv",
    primaryClass = "astro-ph.HE",
    doi = "10.1103/PhysRevLett.132.151001",
    journal = "Phys. Rev. Lett.",
    volume = "132",
    number = "15",
    pages = "151001",
    year = "2024"
}

@article{IceCube:2020fpi,
    author = "Abbasi, R. and others",
    collaboration = "IceCube",
    title = "{Detection of astrophysical tau neutrino candidates in IceCube}",
    eprint = "2011.03561",
    archivePrefix = "arXiv",
    primaryClass = "hep-ex",
    doi = "10.1140/epjc/s10052-022-10795-y",
    journal = "Eur. Phys. J. C",
    volume = "82",
    number = "11",
    pages = "1031",
    year = "2022"
}

\end{document}